\documentclass[11pt]{amsart}

\usepackage{amssymb}
\usepackage{epsfig}
\usepackage{comment}
\usepackage{esint}
\usepackage{algorithm}
\usepackage{algorithmic}
\usepackage{subcaption}

\newtheorem{thm}{Theorem}[section]

\theoremstyle{definition}

\newtheorem{example}[thm]{Example}

\theoremstyle{remark}

\newcommand{\curl}{{\rm curl}\,}

\newcommand{\mbs}[1]{\boldsymbol{#1}}

\graphicspath {
    {./}
    {./Figures/}
}

\begin{document}

\title{A line-free method of monopoles for 3D dislocation dynamics}

\author{A.~Deffo, M.~P.~Ariza and M. Ortiz}

\begin{abstract}
We develop an approximation scheme for three-dimensional dislocation dynamics in which the dislocation line density is concentrated at points, or {\sl monopoles}. Every monopole carries a Burgers vector and an element of line. The monopoles move according to mobility kinetics driven by elastic and applied forces. The divergence constraint, expressing the requirement that the monopoles approximate a boundary, is enforced weakly. The fundamental difference with traditional approximation schemes based on segments is that in the present approach an explicit linear connectivity, or 'sequence', between the monopoles need not be defined. Instead, the monopoles move as an unstructured point set subject to the weak divergence constraint. In this sense, the new paradigm is 'line-free', i.~e., it sidesteps the need to track dislocation lines. This attribute offers significant computational advantages in terms of simplicity, robustness and efficiency, as demonstrated by means of selected numerical examples.
\end{abstract}

\maketitle


\section{Introduction}

The plastic deformation of crystals is the macroscopic effect of the cooperative motion of large ensembles of lattice dislocations. From a geometric point of view, dislocations are line-like lattice defects that demarcate the boundary of areas of constant crystallographic slip on slip planes. By the discrete nature of crystal lattices, crystallographic slip is in turn constrained to occur in quanta of Burgers vector and, therefore, the dislocation lines carry a quantized Burgers vector 'charge'. Since dislocations are boundaries, they must themselves have no boundary, i.~e., they must form closed loops, branch according to Frank's rule or terminate at the boundary of the solid. The motion of dislocation is driven by the Peach-K\"ohler force induced by the applied stresses and by the elastic interaction between dislocation segments and is controlled by dislocation mobility. As they move, dislocations may undergo line stretching, pair annihilation, dislocation reactions, pinning-depinning interactions and other complex geometrical and topological transitions.

Given their fundamental role as agents of plastic deformation, dislocations have been extensively studied both experimentally, analytically and computationally. In three dimensions, the prevailing computational paradigm is to regard dislocations as lines and to discretize them into connected segments (cf., e.~g., \cite{BulatovCai:2006}). The motion of the segments is then governed by mobility and driven by a suitably approximated and regularized elastic force. Complex rules must be implemented in order to account for short-range segment-segment interactions, dislocation reactions and topological transitions.  The complexity of this approach derives largely from the need for dislocation segments to remain linearly connected in order for dislocation lines to remain boundaries, or 'divergence-free'. The resulting dynamics inevitably leads to complex line entanglements that are difficult to track and negotiate effectively.

In this work, we develop a 'line-free' dislocation dynamics paradigm differing fundamentally from traditional line-based schemes in that the dislocation density is concentrated at points, or {\sl monopoles}, and an explicit linear connectivity, or 'sequence', between the monopoles is not defined or enforced. Instead, the monopoles move as an unstructured point set subject to a weak divergence constraint. In this sense, the new paradigm sidesteps the need to track dislocation lines, an attribute that offers significant computational advantages in terms of simplicity and efficiency. In particular, it affords an extension to three dimensions of the wealth of point-dislocation methods that have been developed and extensively applied in two dimensions (cf.~the seminal paper \cite{lubarda:1993d} and derivative works thereof, too numerous to list or even summarize here).

The basis for the new paradigm is a reformulation of dislocation dynamics as a problem of {\sl transport of measures}\footnote{Here and throughout this work, the term {\sl measure}, which is standard in mathematics (cf., e.~g., \cite{AmbrosioFP}), is used simply to emphasize that certain fields, such as the plastic deformation of the dislocation density, are not regular functions but are instead concentrated on surfaces or lines and are characterized by their action on appropriate test functions; cf.~also \cite{RN3} for a rigorous treatment of plastic deformations and dislocation densities as currents.} (cf., e.~g., \cite{Villani:2003} for background on optimal transport theory). Whereas the transport nature of dislocation dynamics has long been appreciated (cf., e.~g., \cite{mura:1987}), the literature to date is largely restricted to 'continuously distributed dislocations', or dislocation densities described by regular functions. However, as already noted, dislocations are line defects and, as such, measures and not functions. This distinction is not insignificant but fundamental. Indeed, the reformulation of dislocation transport theory from functions to measures affords a number of essential extensions and provides the basis for the present work: i) it enables the direct treatment of dislocation as line objects, as opposed to 'diffuse' or 'distributed' functions; ii) it leads to notions of {\sl weak solution} and of weak satisfaction of the divergence constraint that, in particular, open the way for spatial approximation schemes other than segments; iii) it introduces concepts from transport theory such as transport maps and push-forward operations enabling exact geometrical updates; iv) it supports time discretizations resulting in incremental minimum problems for energy-dissipation functionals; and v) it enables discretizations of the dislocation density within spaces of measures, e.~g., by means of Dirac masses or 'monopoles', which would otherwise be undefined in functional spaces.


Within this measure-theoretical framework, monopoles suggest themselves as a canonical approximation owing to the density properties of Dirac masses in spaces of measures. Specifically, every monopole carries a Burgers vector and an element of line. The monopoles then move according to mobility kinetics driven by elastic and applied forces. The divergence constraint, expressing the requirement that the monopoles approximate a boundary, is enforced weakly. Most importantly, at no point in the approximation or in the calculations an explicit linear connectivity, or 'sequence', between the monopoles is defined or enforced. The monopoles instead move as an unstructured point set subject to the weak divergence constraint. In this sense, the new paradigm is 'line-free', i.~e., it sidesteps the need to track dislocation lines, an attribute that offers significant computational advantages in terms of simplicity and efficiency.

The time discretization developed in the present work parallels the pioneering work of Jordan, Kinderlehrer and Otto \cite{JKO:1998} on transport of scalar measures, and reduces the problem to the successive minimization of an incremental energy-dissipation function. The solutions of these minimum problems define a time-wise sequence of incremental transport maps. Pushforward by the incremental transport maps then supplies a geometrically exact update for the dislocation measure that, in particular, preserves the divergence-free constraint. The spatial discretization of the transport maps in turn mirrors similar mesh-free discretization schemes proposed in the context of solid and fluid flows \cite{OTM:2010} and diffusion \cite{FPO:2017}. Because gradients are required by the geometrical updates, the discretization of the transport map must be conforming. We specifically use a max-ent interpolation scheme \cite{ArroyoOrtiz:2006} that, in keeping with the 'line-free' character of the monopole approximation, does not require sequencing of the monopoles.

The paper is organized as follows. In Section~\ref{zLu8oa}, we begin with a succinct review of dislocation dynamics as a problem of transport of measures. The representation of dislocations as line currents\footnote{Currents arise in geometrical measure theory (cf., e.~g., \cite{Morgan:1988, Giaquinta:1998}) as special measures, indeed distributions, which generalize the Dirac delta distribution to lines and surfaces and carry vector or tensor-valued charge capable of acting on general forms.} \cite{AO:2005, RN1, RN3} is summarized in Section~\ref{hya0KL}. Of particular concern is the formulation of transport equations in a weak form that is applicable to dislocation densities that are concentrated on lines or points and that are not differentiable in the sense of ordinary functions, cf.~Section~\ref{7vXTHq}. An additional focus concerns the reformulation of the transport problem in terms of transport maps and geometrically-exact push-forward operations that pave the way for time discretization, cf.~Section~\ref{bR5MnR}. In order to close the transport problem, a mobility law delivering the instantaneous dislocation velocity needs to be specified. Section~\ref{197IIa} develops the conventional energetic viewpoint that the dislocation motion is a {\sl gradient flow} driven by energetic driving forces and governed by kinetics. We specifically focus on variational formulations of mobility, Section~\ref{1KnY2M}, and energy, Sections~\ref{m0X7vX} and~\ref{fo0Fle}, that provide the basis for the incremental minimum problems developed subsequently. Section~\ref{m0X7vX} focuses on the representation of the energy as a function of the dislocation density. In particular, we present a derivation based on the Helmholtz decomposition that generalizes an earlier derivation of Mura \cite{RN2} to arbitrary domains and nonlinear behavior. In Section~\ref{fo0Fle} we deal with the logarithmic divergence of the energy by means of an explicit core regularization based on gradient elasticity. In Section~\ref{3x2efD}, we address issues of approximation, including time discretization, discretization of the dislocation measure and discretization of the transport map. Following \cite{JKO:1998, OTM:2010, FPO:2017}, time discretization is effected by defining an incremental energy-dissipation function for the transport map, with the update of the dislocation measure following as a by-product, cf.~Section~\ref{BL9NY0}. Exploiting the property that the incremental energy-dissipation function is well-defined for general dislocation measures, we proceed to discretize the dislocation measure by means of Dirac masses, or monopoles, cf.~Section~\ref{kvW88u}. We additionally discretize the transport map by means of mesh-free max-ent interpolation \cite{ArroyoOrtiz:2006},  cf.~Section~\ref{KrIth5}, in keeping with the line-free character of the approach. The incremental equations of equilibrium finally follow by rendering the incremental energy-dissipation functional stationary with respect to the monopole positions, cf.~Section~\ref{5UQCjl}. The general structure of the resulting dislocation dynamics solver and selected issues of implementation are discussed in Section~\ref{Xe0cjC}. The developments to this point are based on the assumption that the incremental transport map is continuous and, therefore, the dislocation measure undergoes no topological transitions. Section~\ref{O5HY85} addresses two common topological transitions, in the context of the monopole approximation, namely, dislocation reactions and dislocation nucleation. Finally, selected verification examples are presented in Section~\ref{p9T9VL} that demonstrate the properties, range and scope of the method.

\section{Dislocation dynamics as a problem of transport of measures}\label{zLu8oa}

The formulation of dislocation dynamics as a transport problem is well-known \cite{mura:1987}, but may stand a brief review as it provides the basis for all subsequent developments. We specifically call attention to the representation of dislocations as measures, or, more specifically, as line currents \cite{AO:2005, RN1, RN3}, that opens the way for particle-like approximation schemes such as the method of monopoles developed in this work.

\subsection{Dislocation geometry} \label{hya0KL}

We consider throughout a crystal occupying a region $\Omega \subset \mathbb{R}^3$. A continuum {\sl plastic deformation} generated by crystallographic slip is a tensor-valued measure ${\mbs{\beta}^p}$ supported on a {\sl slip surface} $\Sigma$, contained within crystallographic planes, characterized by the property that
\begin{equation}\label{Sjs4tT}
    \int_\Omega \eta_{ij} \, d\beta^p_{ij}
    =
    \int_\Sigma \eta_{ij} \delta_i \nu_j \, dS
\end{equation}
for all test functions ${\mbs{\eta}}$, where $\mbs{\nu}(\mbs{x})$ and $dS(\mbs{x})$ are the unit normal and element of area at $\mbs{x} \in \Sigma$, respectively. In addition, the slip surface $\Sigma$ is a surface of discontinuity of the displacement field of the crystal and $\mbs{\delta}(\mbs{x})$ is the displacement jump across $\mbs{x} \in \Sigma$. Within the Volterra theory of dislocations, the value of the displacement jump is constrained to be an integer combination of Burgers vectors characteristic of the crystal class. In particular, the displacement jump is piecewise constant over the slip surface.

The Nye \cite{nye:1953} dislocation measure ${\mbs{\alpha}}$ is given by Kr\"oner's formula \cite{Kroner:1958} as 
\begin{equation}\label{p8DN2n}
    \alpha_{ij} = - \beta^p_{ik,l} e_{lkj} ,
\end{equation}
where $e_{ijk}$ denotes the permutation tensor and the curl is to be interpreted in a distributional sense, i.~e.,
\begin{equation}
    \int_\Omega \eta_{ij} \, d\alpha_{ij}
    =
    \int_\Sigma \eta_{ij,l} e_{jlk} \, d\beta^p_{ik}
\end{equation}
for all test functions ${\mbs{\eta}}$. For Volterra dislocations, the dislocation measure has the representation
\begin{equation}\label{Fh7rTj} 
    \int_\Omega \eta_{ij} \, d\alpha_{ij}
    =
    \int_\Gamma \eta_{ij} b_i t_j \, ds ,
\end{equation}
for all test functions ${\mbs{\eta}}$, where $\Gamma$ is a rectifiable curve, or {\sl dislocation line}, within $\Sigma$, $\mbs{t}(\mbs{x})$ and $ds(\mbs{x})$ are the unit tangent vector and the element of length at $\mbs{x} \in \Gamma$, respectively, and, for every $\mbs{x} \in \Gamma$, $\mbs{b}(\mbs{x})$ is a Burgers vector of the crystal. In particular, the dislocation line separates regions of constant displacement jump within the slip surface.

It follows from Kr\"oner's formula (\ref{p8DN2n}) that
\begin{equation}\label{45svDN}
    \alpha_{ij,j} = 0 ,
\end{equation}
i.~e., the dislocation density is closed, or {\sl divergence-free}. Here, again, the divergence is to be interpreted in a distributional sense, i.~e.,
\begin{equation}\label{phl6jO}
    -
    \int_\Gamma \xi_{i,j} \, d\alpha_{ij}
    =
    0 ,
\end{equation}
for all test functions ${\mbs{\xi}}$. The null-divergence property (\ref{45svDN}) of the dislocation density implies that dislocations cannot terminate in the bulk but must form closed loops or networks or exit through the boundary. It also implies Frank's rule for dislocation branching (cf., e.~g., \cite{hirth:1968}).

Representations (\ref{Sjs4tT}) and (\ref{Fh7rTj}) give measure-theoretical expression to distributions of crystallographic slip and Volterra dislocations. We note that the plastic-deformation and dislocation measures are also (rectifiable integer-valued) {\sl currents} \cite{RN1, RN3} of well-defined dimension, namely, the plastic deformation measure is a two-dimensional current and the dislocation measure is a one-dimensional current. As currents, the plastic deformation and dislocation measures additionally have well-defined {\sl boundaries}. In the sense of currents, Kr\"oner's formula (\ref{p8DN2n}) simply defines the dislocation current as the boundary of the plastic-deformation current. In addition, the divergence-free condition (\ref{45svDN}) simply records the fact that a boundary has itself null boundary.

\subsection{Dislocation transport} \label{7vXTHq}

Next, we consider moving dislocations characterized by a time-dependent dislocation measure ${\mbs{\alpha}}(\mbs{x},t)$. Let $S$ be a fixed, arbitrary oriented surface with boundary $\partial S$. The total Burgers vector crossing $S$ is then given by\footnote{More precisely, $b(S,t)$ is the link of $\mbs{\alpha}$, regarded as a current, and $S$, cf., e.~g., \cite{BottTu:1982}.}
\begin{equation}
    b_i(S,t)
    =
    \int_S
        \nu_j
    d\alpha_{ij} .
\end{equation}
Taking rates, we obtain
\begin{equation}
    \dot{b}_i(S,t)
    =
    \int_S
        \nu_j
    d\dot{\alpha}_{ij} ,
\end{equation}
with all derivatives understood in the distributional sense. But $\dot{\mbs{b}}(S,t)$ must also equal the flux of Burgers vector across the boundary $\partial S$, i.~e.,
\begin{equation}
    \dot{b}_i(S,t)
    =
    \int_{\partial S}
        e_{mnk} v_m  t_k
    \, d\alpha_{in} ,
\end{equation}
where $\mbs{v}(\mbs{x},t)$ is the dislocation velocity. Note that $\mbs{v}(\mbs{x},t)$ has no contribution to $\dot{\mbs{b}}(S,t)$ if it is parallel to the dislocation line or the contour $\partial S$ at $\mbs{x}$, as required. An application of Stoke's theorem then gives
\begin{equation}
    \int_S
        \nu_j
    \, d\dot{\alpha}_{ij}
    =
    \int_S
        e_{jlk} e_{mnk} \nu_j
    (v_m d\alpha_{in}),_l ,
\end{equation}
and, since $S$ is arbitrary,
\begin{equation}\label{3kgrKx} 
    \dot{\alpha}_{ij}
    -
    e_{jlk} e_{mnk} (v_m \alpha_{in}),_l
    =
    0 ,
\end{equation}
which defines a transport equation for the dislocation measure. Using the identity
\begin{equation}
    e_{ijk} e_{imn}
    =
    \delta_{jm} \delta_{kn}
    -
    \delta_{jn} \delta_{km} ,
\end{equation}
the transport equation (\ref{3kgrKx}) can be recast in the equivalent form
\begin{equation}
    \dot{\alpha}_{ij}
    -
    ( \alpha_{il} v_j - \alpha_{ij} v_l ),_l
    =
    0 ,
\end{equation}
or, using (\ref{45svDN}),
\begin{equation}\label{vlaML1}
    \dot{\alpha}_{ij}
    +
    \alpha_{ij,l} v_l
    -
    \alpha_{il} v_{j,l}
    +
    \alpha_{ij} v_{l,l}
    =
    0 .
\end{equation}
Taking the distributional divergence of this equation we additionally find that
\begin{equation}
    \dot{\alpha}_{ij,j}
    =
    {0} ,
\end{equation}
which shows that the transport equation (\ref{3kgrKx}) is indeed consistent with the divergence constraint. More precisely, testing (\ref{3kgrKx}) with $\eta_{ij}$ we obtain
\begin{equation}\label{09ZVMn}
    \int_\Omega \eta_{ij} d\dot{\alpha}_{ij}
    +
    \int_\Omega e_{jlk} e_{mnk} v_m \eta_{ij,l} d\alpha_{in}
    =
    0 .
\end{equation}
Setting $\eta_{ij} = \xi_{i,j}$, this identity further reduces to
\begin{equation}
    \int_\Omega \xi_{i,j} d\dot{\alpha}_{ij}
    =
    0 ,
\end{equation}
which indeed implies (\ref{phl6jO}).

The transport equation (\ref{3kgrKx}) has the effect of restricting the possible rates and variations of the dislocation measure ${\mbs{\alpha}}$. Specifically, for a rate $\dot{{\mbs{\alpha}}}$ to be admissible, there must exist a vector field $\mbs{v}(\mbs{x},t)$ such that the curl of ${\mbs{\alpha}} \times \mbs{v}$ equals $-\dot{{\mbs{\alpha}}}$.

\begin{example}[Expanding circular loop]\label{oF8lo5}
A simple example that illustrates the geometry of dislocation transport concerns an expanding circular loop. In this case, the time-dependent dislocation density $\mbs{\alpha}(\mbs{x},t)$ is characterized by the condition that
\begin{equation}\label{vk2GfC}
    \int \eta_{ij}(x) \, d\alpha_{ij}(x,t)
    =
    \int_0^{2\pi}
    \eta_{ij}(r(t),\theta)
    b_i t_j(\theta)
    \, r(t) \, d\theta ,
\end{equation}
for all test functions $\mbs{\eta}$. Here, $r(t)$ is the radius of the loop and
\begin{equation}
    t_j(\theta)
    =
    -
    \sin\theta \delta_{1j}
    +
    \cos\theta \delta_{2j}
\end{equation}
is the tangent vector expressed in terms of the polar angle $\theta$ in the plane of the loop. Likewise, taking rates in (\ref{vk2GfC}) we find that the rate $\dot{\mbs{\alpha}}$ of the dislocation measure is characterized by the condition that
\begin{equation}\label{HZ9rFY}
\begin{split}
    &
    \frac{d}{dt}
    \int \eta_{ij}(x) d\alpha_{ij}(x,t)
    =
    \int \eta_{ij}(x) d\dot{\alpha}_{ij}(x,t)
    = \\ &
    \int_0^{2\pi}
    \left(
        \frac{\partial\eta_{ij}}{\partial r}(r(t),\theta)
        +
        \frac{1}{r(t)}
        \eta_{ij}(r(t),\theta)
    \right)
    \dot{r}(t)
    b_i t_j(\theta)
    \, r(t) \, d\theta
    = \\ &
    \int
    \left(
        \frac{\partial\eta_{ij}}{\partial r}(r(t),\theta)
        +
        \frac{1}{r(t)}
        \eta_{ij}(r(t),\theta)
    \right)
    \dot{r}(t)
    \, d\alpha_{ij}(x,t) ,
\end{split}
\end{equation}
for all test functions $\mbs{\eta}$. Furthermore, from the representation of the curl in polar coordinates, we have
\begin{equation}\label{uuF7EB}
\begin{split}
    &
    \int e_{jlk} e_{mnk} (v_m d\alpha_{in}),_l \eta_{ij}
    =
    -
    \int e_{jlk} e_{mnk} (v_m d\alpha_{in}) \eta_{ij,l}
    = \\ &
    \int_0^{2\pi}
        \curl {\eta}_{ik}(r(t),\theta) \,
        b_i (v \times t)_k(\theta) \,
    r(t) \, d\theta
    = \\ &
    \int_0^{2\pi}
        \curl {\eta}_{i3}(r(t),\theta) \,
        b_i \dot{r}(t) \,
    r(t) \, d\theta
    = \\ &
    \int_0^{2\pi}
    \left(
        \frac{\partial\eta_{ij}}{\partial r}(r(t),\theta)
        +
        \frac{1}{r(t)}
        \eta_{ij}(r(t),\theta)
    \right)
    \dot{r}(t)
    b_i t_j(\theta)
    \, r(t) \, d\theta
    = \\ &
    \int
    \left(
        \frac{\partial\eta_{ij}}{\partial r}(r(t),\theta)
        +
        \frac{1}{r(t)}
        \eta_{ij}(r(t),\theta)
    \right)
    \dot{r}(t)
    \, d\alpha_{ij}(\mbs{x},t) ,
\end{split}
\end{equation}
for all test functions $\mbs{\eta}$. Comparing (\ref{HZ9rFY}) and (\ref{uuF7EB}), we verify that the transport equation (\ref{3kgrKx}) is indeed verified by $\mbs{\alpha}(x,t)$.
\hfill$\square$
\end{example}

\subsection{Transport maps} \label{bR5MnR}

An equivalent Lagrangian formulation of the transport problem that plays a central role in time discretization can be formulated in terms of a transport map $\mbs{\varphi} : \Omega \times [0,T] \to \Omega$ \cite{Villani:2003}. In this representation, the dislocation measure $\mbs{\alpha}(\cdot,t)$ at time $t$ is the {\sl push-forward} of the initial dislocation measure $\mbs{\alpha}_0(\mbs{x}) = \mbs{\alpha}(\cdot,0)$ by $\mbs{\varphi}(\cdot,t)$. Formally,
\begin{equation}\label{cLa5La}
    \mbs{\alpha}_t = {\mbs{\varphi}_t}_\# \mbs{\alpha}_0 ,
\end{equation}
where we write $d\mbs{\alpha}_t(\mbs{x}) = d\mbs{\alpha}(\mbs{x},t)$, $\mbs{\varphi}_t(\mbs{x}) = \mbs{\varphi}(\mbs{x},t)$ and ${\mbs{\varphi}_t}_\#$ denotes the push-forward by ${\mbs{\varphi}_t}_\#$.

The appropriate notion of push-forward for dislocation measures, regarded as line currents, is that $\mbs{\nu}$ is the push-forward of $\mbs{\mu}$ by $\mbs{\varphi}$ if
\begin{equation}\label{7Hiest}
    \int_\Omega \eta_{ij}(\mbs{y}) d\nu_{ij}(\mbs{y})
    =
    \int_\Omega
        \eta_{ij}(\mbs{\varphi}(\mbs{x}))
        \nabla\varphi_{jp}(\mbs{x})
        d\mu_{ip}(\mbs{x}) ,
\end{equation}
for all test functions $\mbs{\eta}$. We note that
\begin{equation}
\begin{split}
    \int_\Omega \xi_{i,j}(\mbs{y}) d\nu_{ij}(\mbs{y})
    & =
    \int_\Omega
        \xi_{i,j}(\mbs{\varphi}(\mbs{x}))
        \nabla\varphi_{jp}(\mbs{x})
        d\mu_{ip}(\mbs{x})
    \\ & =
    \int_\Omega
        (\xi_{i}(\mbs{\varphi}(\mbs{x}))),_p
        d\mu_{ip}(\mbs{x}) ,
\end{split}
\end{equation}
whence it follows that the push-forward operation preserves the divergence-free condition, i.~e., if $\mbs{\mu}$ is divergence-free then so is $\mbs{\nu}$. The local form of the push-forward is obtained by considering absolutely continuous dislocation measures, or {\sl continuously distributed dislocations}, $d\mu_{ij} = f_{ij} \, dx$ and $d\nu_{ij} = g_{ij} \, dy$, where $\mbs{f}$ and $\mbs{g}$ are regular {\sl dislocation densities}. In this case,
\begin{equation}
\begin{split}
    \int_\Omega \eta_{ij}(\mbs{y}) g_{ij}(\mbs{y}) \, dy
    & =
    \int_\Omega
        \eta_{ij}(\mbs{\varphi}(\mbs{x})) g_{ij}(\mbs{\varphi}(\mbs{x}))
        \det(\nabla\mbs{\varphi}(\mbs{x}))
    \, dx
    \\ & =
    \int_\Omega
        \eta_{ij}(\mbs{\varphi}(\mbs{x}))
        \nabla\varphi_{jp}(\mbs{x})
        f_{ip}(\mbs{x}) \, dx ,
\end{split}
\end{equation}
which requires that
\begin{equation}
    g_{ij}(\mbs{\varphi}(\mbs{x}))
    =
    \frac
    {
        \nabla\varphi_{jp}(\mbs{x}) f_{ip}(\mbs{x})
    }
    {
        \det(\nabla\mbs{\varphi}(\mbs{x}))
    } .
\end{equation}
We observe that the push-forward operation entails reorientation and stretching of the dislocation line.

For completeness, we verify that (\ref{cLa5La}) is equivalent to the transport equation (\ref{3kgrKx}). Using the definition (\ref{7Hiest}) of push-forward, we have
\begin{equation}\label{3rouPi}
    \int_\Omega \eta_{ij}(\mbs{y}) d\alpha_{ij}(\mbs{y},t)
    =
    \int_\Omega
        \eta_{ij}(\mbs{\varphi}(\mbs{x},t))
        \nabla\varphi_{jp}(\mbs{x},t)
    d\alpha_{ip}(\mbs{x},0) .
\end{equation}
For simplicity, we consider the case of absolutely continuous dislocation measures  $d\alpha_{ij}(\mbs{x},t) = \rho_{ij}(\mbs{x},t) \, dx$. In this case, the push-forward (\ref{3rouPi}) reduces to
\begin{equation}
    \rho_{ij}(\mbs{\varphi}(\mbs{x},t),t)
    =
    \frac
    {
        \nabla\varphi_{jp}(\mbs{x},t) \rho_{ip}(\mbs{x},0)
    }
    {
        \det(\nabla\mbs{\varphi}(\mbs{x},t))
    } .
\end{equation}
Taking time derivatives, we obtain
\begin{equation}
\begin{split}
    &
    \dot{\rho}_{ij}(\mbs{\varphi}(\mbs{x},t),t)
    +
    \rho_{ij,k}(\mbs{\varphi}(\mbs{x},t),t) \dot{\varphi}_k(\mbs{x},t)
    = \\ &
    \frac
    {
        \nabla\dot{\varphi}_{jp}(\mbs{x},t)\rho_{ip}(\mbs{x},0)
    }
    {
        \det(\nabla\mbs{\varphi}(\mbs{x},t))
    }
    -
    \frac
    {
        \nabla\varphi_{jp}(\mbs{x},t)\rho_{ip}(\mbs{x},0)
    }
    {
        \det(\nabla\mbs{\varphi}(\mbs{x},t))
    }
    \nabla\dot{\varphi}_{lq}(\mbs{x},t)
    \nabla\varphi^{-1}_{ql}(\mbs{x},t)
    = \\ &
    \nabla\dot{\varphi}_{jp}(\mbs{x},t)
    \nabla\varphi^{-1}_{pk}(\mbs{x},t)
    \rho_{ik}(\mbs{\varphi}(\mbs{x},t),t)
    - \\ &
    \nabla\dot{\varphi}_{lq}(\mbs{x},t)
    \nabla\varphi^{-1}_{ql}(\mbs{x},t)
    \rho_{ij}(\mbs{\varphi}(\mbs{x},t),t) ,
\end{split}
\end{equation}
or, by a suitable change of variables,
\begin{equation}
    \dot{\rho}_{ij}(\mbs{y},t)
    +
    \rho_{ij,k}(\mbs{y},t) v_k(\mbs{y},t)
    =
    v_{j,k}(\mbs{y},t)
    \rho_{ik}(\mbs{y},t)
    -
    v_{k,k}(\mbs{y},t)
    \rho_{ij}(\mbs{y},t) ,
\end{equation}
which is identical to (\ref{vlaML1}) with velocity
\begin{equation}
    v_i(\mbs{y},t)
    =
    \dot{\varphi}_i(\mbs{\varphi}^{-1}(\mbs{y},t),t) ,
\end{equation}
as required.

\section{Mobility and energetics}
\label{197IIa}

In order to close the transport problem (\ref{3kgrKx}) we need to specify a mobility law that supplies the instantaneous dislocation velocity. Whereas the transport problem (\ref{3kgrKx}) concerns the geometry of the dislocations and its evolution in time, the mobility law encodes the {\sl kinetics} of dislocation motion. In this section, we develop the conventional energetic viewpoint that the dislocation motion is a {\sl gradient flow} driven by energetic driving forces and governed by kinetics. We specifically focus on variational formulations that provide the basis for the incremental minimum problems developed subsequently.

\subsection{Dislocation mobility} \label{1KnY2M}

In order to identify the appropriate driving force for dislocation motion, we consider the rate of elastic energy $\dot{E}$ attendant to a plastic deformation rate $\dot{\mbs{\beta}}^p$. Alternatively, we may regard $\dot{\mbs{\beta}}^p$ as a {\sl variation} of the plastic deformation $\mbs{\beta}^p$ and $\dot{E}$ the attendant variation of the energy $E$, as the operations of taking rates and variations are mathematically identical. We have,
\begin{equation}\label{LYw64C}
\begin{split}
    \dot{E}
    & =
    \int_{\Omega\backslash\Sigma}
        \sigma_{ij} \dot{\beta}^e_{ij}
    \, dx
    =
    \int_\Omega
        \sigma_{ij} (\dot{u}_{i,j} - \dot{\beta}^p_{ij})
    \, dx
    \\ & =
    \int_\Omega
        \sigma_{ij} (\dot{u}_{i,j} - \dot{\beta}^p_{ij})
    \, dx
    =
    -
    \int_\Omega
        \sigma_{ij} d\dot{\beta}^p_{ij} ,
\end{split}
\end{equation}
i.~e., at equilibrium and in the absence of body forces and applied tractions, the rate of elastic energy equals the negative of the plastic work rate. We recall that equilibrium stress fields admit the representation
\begin{subequations}
\begin{align}
    &
    \sigma_{ij}
    =
    -
    \chi_{ik,l} e_{lkj}
    =
    \sigma_{ji} ,
    \\ &
    \chi_{ij,j}
    =
    0 ,
\end{align}
\end{subequations}
in terms of an Airy stress potential $\chi$. Inserting this representation into (\ref{LYw64C}), we obtain
\begin{equation}
\begin{split}
    \dot{E}
    & =
    \int_\Omega
        \chi_{ik,l} e_{lkj} d\dot{\beta}^p_{ij}
    =
    \int_\Omega
        \chi_{ij} d\dot{\alpha}_{ij} ,
\end{split}
\end{equation}
which shows that the Airy stress potential and the dislocation measure are {\sl work conjugate}.

In order to proceed further, we need to characterize the admissible rates, or variations, $\dot{\mbs{\alpha}}$. This characterization is non-trivial since the dislocation densities $\mbs{\alpha}$ define a non-linear space.\footnote{cf., e.~g., \cite{Gangbo:1996, Villani:2003} for background on the closely related spaces of probability measures that arise in scalar transport problems. The geometry of optimal transport of vector-value measures appears to be considerably less developed.} Formally, the appropriate notion of variation of $\mbs{\alpha}$ follows from the transport equation (\ref{3kgrKx}), namely, $\dot{\mbs{\alpha}}$ is an admissible rate, or variation, if there exists a velocity field $\mbs{v}$ such that (\ref{3kgrKx}) is satisfied. Using this differential structure, we have
\begin{equation}
\begin{split}
    \dot{E}
    & =
    \int_\Omega
        \chi_{ij} d\dot{\alpha}_{ij}
    =
    \int_\Omega
        \chi_{ij,l}
        e_{jlk} e_{mnk} v_m d\alpha_{in}
    \\ & =
    \int_\Omega
        \sigma_{ik}
        e_{nmk} v_m d\alpha_{in}
    =
    \int_\Gamma
        \sigma_{ik}
        e_{nmk} v_m b_i t_n \, ds
    =
    \int_\Gamma
        f_m v_m \, ds ,
\end{split}
\end{equation}where
\begin{equation}\label{priu1I} 
    f_m = \sigma_{ik} e_{nmk} b_i t_n
\end{equation}
is the Peach-K\"ohler force per unit dislocation length.

In view of (\ref{priu1I}), standard thermodynamic arguments suggest that the dislocation motion is governed by a mobility law of the type
\begin{equation}\label{drIE1o}
    v_i
    =
    D_i \psi^*(\mbs{f}) ,
\end{equation}
where $\psi(\mbs{f})$ is a dual kinetic potential and $D_i$ denotes partial differentiation. Alternatively, we may express the mobility law in inverse form as
\begin{equation}\label{rL5pRl}
    f_i
    =
    D_i \psi(\mbs{v}) ,
\end{equation}
where the kinetic potential $\psi(\mbs{v})$ is the Legendre transform of $\psi^*(\mbs{f})$, provided that it exists.\footnote{We recall that the Legendre transform is well-defined on proper, convex, lower-semicontinuous functions, cf., e.~g., \cite{RN27}.}

The precise form of the mobility law, and the potential $\psi(\mbs{v})$ depends on the physical processes that limit dislocation mobility (cf., e.~g., \cite{hirth:1968}). For instance, if lattice friction is the rate-limiting mechanism, then
\begin{equation}
    \psi(\mbs{v}) = \tau_c | \mbs{v} | ,
\end{equation}
where $\tau_c$ is the critical resolved shear stress. In particular, $\psi(\mbs{v})$ is homogeneous of degree one in the dislocation velocity. If, instead, dislocation motion is controlled by phonon drag, then
\begin{equation}
    \psi(\mbs{v}) = \frac{B}{2} |\mbs{v}|^2 ,
\end{equation}
where $B$ is a phonon-drag coefficient. In this case, $\psi(\mbs{v})$ is quadratic in the dislocation velocity.

\subsection{Dislocation energy} \label{m0X7vX}

As we have seen, within an energetic framework the motion of the dislocations, and the attendant evolution of the dislocation measure, is driven by energetic or Peach-K\"ohler forces. For present purposes, we shall require a representation of the energy that is well-defined for general dislocation measures, including Volterra dislocations and, subsequently, dislocation monopoles. We derive one such representation in two steps. Firstly, we present a general argument based on the Helmholtz decomposition \cite{AO:2005} that shows that, in the absence of Dirichlet boundary conditions, the---possibly nonlinear---elastic energy of the solid depends solely on the dislocation density. This representation generalizes a similar result obtained by Mura \cite{RN2} for the special case of linear elasticity. Unfortunately, conventional linear elasticity is not well-suited to Volterra dislocations due to the well-known logarithmic divergence of the energy (cf., e.~g., \cite{hirth:1968}). In order to sidestep this difficulty, we develop a regularization based on strain-gradient elasticity that renders the energy well-defined for general dislocation measures, including dislocation monopoles.

Suppose that the crystal deforms under the action of body forces $\mbs{f}$, prescribed displacements $\mbs{g}$ over the displacement or Dirichlet boundary $\Gamma_D$ and applied tractions $\mbs{h}$ over the traction or Neumann boundary $\Gamma_N$. We recall that the Helmholtz decomposition of $\mbs{\beta}^p$ is \cite{RN49}
\begin{equation}\label{Z2aphO}
    \beta^p_{ij}
    =
    v_{i,j}
    +
    w_{ik,l} e_{lkj} ,
\end{equation}
where $\mbs{v}$ and $\mbs{w}$ are potentials. To this representation, we additionally append the Lorenz gauge condition
\begin{equation}\label{p0iUxl}
    w_{ij,j} = 0 ,
\end{equation}
and the boundary conditions
\begin{subequations}\label{diuqI7}
\begin{align}
    & \label{viu5lU}
    v_i = 0, & \text{ on } \Gamma_D ,
    \\ &
    w_{ik,l} e_{lkj} n_j = 0, & \text{ on } \Gamma_N .
\end{align}
\end{subequations}
Taking the divergence and the curl of (\ref{Z2aphO}), we obtain
\begin{subequations}\label{Yl5Hou}
\begin{align}
    &
    \beta^p_{ij,j}
    =
    v_{i,jj} ,
    \\ &
    \beta^p_{im,n} e_{nmj}
    =
    -
    \alpha_{ij}
    =
    w_{ik,ln} e_{lkm} e_{nmj}
    =
    -
    w_{ij,kk} ,
\end{align}
\end{subequations}
where we have used the gauge condition (\ref{p0iUxl}) and Kr\"oner's formula (\ref{p8DN2n}). Eqs.~(\ref{Yl5Hou}), together with the boundary conditions (\ref{diuqI7}), uniquely determine the potentials. In particular, we note that the vector potential $\mbs{w}$ is fully determined by the dislocation measure $\mbs{\alpha}$.

Let $\bar{\mbs{\sigma}}$ be a stress field in equilibrium with the body forces and the applied tractions, i.~e.,
\begin{subequations}
\begin{align}
    &
    \bar{\sigma}_{ij,j} + f_i = 0 ,
    & \text{in } \Omega ,
    \\ &
    \bar{\sigma}_{ij} n_j = h_i ,
    & \text{on } \Gamma_N .
\end{align}
\end{subequations}
Then, the potential energy of the crystal takes the form
\begin{equation}\label{8roasO}
    \Phi(\mbs{u},\mbs{\beta}^p)
    =
    \int_\Omega
        \big(
            W({D}\mbs{u} - \mbs{\beta}^p)
            -
            \bar{\sigma}_{ij} ( u_{i,j} - \beta^p_{ij} )
        \big)
    \, dx ,
\end{equation}
where $W(\mbs{\beta}^e)$ is the elastic strain energy density and ${D}\mbs{u}$ is the distributional derivative of the displacement field. For a linear elastic crystal,
\begin{equation}\label{tHO4wr}
    W(\mbs{\beta}^e)
    =
    \frac{1}{2} c_{ijkl} \epsilon^e_{ij} \epsilon^e_{kl}
\end{equation}
where $c_{ijkl}$ are the elastic moduli and $\epsilon^e_{ij} = (\beta^e_{ij} + \beta^e_{ji})/2$ are the elastic strains. However, we emphasize that the present derivation does not require linearity and holds for general strain energy densities. The elastic energy at equilibrium follows as
\begin{equation}\label{wr1uZl}
    E(\mbs{\beta}^p)
    =
    \inf
    \{
        \Phi(\mbs{u},\mbs{\beta}^p),
        \
        \mbs{u} = \mbs{g}
        \text{ on } \Gamma_D
    \} .
\end{equation}
But, inserting the Helmholtz decomposition (\ref{Z2aphO}) into (\ref{8roasO}) gives
\begin{equation}
    \Phi(\mbs{u},\mbs{\beta}^p)
    =
    \int_\Omega
        \big(
            W({D} \mbs{u} - {D} \mbs{v} - \curl \mbs{w})
            -
            \bar{\sigma}_{ij}
            ( v_{i,j} + w_{ik,l} e_{lkj} )
        \big)
    \, dx .
\end{equation}
Absorbing $\mbs{v}$ into $\mbs{u}$, which by (\ref{viu5lU}) leaves $\mbs{u}$ unchanged over $\Gamma_D$, we obtain
\begin{equation}\label{gL1dri}
    E(\mbs{\beta}^p)
    =
    \inf
    \{
        \Phi(\mbs{u},\curl \mbs{w}) ,
        \
        \mbs{u} = \mbs{g}
        \text{ on } \Gamma_D
    \}
    =
    E(\mbs{\alpha}) ,
\end{equation}
since the potential $\mbs{w}$ is fully determined by the dislocation density $\mbs{\alpha}$. Let $\mbs{u}^*$ be the displacement field at equilibrium, also fully determined by $\mbs{\alpha}$. Then,
\begin{equation}
    \int_\Omega
        \big(
            DW_{ij}({D} \mbs{u}^* - \curl \mbs{w})
            -
            \bar{\sigma}_{ij}
        \big)
    \, \eta_{i,j} \, dx
    =
    0 ,
\end{equation}
for all test functions $\mbs{\eta}$, and the stress field at equilibrium follows as
\begin{equation}\label{uRFL85}
    \sigma_{ij}^*
    =
    DW_{ij}({D}\mbs{u}^* - \curl \mbs{w}) ,
\end{equation}
which is also fully determined by the dislocation density $\mbs{\alpha}$.

In cases where the body is subject to traction boundary conditions only, such as an infinite body or a periodic unit cell, a more direct expression for the energy can be obtained as follows. Begin by writing the potential energy (\ref{8roasO}) as
\begin{equation}
    \Phi(\mbs{\beta}^e)
    =
    \int_\Omega
        \big(
            W(\mbs{\beta}^e)
            -
            \bar{\sigma}_{ij} \beta^e_{ij}
        \big)
    \, dx
\end{equation}
where
\begin{equation}
    \beta^e_{ij} = u_{i,j} - \beta^p_{ij}
\end{equation}
is the elastic deformation. From Kr\"oner's formula (\ref{p8DN2n}) we have
\begin{equation}
    \alpha_{ij} = \beta^e_{ik,l} e_{lkj} .
\end{equation}
Thus, in the absence of displacement boundary conditions, the equilibrium elastic deformation $\mbs{\beta}^{e*}$ follows directly from the minimum problem
\begin{equation}
    \mbs{\beta}^{e*}
    \in
    {\rm argmin} \,
    \{
        \Phi(\mbs{\beta}^e), \
        \curl \mbs{\beta}^e = \mbs{\alpha}
    \} .
\end{equation}
Thus, $\mbs{\beta}^{e*}$ minimizes the potential energy of the solid subject to the constraint that it be compatible everywhere except on the support of the dislocation measure, where it must satisfy a curl constraint, e.~g., in the sense of Burgers circuits. Enforcing the curl constraint by means of a Lagrange multiplier $\mbs{\chi}$, or Airy stress potential, results in the Lagrangian
\begin{equation}
    L(\mbs{\beta}^e,\mbs{\chi})
    =
    \int_\Omega
        \big(
            W(\mbs{\beta}^e)
            -
            \bar{\sigma}_{ij} \beta^e_{ij}
            -
            \chi_{ij}
            (\alpha_{ij}- \beta^e_{ik,l} e_{lkj})
        \big)
    \, dx ,
\end{equation}
or, integrating by parts,
\begin{equation}
    L(\mbs{\beta}^e,\mbs{\chi})
    =
    \int_\Omega
        \big(
            W(\mbs{\beta}^e)
            -
            (\bar{\sigma}_{ij} + \chi_{ik,l} e_{lkj}) \beta^e_{ij}
            -
            \chi_{ij} \alpha_{ij}
        \big)
    \, dx ,
\end{equation}
which must be stationary at equilibrium. We note that $\mbs{\beta}^e$ enters the Lagrangian undifferentiated and can, therefore, be minimized pointwise, which results in the complementary energy
\begin{equation}\label{cHiaj8}
    \Phi^*(\mbs{\chi},\mbs{\alpha})
    =
    \int_\Omega
        \big(
            W^*(\curl\mbs{\chi} + \bar{\mbs{\sigma}})
            +
            \chi_{ij} \alpha_{ij}
        \big)
    \, dx ,
\end{equation}
where
\begin{equation}
    W^*(\mbs{\sigma})
    =
    \sup \{ \sigma_{ij} \beta^e_{ij} - W(\mbs{\beta}^e) \}
\end{equation}
is the complementary energy density. For linear elastic solids, eq.~(\ref{tHO4wr}), we explicitly have
\begin{equation}
    W^*(\mbs{\sigma})
    =
    \frac{1}{2} c_{ijkl}^{-1} \sigma_{ij} \sigma_{kl} .
\end{equation}
The elastic energy follows again by minimization with respect to the elastic deformations, i.~e.,
\begin{equation}\label{HlaT9o}
    E(\mbs{\alpha}) = \inf \Phi^*(\cdot,\mbs{\alpha}) .
\end{equation}
From (\ref{HlaT9o}) we conclude that, in the absence of Dirichlet boundary conditions, the elastic energy at equilibrium is a function solely of the dislocation measure, as advertised.

\subsection{Core regularization}\label{fo0Fle}

One key advantage of the representation (\ref{HlaT9o}) of the energy is that $\mbs{\alpha}$ enters in $\Phi^*(\mbs{\chi},\mbs{\alpha})$ linearly and can therefore be treated as a general measure, as required by the monopole approximations pursued subsequently. However, as already noted, a direct application of (\ref{HlaT9o}) to Volterra dislocations is not possible due to the well-known logarithmic divergence of the energy. This type of energy divergence is well-known in connection with elliptic problems with measure data, e.~g., the Laplace equation with a point source \cite{Stampacchia:1965}. In these problems, equilibrium solutions exist but have infinite energy, which precludes an energetic characterization of the solutions and attendant configurational forces.

A number of regularizations of linear elasticity have proposed in order to eliminate the logarithmic divergence of the energy of Volterra dislocations (cf., e.~g., \cite{BulatovCai:2006}), including discrete elasticity \cite{AO:2005,RN47}, core cut-offs \cite{hirth:1968} and nonlinear elasticity \cite{RN64, RN57}. Yet another regularization that is particularly well-suited to general dislocation measures consists of endowing dislocation lines with a {\sl core profile}, e.~g., by mollifying the dislocation measure on the scale of the lattice parameter $\epsilon$ \cite{RN1}. In this approach, the dislocation density is given the representation
\begin{equation}\label{v6UpHi}
    \mbs{\alpha}^\epsilon = \phi^\epsilon *\emph{} \mbs{\alpha} ,
\end{equation}
where $\mbs{\alpha}$ is a collection of Volterra dislocation lines, $\phi^\epsilon$ is a mollifier and $*$ denotes convolution\footnote{By a {\sl mollifier} here we understand a sequence $\phi^\epsilon$ of smooth positive functions of total mass $1$ defining a Dirac-sequence. We also recall that the convolution of two functions is defined as $f*g = \int f(\mbs{x}-\mbs{x}') g(\mbs{x}') \, dx'$ (cf., e.~g., \cite{RN66}).}. The regularized energy is then
\begin{equation}\label{6iaSwl}
    E^\epsilon(\mbs{\alpha})
    =
    E(\mbs{\alpha}^\epsilon) ,
\end{equation}
with $E(\cdot)$ given by (\ref{HlaT9o}).

A connection between mollification of the dislocation density and strain-gradient elasticity can be established as follows. Begin by regularizing the complementary energy (\ref{cHiaj8}) as
\begin{equation}\label{tRi9gL}
    \Phi^{*\epsilon}(\mbs{\chi},\mbs{\alpha})
    =
    \int_\Omega
        \big(
            W^*(\curl(1-\epsilon^2\Delta)\mbs{\chi} + \bar{\mbs{\sigma}})
            +
            \chi_{ij} \alpha_{ij}
        \big)
    \, dx ,
\end{equation}
where $\Delta$ denotes the Laplacian operator. Changing variables to
\begin{equation}
    \mbs{\chi}^\epsilon
    =
    (1-\epsilon^2\Delta)\mbs{\chi} ,
\end{equation}
the regularized complementary energy (\ref{tRi9gL}) becomes
\begin{equation}
    \Phi^{*\epsilon}(\mbs{\chi},\mbs{\alpha})
    =
    \int_\Omega
        \big(
            W^*({\rm curl} \, \mbs{\chi}^\epsilon + \bar{\mbs{\sigma}})
            +
            \chi^\epsilon_{ij} \alpha^\epsilon_{ij}
        \big)
    \, dx
    =
    \Phi^*(\mbs{\chi}^\epsilon,\mbs{\alpha}^\epsilon) ,
\end{equation}
with
\begin{equation}
    \mbs{\alpha}^\epsilon
    =
    (1-\epsilon^2\Delta)^{-1} \mbs{\alpha}
    =
    \phi^\epsilon * \mbs{\alpha} ,
\end{equation}
and
\begin{equation}\label{tH3epR}
    \phi^\epsilon(\mbs{x})
    =
    \frac{1}{4\pi \epsilon^2 r}
    \,
    {\rm e}^{- |\mbs{x}|/\epsilon } ,
\end{equation}
which identifies the mollifier and the core structure of the dislocations.

For an infinite linear isotropic solid, the elastic energy (\ref{HlaT9o}) of a sufficiently regular dislocation measure follows as (cf.~\cite{hirth:1968}, eq.~(4-44))
\begin{equation}\label{xOU1ie}
\begin{split}
    E(\mbs{\alpha})
    =
    & -
    \frac{\mu}{4\pi}
    \int \int
        \frac{1}{R(\mbs{x}, \mbs{x}')}
        e_{ikm} e_{jln}
        d\alpha_{ij}(\mbs{x})
        d\alpha_{kl}(\mbs{x}')
    \\ & +
    \frac{\mu}{8\pi}
    \int \int
        \frac{1}{R(\mbs{x}, \mbs{x}')}
        d\alpha_{ii}(\mbs{x})
        d\alpha_{jj}(\mbs{x}')
    \\ & +
    \frac{\mu}{8\pi(1-\nu)}
    \int \int
        T_{mn}(\mbs{x}, \mbs{x}')
        e_{ijm} e_{kln}
        d\alpha_{ij}(\mbs{x})
        d\alpha_{kl}(\mbs{x}')
\end{split}
\end{equation}
where
\begin{equation}\label{qoaW3a}
    R(\mbs{x}, \mbs{x}') = | \mbs{x} - \mbs{x}' | ,
    \qquad
    T_{ij}(\mbs{x}, \mbs{x}')
    =
    \frac{\partial^2R}{\partial x_i \partial x'_j}(\mbs{x} - \mbs{x}') ,
\end{equation}
$\mu$ is the shear modulus and $\nu$ Poisson's ratio. Inserting (\ref{v6UpHi}) into (\ref{xOU1ie}), we obtain
\begin{equation}\label{0riEst}
\begin{split}
    E^\epsilon(\mbs{\alpha})
    =
    & -
    \frac{\mu}{4\pi}
    \int \int
        S^\epsilon(\mbs{x}, \mbs{x}')
        e_{ikm} e_{jln}
        d\alpha_{ij}(\mbs{x})
        d\alpha_{kl}(\mbs{x}')
    \\ & +
    \frac{\mu}{8\pi}
    \int \int
        S^\epsilon(\mbs{x}, \mbs{x}')
        d\alpha_{ii}(\mbs{x})
        d\alpha_{jj}(\mbs{x}')
    \\ & +
    \frac{\mu}{8\pi(1-\nu)}
    \int \int
        T_{mn}^\epsilon(\mbs{x}, \mbs{x}')
        e_{ijm} e_{kln}
        d\alpha_{ij}(\mbs{x})
        d\alpha_{kl}(\mbs{x}') ,
\end{split}
\end{equation}
where we write
\begin{equation}
    S = 1/R ,
    \quad
    S^\epsilon = \phi^\epsilon * \phi^\epsilon * S ,
\end{equation}
and
\begin{equation}
    R^\epsilon = \phi^\epsilon * \phi^\epsilon * R ,
    \quad
    T^\epsilon_{ij}
    =
    \phi^\epsilon * \phi^\epsilon * T_{ij}
    =
    \frac{\partial^2R^\epsilon}{\partial x_i \partial x'_j} .
\end{equation}
By virtue of the regularization of the kernels, the energy (\ref{0riEst}) is now finite for general dislocation measures. In particular, for Volterra dislocations (\ref{0riEst}) specializes to
\begin{equation}\label{thieV5}
\begin{split}
    &
    E^\epsilon(\mbs{\alpha})
    = \\ &
    -
    \frac{\mu}{4\pi}
    \int_\Gamma \int_\Gamma
        S^\epsilon(\mbs{x}(s), \mbs{x}(s'))
        (\mbs{b}(s)\times \mbs{b}(s'))\cdot(\mbs{t}(s)\times \mbs{t}(s'))
    \, ds \, ds'
    \\ & +
    \frac{\mu}{8\pi}
    \int_\Gamma \int_\Gamma
        S^\epsilon(\mbs{x}(s), \mbs{x}(s'))
        (\mbs{b}(s)\cdot \mbs{t}(s))(\mbs{b}(s')\cdot \mbs{t}(s'))
    \, ds \, ds'
    + \\ &
    \frac{\mu}{8\pi(1-\nu)}
    \int_\Gamma \int_\Gamma
        (\mbs{b}(s)\times \mbs{t}(s))
        \cdot
        \mbs{T}^\epsilon(\mbs{x}(s), \mbs{x}(s'))
        \cdot
        (\mbs{b}(s')\times \mbs{t}(s'))
    \, ds \, ds' .
\end{split}
\end{equation}
For the specific mollifier (\ref{tH3epR}), straightforward calculations using the Fourier transform give, explicitly,
\begin{equation}
    S^\epsilon(r)
    =
    \frac{2 \epsilon - (r + 2 \epsilon) {\rm e}^{-r/\epsilon}}{2 \epsilon r} ,
    \quad
    R^\epsilon(r)
    =
    \frac{r^2 + 4 \epsilon^2 - \epsilon (r + 4 \epsilon){\rm e}^{-r/\epsilon}}{r} ,
\end{equation}
with $r = | \mbs{x} - \mbs{x}' |$.

\begin{figure}
\begin{center}
	\begin{subfigure}{0.49\textwidth}\caption{}
        \includegraphics[width=0.95\linewidth]{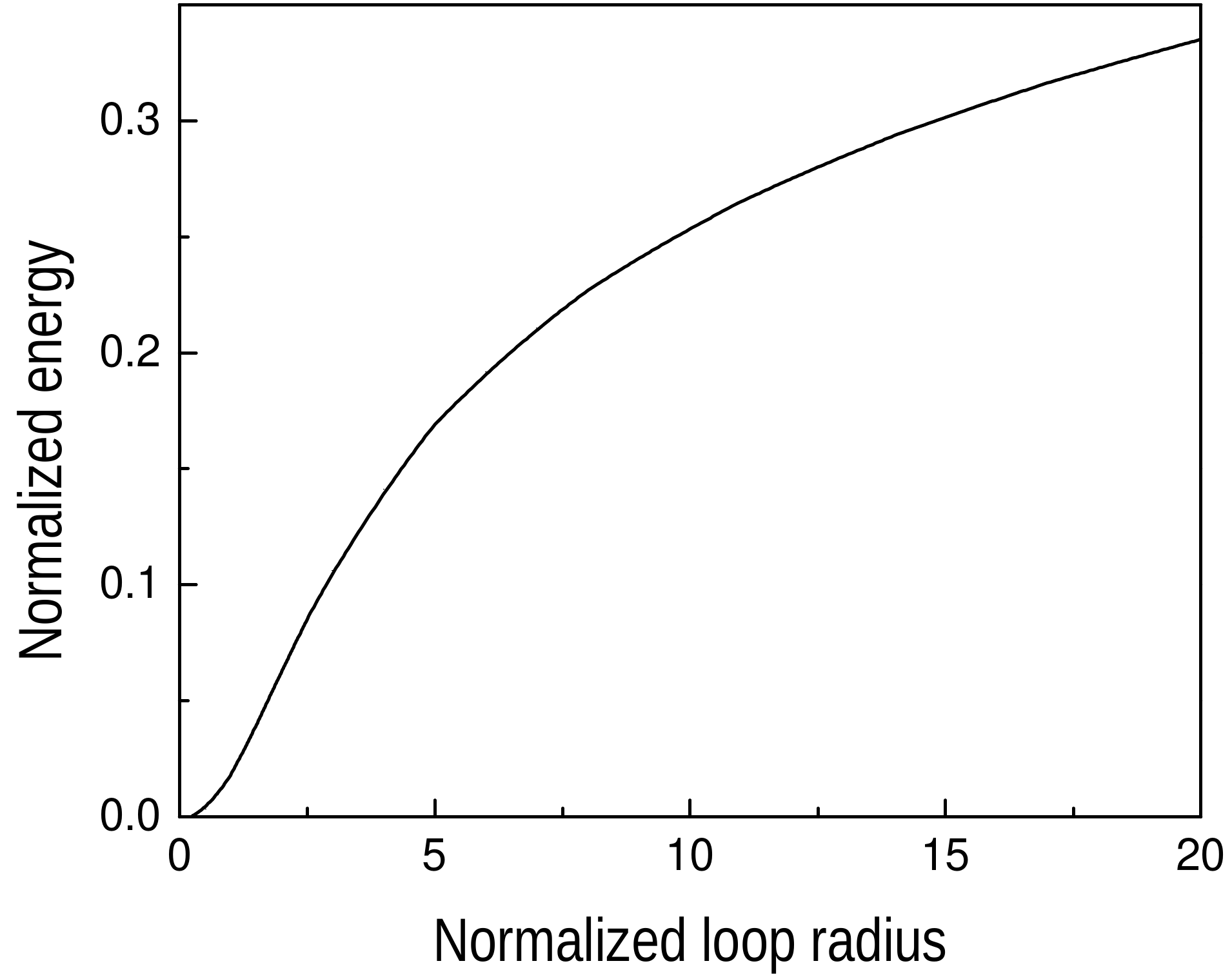}
	\end{subfigure}
	\begin{subfigure}{0.49\textwidth}\caption{}
        \includegraphics[width=0.99\linewidth]{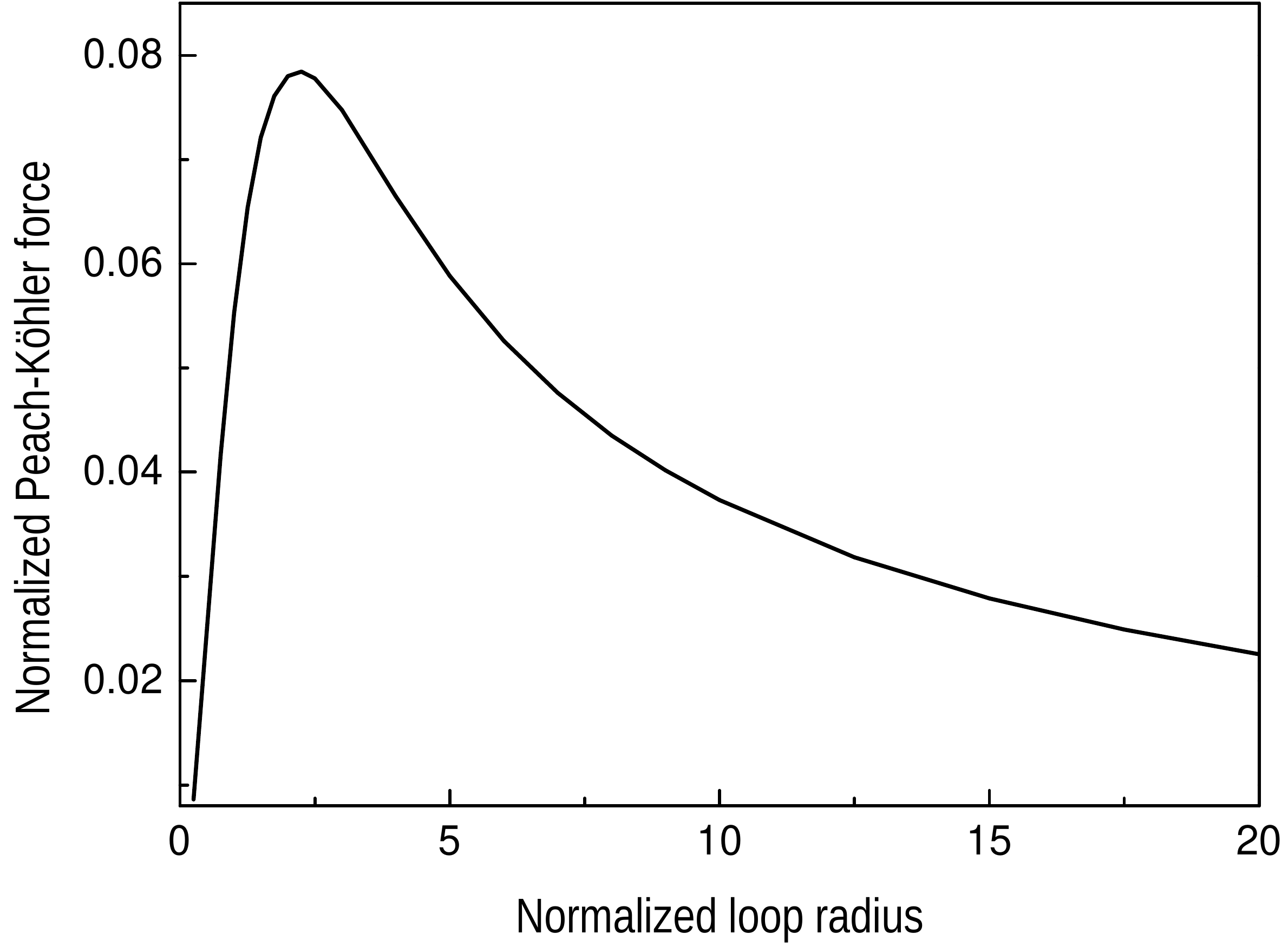}
	\end{subfigure}
\caption{\small Circular prismatic loop. a) Regularized energy normalized by $\frac{\mu b^2 \rho^2}{8\pi(1-\nu)\epsilon}$. b) Regularized Peach-K\"ohler force per unit length normalized by $\frac{\mu b^2 }{8\pi(1-\nu)\epsilon}$. Loop radius normalized by $\epsilon$.}
\label{6hounL}
\end{center}
\end{figure}

\begin{example}[Circular prismatic loop]\label{3Roabo}
We illustrate the logarithmic divergence of linearly elastic Volterra dislocations and the effect of regularization by means of the simple example of a circular prismatic loop. Assume that the loop is in the $(x_1,x_2)$-plane, has radius $\rho$ and its Burgers vector is $\mbs{b} = b \mbs{e}_3$. Under these conditions, (\ref{thieV5}) reduces to
\begin{equation}
    E^\epsilon(\rho)
    =
    \frac{\mu b^2 \rho^2}{8\pi(1-\nu)}
    \int_0^{2\pi} \int_0^{2\pi}
        \mbs{e}_r(\theta)
        \cdot
        \mbs{T}^\epsilon(\rho\mbs{e}_r(\theta) - \rho\mbs{e}_r(\theta'))
        \cdot
        \mbs{e}_r(\theta')
    \, d\theta \, d\theta' ,
\end{equation}
where $\theta$ is the polar angle and $\mbs{e}_r(\theta)$ is the radial unit vector on the plane $x_3=0$. The corresponding Peach-K\"ohler force acting on the loop is
\begin{equation}
    f^\epsilon(\rho)
    =
    \frac{1}{2\pi\rho}
    \frac{\partial E^\epsilon}{\partial \rho}(\rho) .
\end{equation}
The dependence of $E^\epsilon(\rho)$ and $f^\epsilon(\rho)$ on the loop radius $\rho$ is shown in Fig.~\ref{6hounL}. As may be seen from the figure, both the energy and the Peach-K\"ohler force are finite for all $0 \leq \rho < +\infty$. Thus, the regularization eliminates the divergence of energy and the Peach-K\"ohler force as $\rho \to 0$. Specifically, we observe that both the energy and the Peach-K\"ohler force decrease to zero as $\rho \to 0$, at which point the loop annihilates. For large $\rho$, the energy grows as $\rho \log(\rho/\epsilon)$ and the Peach-K\"ohler force decays as $\log(\rho/\epsilon)/\rho$, in agreement with linear elasticity (cf.~\cite{hirth:1968}, eqs.~(5-28) and (6-52)).
\hfill$\square$
\end{example}

\section{Variational formulation and approximation} \label{3x2efD}

We note from (\ref{uRFL85}) that the Peach-K\"ohler driving force (\ref{priu1I}) is a function of the dislocation measure $\mbs{\alpha}$. Therefore, the transport equation (\ref{3kgrKx}) and the mobility law (\ref{drIE1o}) define a closed transport problem governing the evolution of $\mbs{\alpha}$ in time. The study of transport problems for  measures was pioneered by Jordan, Kinderlehrer and Otto (JKO) \cite{JordanKinderlehrerOtto1997, JordanKinderlehrerOtto1998, JordanKinderlehrerOtto1999} in the context of scalar measures. They recognized that such problems can be given a natural variational structure by recourse to time discretization. This incremental approach characterizes the time evolution as a competition between dissipation, which penalizes departures from the current configuration, and energy, which favors low-energy configurations. Quite crucially, JKO quantify the incremental dissipation by means of a Wasserstein-like distance between two consecutive measures. As we shall see, the overwhelming advantage of such measure-theoretical and time-discrete variational approaches is that they are {\sl geometrically exact} in the sense of the incremental push-forward operation. In the present setting, the resulting incremental dislocation updates are exact with respect to dislocation advection, stretching of null-divergence constraint. Another crucial advantage of measure-theoretical approaches is that they supply a suitable mathematical framework for the formulation of particle methods such as the method of monopoles proposed here.

\subsection{Time discretization}
\label{BL9NY0}

We begin by discretizing the transport problem (\ref{3kgrKx}) in time. To this end, let $t_0=0 < t_1 <$ $\cdots$ $t_\nu$ $<$ $t_{\nu+1}$ $\dots$ $<t_N = T$ be a discretization of the time interval $[0,T]$. We wish to determine corresponding discrete approximations $\mbs{\alpha}_0$, $\mbs{\alpha}_1$ $\dots$ $\mbs{\alpha}_N$ of the dislocation measure of a collection of Volterra dislocations and discrete approximations $\mbs{\varphi}_0$, $\mbs{\varphi}_1$ $\dots$ $\mbs{\varphi}_N$ of the transport maps.

We begin by defining an incremental dissipation as
\begin{equation}\label{mo6kIu}
    D(\mbs{\varphi}_\nu, \mbs{\varphi}_{\nu+1})
    =
    \qquad\qquad\qquad\qquad\qquad\qquad
\end{equation}
\begin{equation}\nonumber
    \min
    \Big\{
        \int_{t_\nu}^{t_{\nu+1}}
            \int_{\Gamma_0}
                \psi\big(\dot{\mbs{\varphi}}(s,t)\big)
                |\mbs{\varphi}'(s,t)|
            \, ds
        \, dt
        \, : \, 
        \mbs{\varphi}(t_\nu) = \mbs{\varphi}_\nu ,
        \
        \mbs{\varphi}(t_{\nu+1}) = \mbs{\varphi}_{\nu+1}
    \Big\} ,
\end{equation}
where $\Gamma_0$ is the initial dislocation line parameterized by its arc-length $s$ with unit tangent vector $\mbs{t}(s)$, we write
\begin{equation}
    \mbs{\varphi}(s,t) = \mbs{\varphi}(\mbs{x}(s),t) ,
    \qquad
    \mbs{\varphi}'(s,t)
    =
    \nabla\mbs{\varphi}(\mbs{x}(s),t) \mbs{t}(s) ,
\end{equation}
and the minimum is taken over all transport paths taking values $\mbs{\varphi}_\nu$ at time $t_\nu$ and $\mbs{\varphi}_{\nu+1}$ at time $t_{\nu+1}$. In addition, let $E(t,\mbs{\alpha})$ denote the elastic energy of the dislocation measure $\mbs{\alpha}$ at time $t$, where the explicit dependence on time derives from the time dependence of the applied loads.

On this basis, we introduce the incremental energy-dissipation functional
\begin{equation}\label{tlATh3}
    F(\mbs{\varphi}_\nu, \mbs{\varphi}_{\nu+1})
    =
    D(\mbs{\varphi}_\nu, \mbs{\varphi}_{\nu+1})
    +
    E(t_{\nu+1}, (\mbs{\varphi}_{\nu+1})_\#\mbs{\alpha}_0)
    -
    E(t_\nu, (\mbs{\varphi}_\nu)_\#\mbs{\alpha}_0) ,
\end{equation}
and the incremental minimum problem
\begin{equation}\label{T4iUsw}
    \mbs{\varphi}_{\nu+1} \in {\rm argmin} \, F(\mbs{\varphi}_\nu, \, \cdot \, ) .
\end{equation}
We verify that the solution of this problem indeed approximates the mobility law. Taking variations in (\ref{tlATh3}) with respect to $\mbs{\varphi}_{\nu+1}$ and using the path-optimality of the transport map, we obtain
\begin{equation}\label{w2aTRo}
    \int_{\Gamma_{\nu+1}}
        \Big(
            D_i\psi\big(\mbs{v}(\mbs{x}(s),t_{\nu+1})\big)
            -
            f_i (\mbs{x}(s),t_{\nu+1})
        \Big)
        \,
        \eta_i(s)
    \, ds
    =
    0 ,
\end{equation}
which is a weak statement of the mobility law (\ref{rL5pRl}).

In summary, the incremental minimum problem (\ref{T4iUsw}) determines the updated transport map $\mbs{\varphi}_{\nu+1}$, whereupon the updated dislocation measure $\mbs{\alpha}_{\nu+1}$ follows from the {\sl exact geometric update} (\ref{3rouPi}). Specifically, we see from (\ref{tlATh3}) that the updated transport map $\mbs{\varphi}_{\nu+1}$ follows from a competition between the incremental dissipation $D(\mbs{\varphi}_\nu, \mbs{\varphi}_{\nu+1})$, which penalizes departures from $\mbs{\varphi}_\nu$, and the energy $E(t_{\nu+1}, (\mbs{\varphi}_{\nu+1})_\#\mbs{\alpha}_0)$, which drives $\mbs{\varphi}_{\nu+1}$ towards energy minima.

We note the similarity between the incremental dissipation (\ref{mo6kIu}) and the Wasserstein distance between scalar measures \cite{Villani:2003}. It is easy to see that the incremental dissipation $D(\mbs{\varphi}_\nu, \mbs{\varphi}_{\nu+1})$ also defines a distance between dislocation measures. The paths for which the minimum in (\ref{mo6kIu}) is attained are known as {\sl minimizing paths} and arise in theories of inelasticity including plasticity, where they also supply a nexus between time discretization and incremental variational principles \cite{ortiz:1989b}. The minimizing path definition (\ref{mo6kIu}) of the incremental dissipation has the important property that it results in {\sl a priori} energy bounds that in turn ensure the weak convergence of the time-discretized solutions (cf., e.~g., \cite{JordanKinderlehrerOtto1997, JordanKinderlehrerOtto1998, JordanKinderlehrerOtto1999}).

Unfortunately, because of the geometrical evolution of the dislocation line, the minimizing paths that deliver the incremental dissipation (\ref{mo6kIu}) cannot be characterized in closed form and additional approximations are required. A simple scheme consists of restricting the incremental paths to a convenient class, e.~g., piecewise linear paths of the form
\begin{equation}
    \mbs{\varphi}(s,t)
    \approx
    \frac{t_{\nu+1}-t}{t_{\nu+1}-t_\nu}
    \mbs{\varphi}_\nu(s)
    +
    \frac{t - t_\nu}{t_{\nu+1}-t_\nu}
    \mbs{\varphi}_{\nu+1}(s) ,
    \qquad
    t \in [t_\nu, t_{\nu+1}-t],
\end{equation}
whereupon (\ref{mo6kIu}) reduces to
\begin{equation}
    D(\mbs{\varphi}_\nu, \mbs{\varphi}_{\nu+1})
    \approx
    \int_{t_\nu}^{t_{\nu+1}}
        \int_{\Gamma_0}
        \psi
        \Big(
            \frac
            {
                \mbs{\varphi}_{\nu+1}(s)-\mbs{\varphi}_\nu(s)
            }
            {
                t_{\nu+1}-t_\nu
            }
        \Big)
        |\mbs{\varphi}'(s,t)|
        \, ds
    \, dt ,
\end{equation}
or, exchanging the order of integration,
\begin{equation}\label{pOe6oe}
    D(\mbs{\varphi}_\nu, \mbs{\varphi}_{\nu+1})
    \approx
    (t_{\nu+1}-t_\nu)
    \int_{\Gamma_0}
        \psi
        \Big(
            \frac
            {
                \mbs{\varphi}_{\nu+1}(s)-\mbs{\varphi}_\nu(s)
            }
            {
                t_{\nu+1}-t_\nu
            }
        \Big)
        \lambda_{\nu\to\nu+1}(s)
    \, ds ,
\end{equation}
where
\begin{equation}
    \lambda_{\nu\to\nu+1}(s)
    =
    \frac{1}{t_{\nu+1}-t_\nu}
    \int_{t_\nu}^{t_{\nu+1}}
        |\mbs{\varphi}'(s,t)|
    \, dt
\end{equation}
is the average stretch ratio of the dislocation line over the interval $[t_\nu, t_{\nu+1}]$. A further approximation by recourse to the generalized trapezoidal rule gives, explicitly,
\begin{equation}\label{glAt8l}
    \lambda_{\nu\to\nu+1}(s)
    \approx
    (1-\gamma)
    |\mbs{\varphi}'_\nu(s)|
    +
    \gamma
    |\mbs{\varphi}'_{\nu+1}(s)| ,
\end{equation}
with $\gamma \in [0,1]$.

\subsection{Monopole discretization of the dislocation measure}
\label{kvW88u}

Next, we turn to the question of spatial discretization of the incremental minimum problem (\ref{T4iUsw}) and the weak form of the transport equation (\ref{3rouPi}). The structure of these problems reveals the need for two types of approximations: i) the discretization of the dislocation measure $\mbs{\alpha}_{\nu+1}$, and ii) the discretization of the transport map $\mbs{\varphi}_{\nu+1}$. We consider these two approximations in turn.

As already noted, the dislocation measure $\mbs{\alpha}_{\nu+1}$ enters (\ref{T4iUsw}) and (\ref{tlATh3}) linearly and undifferentiated. In addition, the regularized energy (\ref{6iaSwl}) is finite for general measures, including Dirac atoms. Therefore, a natural and computationally convenient spatial discretization of the dislocation measure is as a linear combination of {\sl dislocation monopoles}, i.~e.,
\begin{equation}\label{sti6pH}
    \mbs{\alpha}_\nu
    =
    \sum_{a=1}^M
        \mbs{b}_{a,\nu}\otimes \mbs{\xi}_{a,\nu}
        \, \delta_{\mbs{x}_{a,\nu}} ,
\end{equation}
where $\mbs{x}_{a,\nu}$ is the position of monopole $a$ at time $t_\nu$, $\mbs{b}_{a,\nu}$ is its Burbers vector, $\mbs{\xi}_{a,\nu}$ its element of line, $\delta_{\mbs{x}_{a,\nu}}$ is the Dirac-delta distribution centered at $\mbs{x}_{a,\nu}$, and $M$ is the number of dislocation monopoles. It bears emphasis that (\ref{sti6pH}) represents a totally {\sl unstructured} monopole ensemble and that no connectivity or sequencing between the monopoles is implied by the representation.

For dislocation measures of the form (\ref{sti6pH}), the push-forward (\ref{3rouPi}) reduces to
\begin{equation}
    \sum_{p=1}^M
    (\mbs{b}_{a,\nu+1}\otimes \mbs{\xi}_{a,\nu+1}) \cdot \mbs{\eta}(\mbs{x}_{a,\nu})
    =
    \sum_{p=1}^M
    (\mbs{b}_{a,\nu}\otimes \nabla\mbs{\varphi}_{a,\nu} \, \mbs{\xi}_{a,\nu}) \cdot \mbs{\eta}(\mbs{x}_{a,\nu})
\end{equation}
which must be satisfied for all test functions $\mbs{\eta}$. Hence, we must have
\begin{subequations}\label{0LetiE}
\begin{align}
    & \label{Qk3WSo}
    \mbs{b}_{a,\nu+1} = \mbs{b}_{a,\nu} ,
    \\ & \label{3hIado}
    \mbs{\xi}_{a,\nu+1}
    =
    \nabla\mbs{\varphi}_{\nu\to \nu+1} (\mbs{x}_{a,\nu})
    \mbs{\xi}_{a,\nu} ,
\end{align}
\end{subequations}
i.~e., the monopoles carry a constant Burgers vector and the element of line of every monopole is advected by the local gradient of the incremental transport map
\begin{equation}
    \mbs{\varphi}_{\nu\to \nu+1}
    =
    \mbs{\varphi}_{\nu+1}
    \circ
    \mbs{\varphi}_{\nu}^{-1} .
\end{equation}
Thus, in the absence of topological transitions, i.~e., if the incremental transport map is continuous, the weak reformulation of the dislocation transport problem results trivially in {\sl Burgers vector conservation}, simply by keeping the Burgers vector of all monopoles constant. In addition, the requisite null-divergence property of the dislocation measure is ensured by the geometrically-exact character of the push-forward operations (\ref{0LetiE}).

\subsection{Spatial discretization of the incremental transport map}
\label{KrIth5}

A full spatial discretization additionally requires the interpolation of the incremental transport map $\mbs{\varphi}_{\nu\to \nu+1}$. Since $\mbs{\varphi}_{\nu\to \nu+1}$ and its variations enter the governing equations (\ref{T4iUsw}) and (\ref{3rouPi}) differentiated, its interpolation must be conforming. To this end, we consider general linear interpolation schemes of the form
\begin{equation}\label{eq:SD:Varphih}
    \mbs{\varphi}_{\nu\to \nu+1}(\mbs{x})
    =
    \mbs{x}
    +
    \sum_{a=1}^M (\mbs{x}_{a,\nu+1}-\mbs{x}_{a,\nu}) N_{a,\nu}(\mbs{x}) ,
\end{equation}
with gradient
\begin{equation}\label{c5oAco}
    \nabla \mbs{\varphi}_{\nu\to \nu+1}(\mbs{x})
    =
    \mbs{I}
    +
    \sum_{a=1}^M
    (\mbs{x}_{a,\nu+1}-\mbs{x}_{a,\nu}) \otimes \nabla N_{a,\nu}(\mbs{x}) ,
\end{equation}
where $a$ again indexes the dislocation monopoles, $\{N_{a,\nu}\}_{a=1}^M$ are consistent shape functions at time $t_\nu$ and $\{\mbs{x}_{a,\nu}\}_{a=1}^M$ and $\{\mbs{x}_{a,\nu+1}\}_{a=1}^M$ are the arrays of monopole coordinates at time $t_{\nu}$ and $t_{\nu+1}$, respectively. Consistency here means, specifically, that the shape functions satisfy the identity
\begin{subequations}\label{eq:SD:Cons}
\begin{align}
    & \label{eq:SD:Cons1}
    \sum_{a=1}^M N_{a,\nu}(\mbs{x}) = 1 ,
\end{align}
\end{subequations}
ensuring an exact dislocation update for a uniform translation of all the monopoles. An example of consistent mesh-free interpolation is given in Appendix~\ref{riePr8}.

\subsection{Incremental equilibrium equations}
\label{5UQCjl}

Inserting interpolation (\ref{c5oAco}) into (\ref{3hIado}), we obtain the relation
\begin{equation}\label{8lapOa}
    {\mbs{\xi}}_{a,\nu+1}
    =
    {\mbs{\xi}}_{a,\nu}
    +
    \left(
        \sum_{b=1}^M
        (\mbs{x}_{b,\nu+1}-\mbs{x}_{b,\nu})
        \nabla N_{b,\nu}(\mbs{x}_{a,\nu})
    \right)
    \cdot {\mbs{\xi}}_{a,\nu} ,
\end{equation}
which defines a geometrical update for the monopole elements of line. This relation in turn reveals that the updated elements of line $\{\mbs{\xi}_{a,\nu+1}\}_{a=1}^M$ are fully determined by the updated monopole positions $\{\mbs{x}_{a,\nu+1}\}_{a=1}^M$. Thus, the updated elements of line are not independent variables but are tied to the updated monopole positions. We may therefore render the incremental energy-dissipation function $F$ a sole function of the updated monopole positions by inserting interpolation (\ref{sti6pH}) into (\ref{tlATh3}) with all elements of line updated as in (\ref{8lapOa}). The corresponding incremental equilibrium equations then follow as
\begin{equation}\label{blATl1}
\begin{split}
    \mbs{f}_{a,\nu+1}
    & =
    \frac{\partial F}{\partial \mbs{x}_{a,\nu+1}}
    =
    \sum_{b=1}^M
    \Big(
        \frac{\partial F}{\partial {\mbs{x}}_{a,\nu+1}}
        +
        \frac{\partial F}{\partial {\mbs{\xi}}_{b,\nu+1}}
        \frac{\partial {\mbs{\xi}}_{b,\nu+1}}{\partial \mbs{x}_{a,\nu+1}}
    \Big)
    \\ & =
    \sum_{b=1}^M
    \Big(
        \frac{\partial F}{\partial {\mbs{x}}_{a,\nu+1}}
        +
        \frac{\partial F}{\partial {\mbs{\xi}}_{b,\nu+1}}
        \nabla N_{a,\nu}(\mbs{x}_{b,\nu})\cdot {\mbs{\xi}}_{b,\nu}
    \Big)
    =
    0 ,
\end{split}
\end{equation}
where we have made use of the update (\ref{8lapOa}). We note that the effective forces $\{\mbs{f}_{a,\nu+1}\}_{a=1}^M$ on the monopoles comprise a direct term, corresponding to the direct dependence of $F$ on the updated monopole positions, and a geometrical term resulting from the dependence of $F$ on the updated monopole elements of line.

\subsubsection{Incremental dissipation}

Inserting the monopole approximation (\ref{sti6pH}) into the incremental dissipation (\ref{pOe6oe}), we obtain
\begin{equation}\label{br1eTr}
\begin{split}
    &
    D(\{\mbs{x}_{a,\nu}\}_{a=1}^M, \{\mbs{x}_{a,\nu+1}\}_{a=1}^M)
    \approx \\ &
    (t_{\nu+1}-t_\nu)
    \sum_{a=1}^M
        \psi
        \Big(
            \frac
            {
                \mbs{x}_{a,\nu+1}-\mbs{x}_{a,\nu}
            }
            {
                t_{\nu+1}-t_\nu
            }
        \Big)
        \Big(
            (1-\gamma)
            |\mbs{\xi}_{a,\nu}|
            +
            \gamma
            |\mbs{\xi}_{a,\nu+1}|
        \Big)
    \, ds ,
\end{split}
\end{equation}
where we have used (\ref{glAt8l}) and $\{\mbs{\xi}_{a,\nu+1}\}_{a=1}^M$ is tied to $\{\mbs{x}_{a,\nu+1}\}_{a=1}^M$ through the geometrical update (\ref{8lapOa}). It follows from (\ref{blATl1}) that the corresponding monopole forces (\ref{blATl1}) consist of a direct term and a geometrical term. The direct term encodes the dilocation mobility law whereas the geometrical term takes into account the advection and stretching of the dislocation line. We note that the geometrical term vanishes for the particular choice $\gamma=0$.

\subsubsection{Regularized linear elasticity}

Inserting the monopole representation (\ref{sti6pH}) into the regularized energy (\ref{thieV5}), we obtain
\begin{equation}\label{qoU8Le}
    E^\epsilon(\{\mbs{x}_{a,\nu+1}\}_{a=1}^M)
    =
    \sum_{a=1}^M E^\epsilon_{a,\nu+1}
    +
    \sum_{a=1}^M \sum_{b=1 \atop b\neq a}^M
    E^\epsilon_{ab,\nu+1} ,
\end{equation}
where
\begin{equation}\label{pI5kle}
\begin{split}
    &
    E^\epsilon_{ab,\nu+1}
    =
    -
    \frac{\mu}{4\pi}
        S^\epsilon(\mbs{x}_{a,\nu+1}, \mbs{x}_{b,\nu+1})
        (\mbs{b}_{a,\nu+1}\times \mbs{b}_{b,\nu+1})
        \cdot
        (\mbs{\xi}_{a,\nu+1}\times \mbs{\xi}_{b,\nu+1})
    \\ & +
    \frac{\mu}{8\pi}
        S^\epsilon(\mbs{x}_{a,\nu+1}, \mbs{x}_{b,\nu+1})
        (\mbs{b}_{a,\nu+1}\cdot \mbs{\xi}_{a,\nu+1})
        (\mbs{b}_{b,\nu+1}\cdot \mbs{\xi}_{b,\nu+1})
    \\ & +
    \frac{\mu}{8\pi(1-\nu)}
        (\mbs{b}_{a,\nu+1}\times \mbs{\xi}_{a,\nu+1})
        \cdot
        \mbs{T}^\epsilon(\mbs{x}_{a,\nu+1}, \mbs{x}_{b,\nu+1})
        \cdot
        (\mbs{b}_{b,\nu+1}\times \mbs{\xi}_{b,\nu+1}) ,
\end{split}
\end{equation}
is the interaction energy between monopoles $a$ and $b$. In addition, the self-energy of the monopoles is obtained by taking the limit of $\mbs{x}_b \to \mbs{x}_a$, with the explicit result
\begin{equation}\label{jlArI1}
    E^\epsilon_{a,\nu+1}
    =
    \frac{\mu}{8\pi}
    \frac{1}{2\epsilon} (\mbs{b}_{a,\nu+1}\cdot \mbs{\xi}_{a,\nu+1})^2
    +
    \frac{\mu}{8\pi(1-\nu)}
    \frac{1}{3\epsilon} |\mbs{b}_{a,\nu+1}\times \mbs{\xi}_{a,\nu+1}|^2 .
\end{equation}
The essential role of the regularization of the elastic energy is clear in these expressions. In particular, the self-energy of the monopoles is finite but diverges as $\epsilon \to 0$, as expected.

We note that the self-energy (\ref{jlArI1}) of the monopoles depends on the angle subtended by the Burgers vector and the element of line. This dependence introduces a {\sl line-tension anisotropy} that favors certain monopole directions over others. For instance, in the usual range of $\nu > 0$, screw monopoles, $\mbs{b} \times \mbs{\xi} = \mbs{0}$, have lower energy than---and therefore are favored over---edge monopoles $\mbs{b} \cdot \mbs{\xi} = 0$. In BCC crystals, this line-tension anisotropy is specially pronounced, resulting is a proliferation of long screw segments.

Applications are often concerned with the motion of dislocations under the action of an applied stress $\mbs{\sigma}^\infty$. The effect of the applied stress is to add the term
\begin{equation}\label{8hIuqi}
    E^{\rm ext}(\{\mbs{x}_{a,\nu+1}\}_{a=1}^M)
    =
    \sum_{a=1}^M
        ((\mbs{\sigma}^\infty \mbs{b}_{a,\nu+1})
        \times
        {\mbs{\xi}}_{a,\nu+1} ) \cdot \mbs{x}_{a,\nu+1}
\end{equation}
to the total energy. Rearranging terms, we can alternatively write (\ref{8hIuqi}) in the form
\begin{equation}
    E^{\rm ext}(\{\mbs{x}_{a,\nu+1}\}_{a=1}^M)
    =
    -
    V \,
    \mbs{\sigma}^\infty
    \cdot
    \mbs{\epsilon}^p_{\nu+1} ,
\end{equation}
cf.~eq.~(\ref{LYw64C}), where
\begin{equation}
    \mbs{\epsilon}^p_{\nu+1}
    =
    -
    \frac{1}{V}
    \sum_{a=1}^M
        \mbs{b}_{a,\nu+1} \odot (\mbs{\xi}_{a,\nu+1} \times \mbs{x}_{a,\nu+1})
\end{equation}
is the effective or {\sl macroscopic plastic strain}, $V$ is a macroscopic volume and $\mbs{a} \odot \mbs{b} = (\mbs{a} \otimes \mbs{b} + \mbs{b} \otimes \mbs{a})/2$ denotes the symmetric dyadic product of vectors $\mbs{a}$ and $\mbs{b}$.

In computing all contributions to the energy, we regard the updated monopoles element of line $\{\mbs{\xi}_{a,\nu+1}\}_{a=1}^M$ as tied to updated monopole positions $\{\mbs{x}_{a,\nu+1}\}_{a=1}^M$ through the geometrical update (\ref{8lapOa}). The corresponding energetic forces on the monopoles then comprise direct terms, resulting from the dependence of the energy on the updated monopole positions, and geometrical terms, resulting from the dependence of the energy on the updated monopole elements of line.

\subsection{Summary of update algorithm}
\label{Xe0cjC}

\begin{algorithm}
\begin{algorithmic}[1]
\STATE Compute shape functions $\{N_{a,\nu}\}_{a=1}^M$ and $\{\nabla N_{a,\nu}\}_{a=1}^M$ from $\{\mbs{x}_{a,\nu}\}_{a=1}^M$.
\STATE Solve incremental equilibrium equations: $\mbs{f}_{a,\nu+1}=0$ for $\{\mbs{x}_{a,\nu+1}\}_{a=1}^M$.
\STATE Update monopole line elements $\{\mbs{\xi}_{a,\nu}\}_{a=1}^M$, including splitting.
\STATE Reset $\nu\leftarrow \nu+1$, go to (1).
\end{algorithmic}
\caption{\small {\sc Optimal transport of dislocation monopoles.}}
\label{Al:OTM:SF}
\end{algorithm}

The monopole time-stepping algorithm is summarized in Algorithm~\ref{Al:OTM:SF}. The forward solution has the usual structure of implicit time-integration and updated-Lagrangian schemes. The updated monopole positions are computed by solving the incremental equilibrium equations (\ref{blATl1}). The update of the monopole line elements is then effected explicitly through the push-forward operations (\ref{0LetiE}). In calculations, we solve the equilibrium equations (\ref{blATl1}) using the Polak-Ribi\`ere iterative solver \cite{Polak1969} or Scalable Nonlinear Equations Solvers (SNES) in the PETSc library of the Argonne National Laboratory \cite{petsc-efficient}.

The scheme leaves considerable latitude as regards the choice of shape functions for the interpolation of the transport maps. A particularly powerful method for formulating interpolation schemes of any order is provided by maximum-entropy inference \cite{ArroyoOrtiz:2006}. The details of this approach, as it applies in the present context, are summarized in Appendix A. We note that max-ent interpolation introduces a range of interaction $h_a=1/\beta_a^2$ for every monopoles, where $\{\beta_a\}_{a=1}^M$ are parameters of the interpolation. Specifically, the transport map at monopole $a$, and derivatives thereof, depends predominantly on the cluster of monopoles in the $h_a$-neighborhood. A simple form of {\sl adaptivity} is to tie the parameters $\{\beta_a\}_{a=1}^M$ to the length of the corresponding line elements through the constraint
\begin{equation}
    \beta_a |\mbs{\xi}_a|^2 = {\rm constant}.
\end{equation}
In all calculations presented subsequently, we set the constant to $1/2$.


\begin{figure}
\begin{center}
	\begin{subfigure}{0.49\textwidth}\caption{}
        \includegraphics[width=0.95\linewidth]{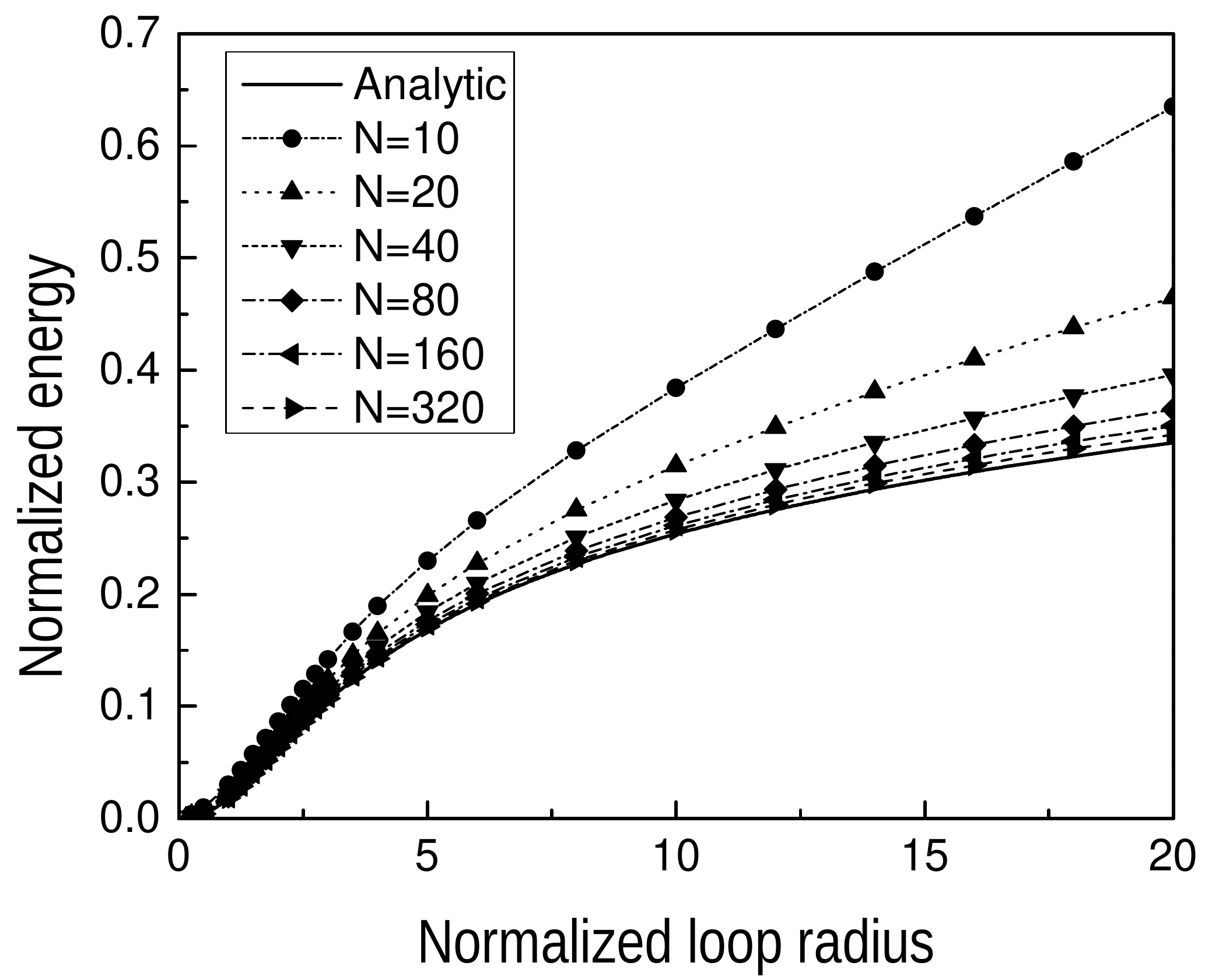}
	\end{subfigure}
	\begin{subfigure}{0.49\textwidth}\caption{}
        \includegraphics[width=0.99\linewidth]{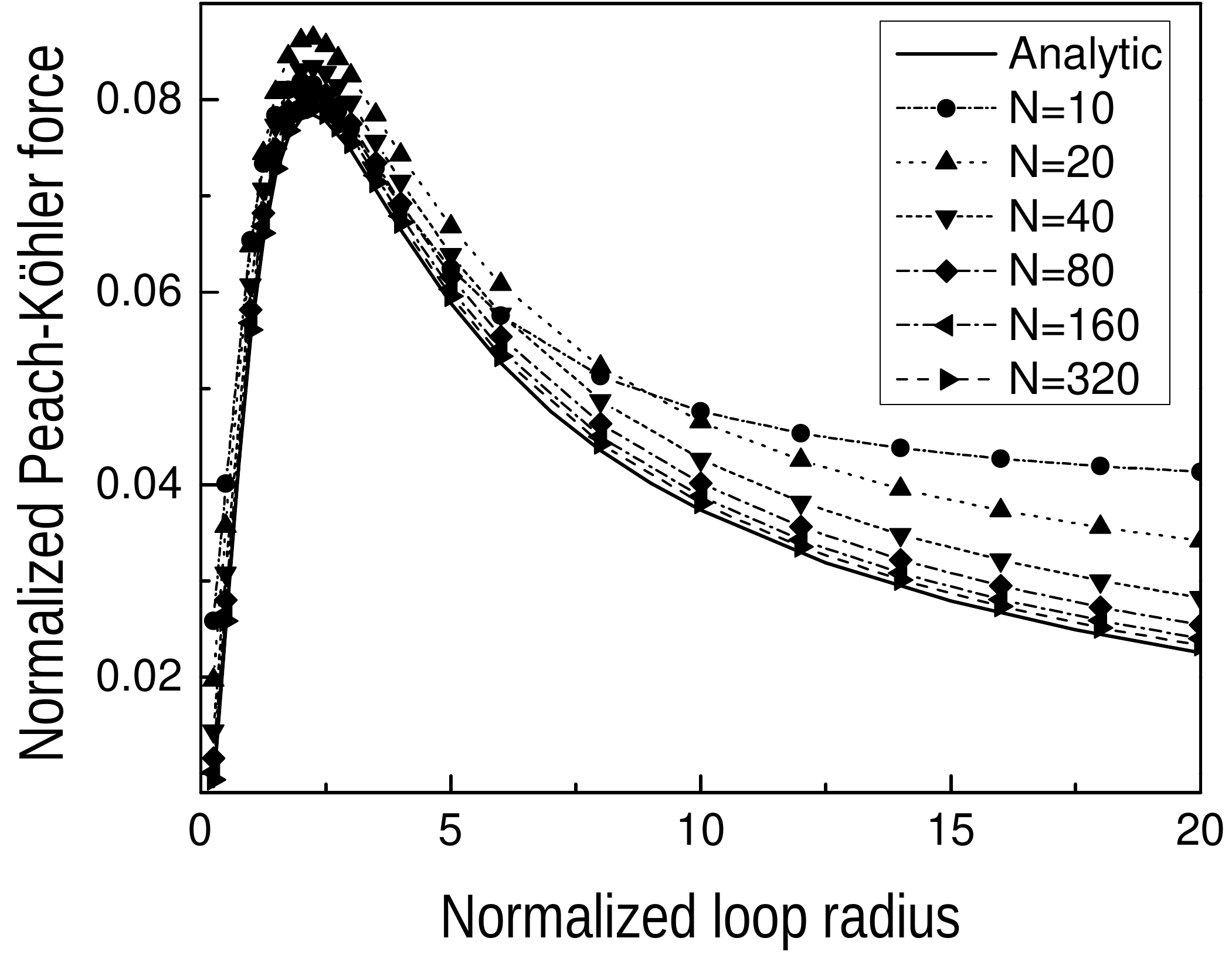}
	\end{subfigure}
\caption{\small Convergence with respect to the number of monopoles for a circular prismatic loop. a) Regularized elastic energy normalized by $\frac{\mu b^2 \rho^2}{8\pi(1-\nu)\epsilon}$. b) Regularized Peach-K\"ohler force normalized by $\frac{\mu b^2 }{8\pi(1-\nu)\epsilon}$. Loop radius normalized by $\epsilon$.}
\label{9oeMlA}
\end{center}
\end{figure}

While a complete analysis of convergence is beyond the scope of this paper,
we illustrate the convergence properties of the monopole approximation by means of the simple example of a circular prismatic loop. Fig.~\ref{9oeMlA} illustrates the convergence of the regularized energy and Peach-K\"ohler force per unit length with respect to the number of monopoles. As may be seen from the figure, coarse discretizations of the loop tend to be overly stiff and overestimate the energy and Peach-K\"ohler force per unit length. The convergence of the monopole approximation with increasing number of monopoles is also evident in the figure.


\begin{figure}
\begin{center}
	\begin{subfigure}{0.49\textwidth}\caption{}
        \includegraphics[width=0.99\linewidth]{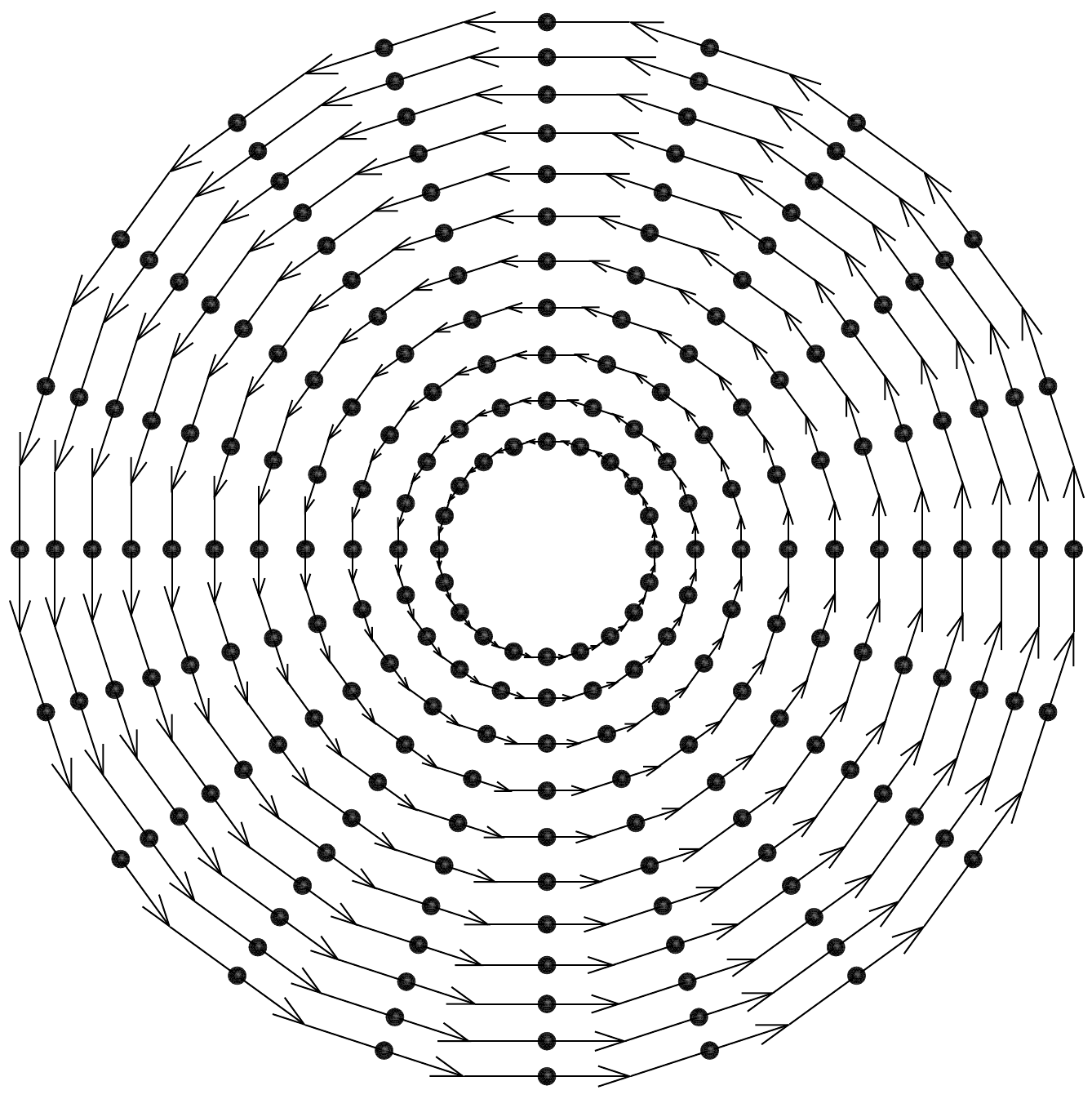}
	\end{subfigure}
	\begin{subfigure}{0.49\textwidth}\caption{}
        \includegraphics[width=0.99\linewidth]{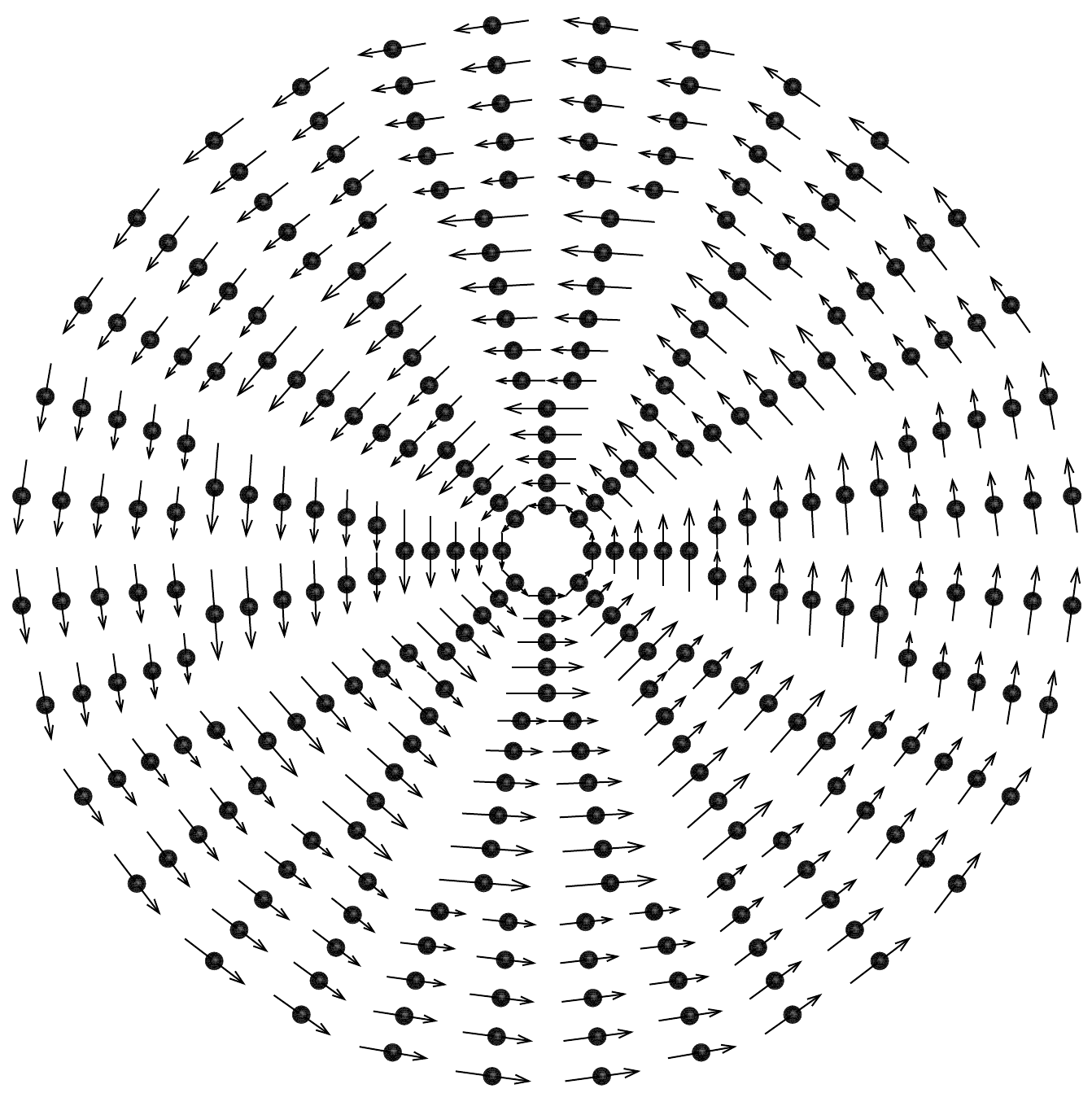}
	\end{subfigure}
\caption{\small Planar circular prismatic loop. a) Evolution in the absence of applied load, illustrating the geometrical update of the monopole line elements in a converging geometry. b) Expansion of the loop under a constant applied normal stress $\sigma^\infty$, illustrating the monopole line-element splitting scheme.}
\label{qSJ46d}
\end{center}
\end{figure}

The geometrical update (\ref{0LetiE}) is illustrated in Fig.~\ref{qSJ46d}a with the aid of a converging circular prismatic loop example. In this example, the initial values of the monopole positions and line segments is prescribed and the loop subsequently shrinks in the absence of an applied stress under the action of its Peach-K\"ohler self-force and linear kinetics. As may be seen from the figure, the length of the monopoles decreases proportionally to the radius, as required, cf.~Example~\ref{oF8lo5}. It bears emphasis that no aligment or compatibility between the monopoles is enforced at any time during the calculations. Instead, the monopoles align spontaneously in order to attain a low-energy configuration. In addition, the head-to-toe compatibility between adjacent monopoles is a direct consequence of the divergence-free character of the initial dislocation measure and the exactness of the geometrical update.


For expanding geometries, the geometrical update (\ref{0LetiE}) results in line stretching and, potentially, in excessively long monopole line elements, with a deleterious effect on accuracy. We prevent this loss of accuracy by splitting the monopoles when they exceed a prespecified length or self-energy. Fig.~\ref{qSJ46d}b illustrates the monopole splitting scheme in the case of a prismatic circular loop expanding under the action of a stress normal to its plane and linear kinetics. As may be seen from the figure, the splitting scheme ensures that the length of the monopoles remains within acceptable bounds as the loop expands.

\section{Topological transitions}
\label{O5HY85}

The preceding developments are predicated on the assumption that the transport maps are continuous. Under such conditions, the topology of the dislocation measure remains invariant. In particular, the Burgers vectors of the monopoles remain constant through the motion, eq.~(\ref{Qk3WSo}). In actual dislocation dynamics, topological transitions occur due to a number of processes, including dislocation nucleation, dislocation reactions and junction formation, among others (cf., e.~g., \cite{BulatovCai:2006}). Topological transitions require additional logic, or 'rules', to be added to the monopole dynamics. Some basic topological transitions that play a role in subsequent calculations are discussed next.


\subsection{Dislocation reactions}

\begin{figure}
\begin{center}
	\begin{subfigure}{0.49\textwidth}\caption{}
        \includegraphics[width=0.99\linewidth]{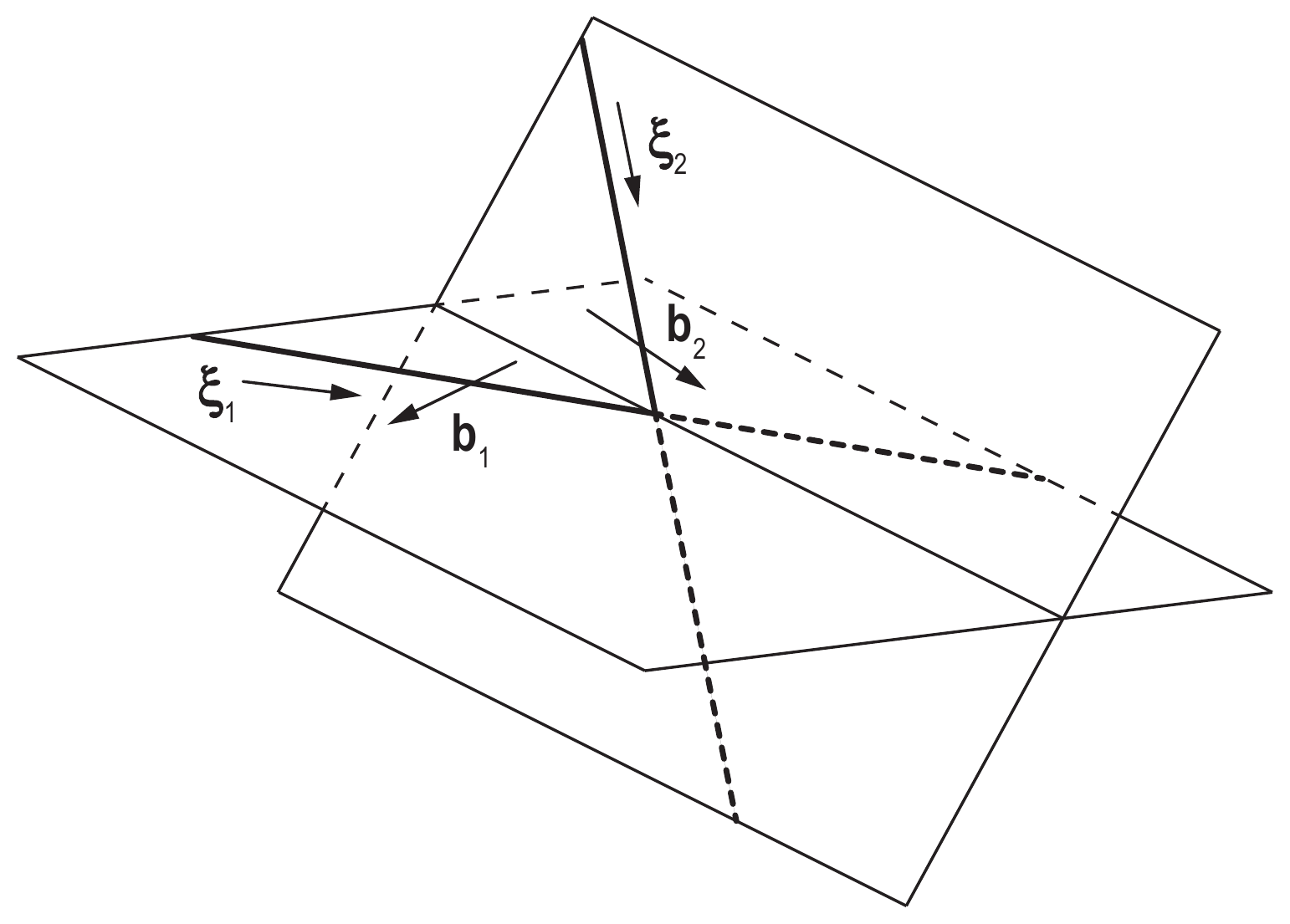}
	\end{subfigure}
	\begin{subfigure}{0.49\textwidth}\caption{}
        \includegraphics[width=0.99\linewidth]{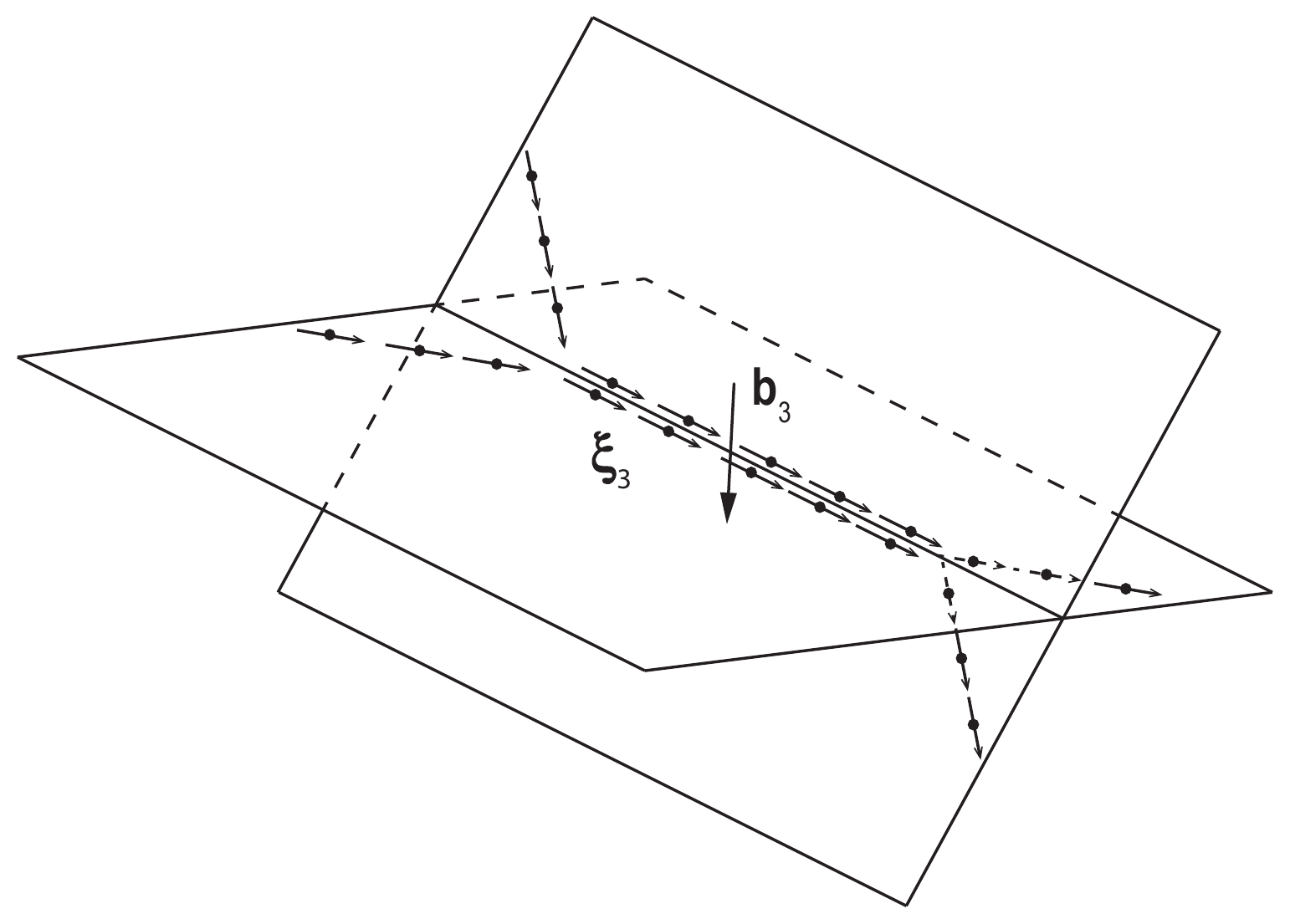}
	\end{subfigure}
\caption{\small Schematic of junction formation. Two intersecting glissile dislocation lines of Burgers vectors $\mbs{b}_1$ and $\mbs{b}_2$ {\sl zip up} along a line of direction $\mbs{\xi}$ on the intersection between their slip planes to form a sessile segment of Burgers vector $\mbs{b}_3=\mbs{b}_1+\mbs{b}_2$. }\label{03leSI}
\end{center}
\end{figure}

Topological transitions may result from {\sl monopole-monopole reactions} of the type
\begin{equation}
    \mbs{b}_1\otimes \mbs{\xi}_1 \delta_{\mbs{x}}
    +
    \mbs{b}_2\otimes \mbs{\xi}_2 \delta_{\mbs{x}}
    \to
    \mbs{b}_3\otimes \mbs{\xi}_3 \delta_{\mbs{x}} ,
\end{equation}
subject to the Burgers-vector conservation constraint
\begin{equation}\label{ZeP9kT}
    \mbs{b}_1
    +
    \mbs{b}_2
    =
    \mbs{b}_3 ,
\end{equation}
where $\mbs{b}_1\otimes \mbs{\xi}_1 \delta_{\mbs{x}}$ and $\mbs{b}_2\otimes \mbs{\xi}_2 \delta_{\mbs{x}}$ are the precursor monopoles and $\mbs{b}_3\otimes \mbs{\xi}_3 \delta_{\mbs{x}}$ is the product monopole. {\sl Pair-annihilation} represents a special type of dislocation reaction in which the reacting monopoles have equal and opposite Burgers vectors and the reaction product is a null monopole. {\sl Junction formation}, Fig.~\ref{03leSI}, entails another special type of dislocation reaction in which two intersecting glissile dislocation lines of Burgers vectors $\mbs{b}_1$ and $\mbs{b}_2$ {\sl zip up} along a line of direction $\mbs{\xi}_3$ on the intersection between their slip planes to form a sessile segment of Burgers vector $\mbs{b}_3=\mbs{b}_1+\mbs{b}_2$.

The stability of monopole-monopole reactions may be elucidated by comparing the energies before and after the reaction. Thus, before the reaction we have from (\ref{qoU8Le})
\begin{equation}
    E^\epsilon
    =
    E^\epsilon_{\rm self}
    +
    E^\epsilon_{\rm int} ,
\end{equation}
where
\begin{equation}
\begin{split}
    E^\epsilon_{\rm self}
    & =
    \frac{\mu}{8\pi}
    \frac{1}{2\epsilon} (\mbs{b}_1\cdot \mbs{\xi}_1)^2
    +
    \frac{\mu}{8\pi(1-\nu)}
    \frac{1}{3\epsilon} |\mbs{b}_1\times \mbs{\xi}_1|^2
    \\ & +
    \frac{\mu}{8\pi}
    \frac{1}{2\epsilon} (\mbs{b}_2\cdot \mbs{\xi}_2)^2
    +
    \frac{\mu}{8\pi(1-\nu)}
    \frac{1}{3\epsilon} |\mbs{b}_2\times \mbs{\xi}_2|^2 ,
\end{split}
\end{equation}
is the self-energy of the precursor monopoles, eq.~(\ref{jlArI1}), and
\begin{equation}\label{58JWSD}
\begin{split}
    &
    E^\epsilon_{\rm int}
    =
    \frac{\mu}{48\pi(1-\nu)\epsilon}
    \Big(
        3(1-\nu)
        (\mbs{b}_1\cdot\mbs{\xi}_1)
        (\mbs{b}_2\cdot\mbs{\xi}_2)
        \\ & -
        6(1-\nu)
        (\mbs{b}_1\times\mbs{b}_2)
        \cdot
        (\mbs{\xi}_1\times\mbs{\xi}_2)
        +
        2
        (\mbs{b}_1\times\mbs{\xi}_1)
        \cdot
        (\mbs{b}_2\times\mbs{\xi}_2)
    \Big)
\end{split}
\end{equation}
is the interaction energy of the precursor monopoles, obtained by taking the limit of $|\mbs{x}_1-\mbs{x}_2 | \to 0$ in eq.~(\ref{pI5kle}). After the reaction, the energy of the product monopole is
\begin{equation}
    E^\epsilon
    =
    \frac{\mu}{8\pi}
    \frac{1}{2\epsilon} (\mbs{b}_3\cdot \mbs{\xi}_3)^2
    +
    \frac{\mu}{8\pi(1-\nu)}
    \frac{1}{3\epsilon} |\mbs{b}_3\times \mbs{\xi}_3|^2 .
\end{equation}
The {\sl stability diagram} of the monopole-monopole reaction is
\begin{subequations}
\begin{align}
    &
    \Delta E^\epsilon < 0
    \Rightarrow \text{stable} ,
    \\ &
    \Delta E^\epsilon = 0
    \Rightarrow \text{indifferent} ,
    \\ &
    \Delta E^\epsilon > 0
    \Rightarrow \text{unstable} .
\end{align}
\end{subequations}
where $\Delta E^\epsilon$ is the difference between the energies after and before the reaction.

In the particular case $\mbs{\xi}_1 = \mbs{\xi}_2 = \mbs{\xi}$, a straightforward calculation using (\ref{ZeP9kT}) gives
\begin{equation}\label{3LKtS6}
    \Delta E^\epsilon
    =
    \frac{\mu}{16\pi\epsilon}
    (\mbs{b}_1\cdot\mbs{\xi})
    (\mbs{b}_2\cdot\mbs{\xi})
    +
    \frac{\mu}{24\pi(1-\nu)\epsilon}
    (\mbs{b}_1\times\mbs{\xi})
    \cdot
    (\mbs{b}_2\times\mbs{\xi}) .
\end{equation}
In addition, for this particular geometry the sequence of dipoles
\begin{equation}
    \mbs{\alpha}_h
    =
    \mbs{b}_1\otimes \mbs{\xi} \delta_{\mbs{x}+\epsilon_h \mbs{e}}
    +
    \mbs{b}_2\otimes \mbs{\xi} \delta_{\mbs{x}-\epsilon_h \mbs{e}} ,
\end{equation}
where $\epsilon_h \downarrow 0$ and $\mbs{e}$ is a direction of approach, converges to the reaction product
\begin{equation}\label{YZ9vbx}
    \mbs{\alpha}
    =
    \mbs{b}_3\otimes \mbs{\xi}
    \delta_{\mbs{x}}
\end{equation}
weakly in the sense of measures, i.~e., $\mbs{\alpha}_h \rightharpoonup \mbs{\alpha}$. However, we see from (\ref{3LKtS6}) that $\Delta E^\epsilon \neq 0$ in general. This shows that, as expected, the energy $E^\epsilon(\mbs{\alpha})$ is not weakly continuous with respect to the dislocation measure $\mbs{\alpha}$. Eq.~(\ref{3LKtS6}) also shows that $\Delta E^\epsilon$ can be positive for some reactions, which additionally shows that $E^\epsilon$ is not weakly lower-semicontinuous. Thus, whereas pair annihilation, $ \mbs{b} - \mbs{b} \to \mbs{0}$, and monopole splitting, $2\mbs{b} \to \mbs{b} + \mbs{b}$, lower the energy, monopole pairing, $\mbs{b} + \mbs{b} \to 2\mbs{b}$, increases the energy. This lack of weak lower-semicontinuity has far-reaching consequences for microstructural evolution, as the crystal can lower its energy, or {\sl relax}, through microstructural rearrangements involving annihilation, splitting, network formation and other mechanisms \cite{Conti:2016}.

In calculations, monopole reactions can be accounted for simply by introducing a {\sl capture distance} and replacing the approaching monopoles by their reaction product if the energy is decreased. However, we note from (\ref{YZ9vbx}) that general reaction products can be {\sl rank-two} monopoles, which adds a certain complexity to the implementation.

\subsection{Loop nucleation}
\label{u7HnBd}

Dislocations are nucleated during plastic slip through a number of mechanisms including Frank-Read sources, double cross-slip and others \cite{hirth:1968}. In calculations, we model nucleation simply by introducing small loops of fixed radius $\rho_0$, e.~g., commensurate with the radius of operation of a Frank-Read source, at prespecified source locations provided that the total energy of the system is decreased, i.~e., provided that
\begin{equation}
    \Delta E^\epsilon \leq b \tau^\infty \pi \rho_0^2 ,
\end{equation}
where $\Delta E^\epsilon$ is the increase in elastic energy due to the introduction of the loop and $\tau^\infty$ is the applied resolved shear stress on the slip system into which the loop is introduced. After nucleation, the new loop shields the source and its operation is shut off until the loop becomes sufficiently large. This transient shielding results in the intermittent emission of loops from the sources.

\section{Numerical examples}
\label{p9T9VL}

In this section, we present selected examples of application that illustrate the range and scope of the method of monopoles presented in the foregoing. We specifically consider the case of a single BCC grain embedded in an elastic matrix. The grain has the shape of a truncated octahedron and the grain boundary is assumed to be impenetrable to dislocations. The impenetrability condition is enforced by means of a potential that penalizes monopole excursions outside the grain. The grain deforms by crystallographic slip on the $12$ slip systems in the classs $\{110\}\langle 111 \rangle$ under the action of a remotely applied uniaxial stress and the dislocation motion obeys linear kinetics. The calculations are carried out as described in Section~\ref{Xe0cjC} and with $\gamma=0$ in (\ref{glAt8l}). Several scenarios of increasing complexity are considered. We emphasize that these scenarios are intended to demonstrate numerical capability and not to provide physically accurate quantitative predictions of material behavior.

\subsection{Activation of a single slip plane}

\begin{figure}
\begin{center}
	\begin{subfigure}{0.30\textwidth}\caption{}
        \includegraphics[width=0.99\linewidth]{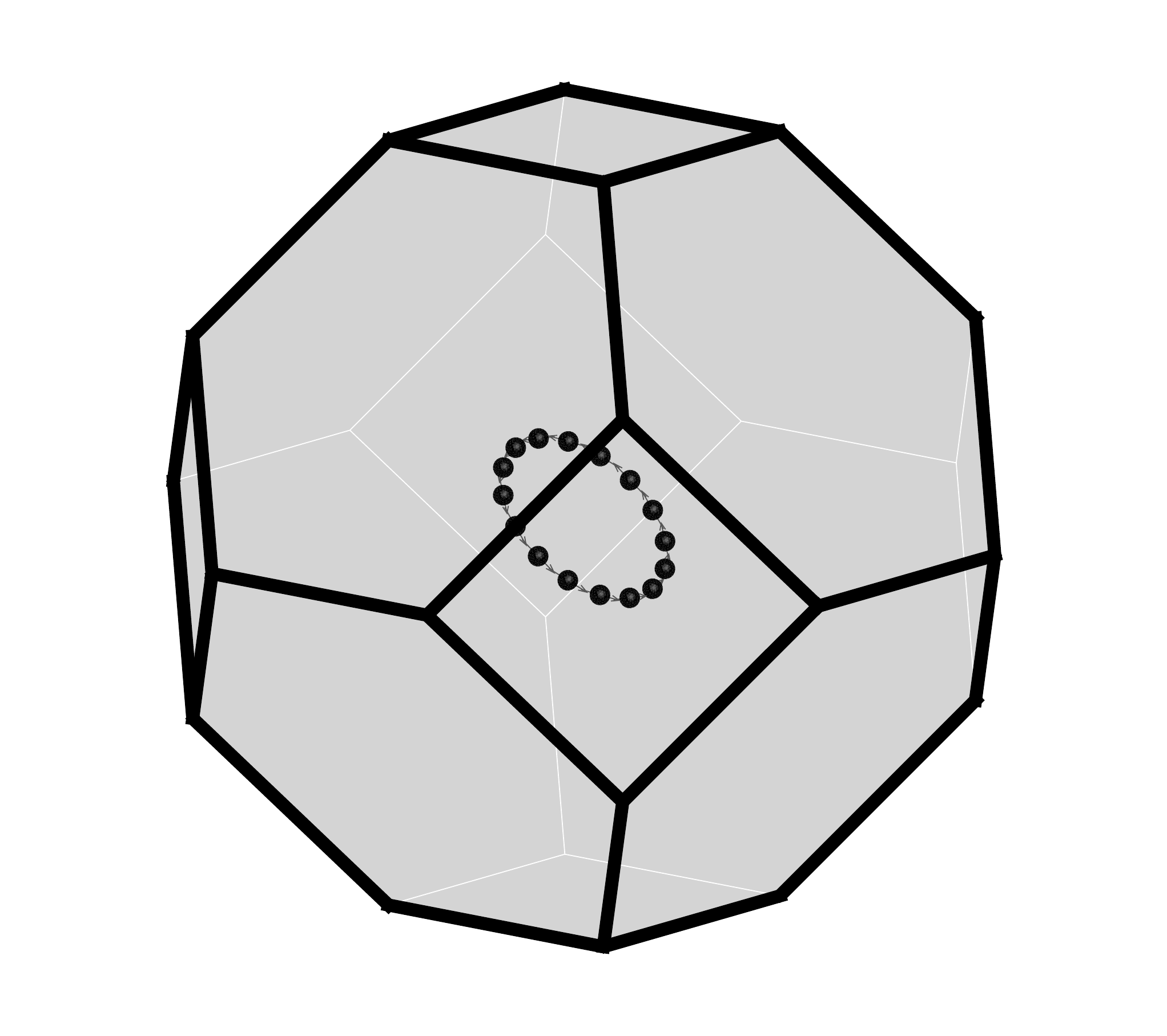}
	\end{subfigure}
	\begin{subfigure}{0.30\textwidth}\caption{}
        \includegraphics[width=0.99\linewidth]{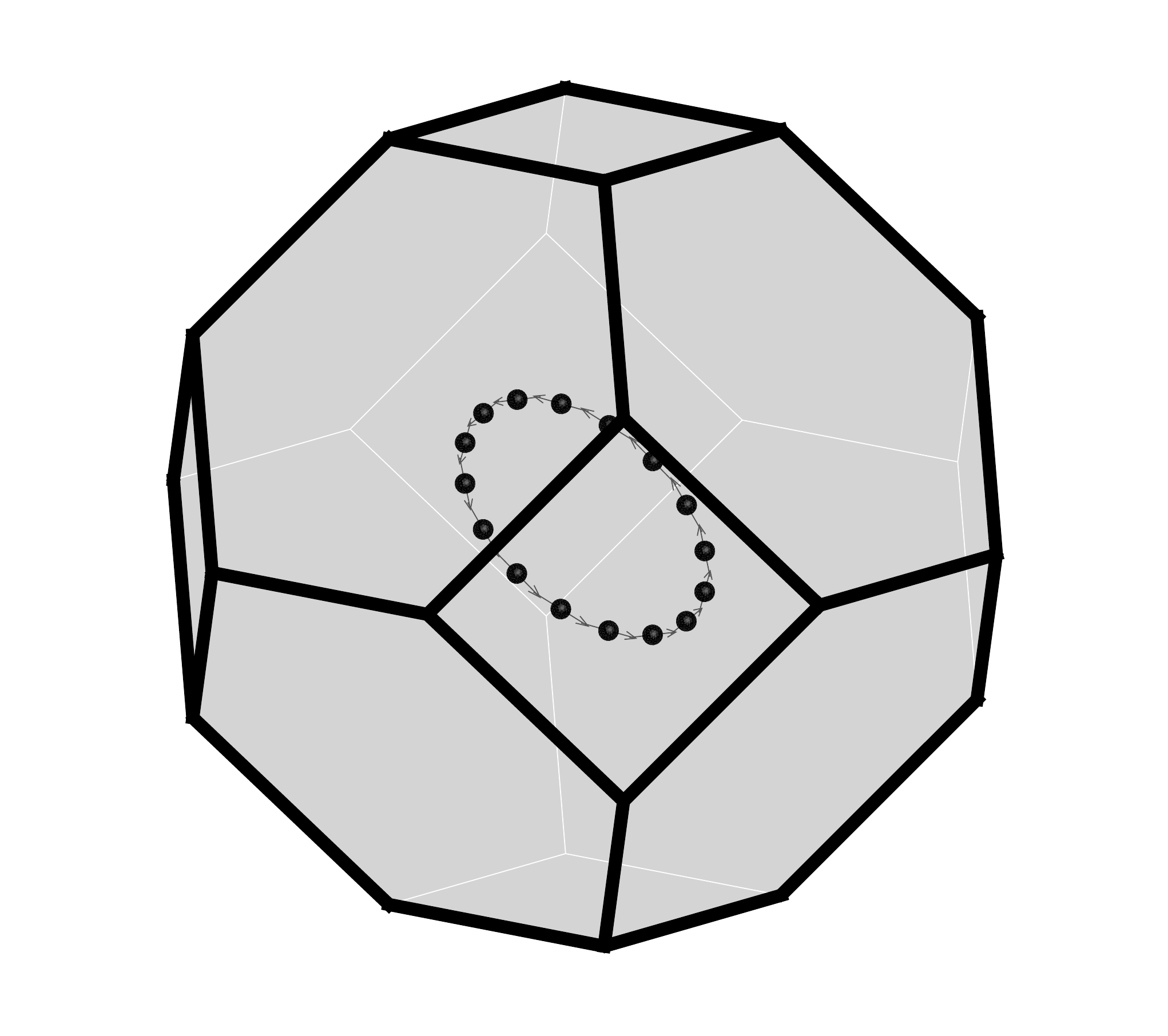}
	\end{subfigure}
	\begin{subfigure}{0.30\textwidth}\caption{}
        \includegraphics[width=0.99\linewidth]{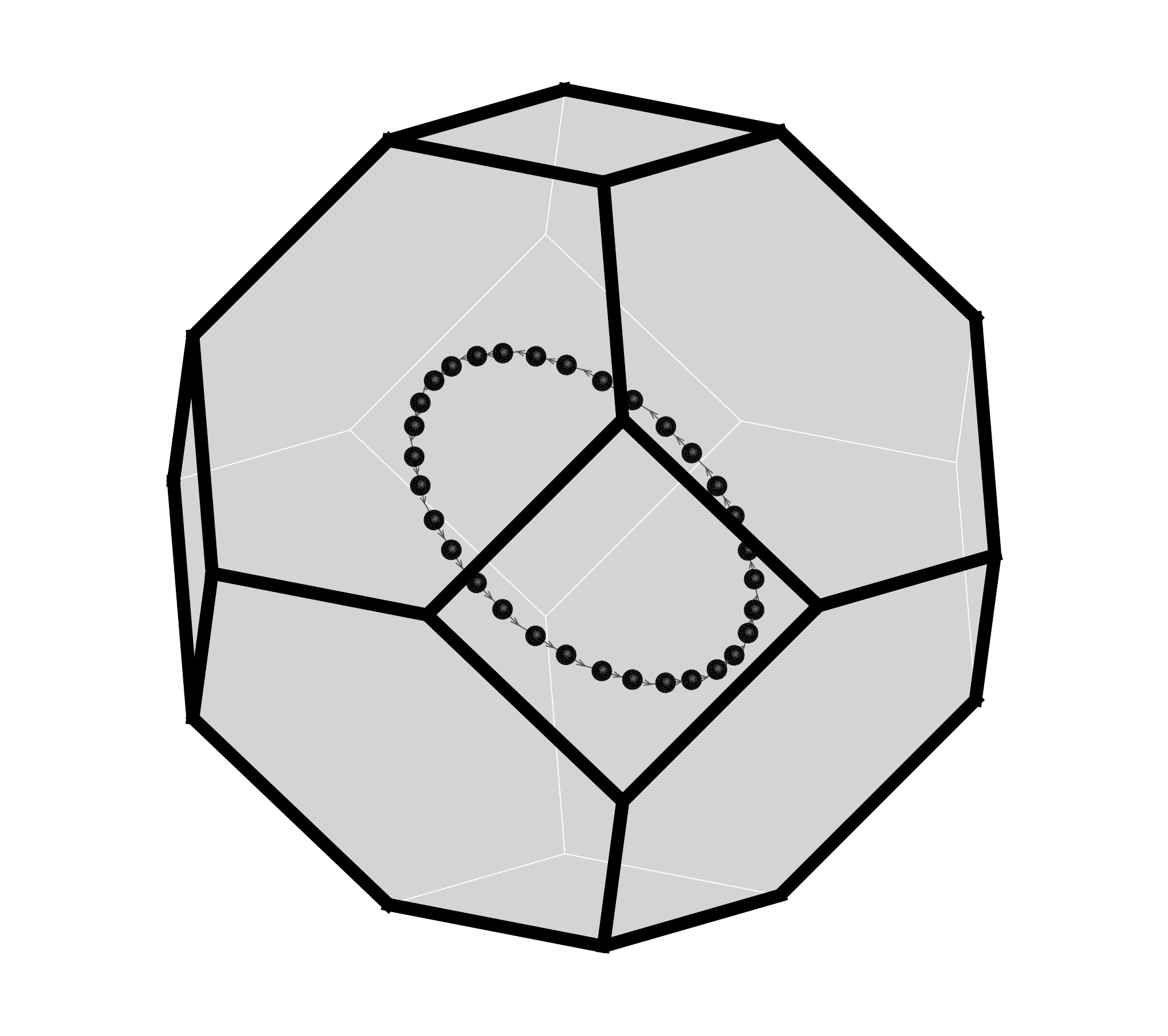}
	\end{subfigure}
	\begin{subfigure}{0.30\textwidth}\caption{}
        \includegraphics[width=0.99\linewidth]{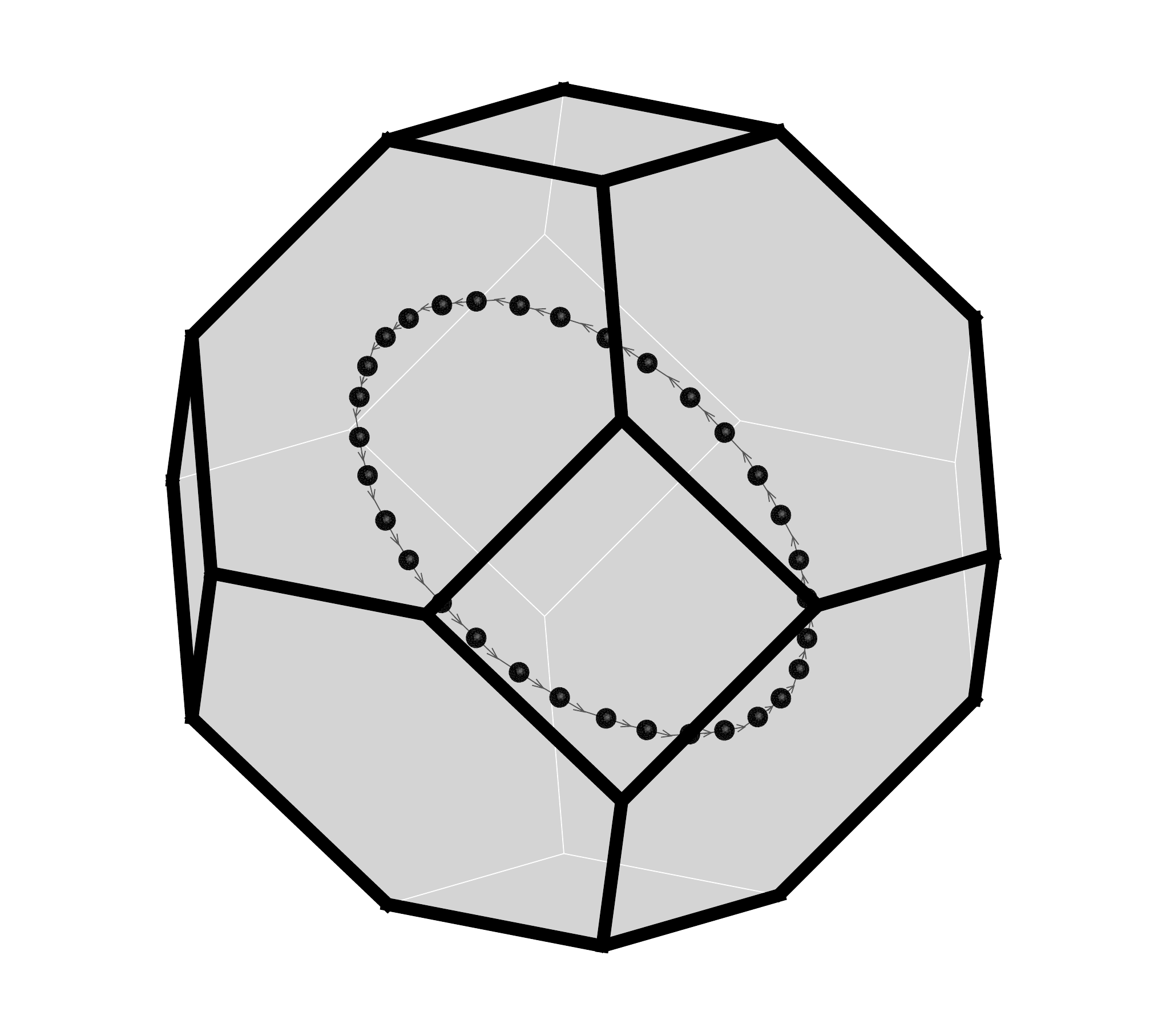}
	\end{subfigure}
	\begin{subfigure}{0.30\textwidth}\caption{}
        \includegraphics[width=0.99\linewidth]{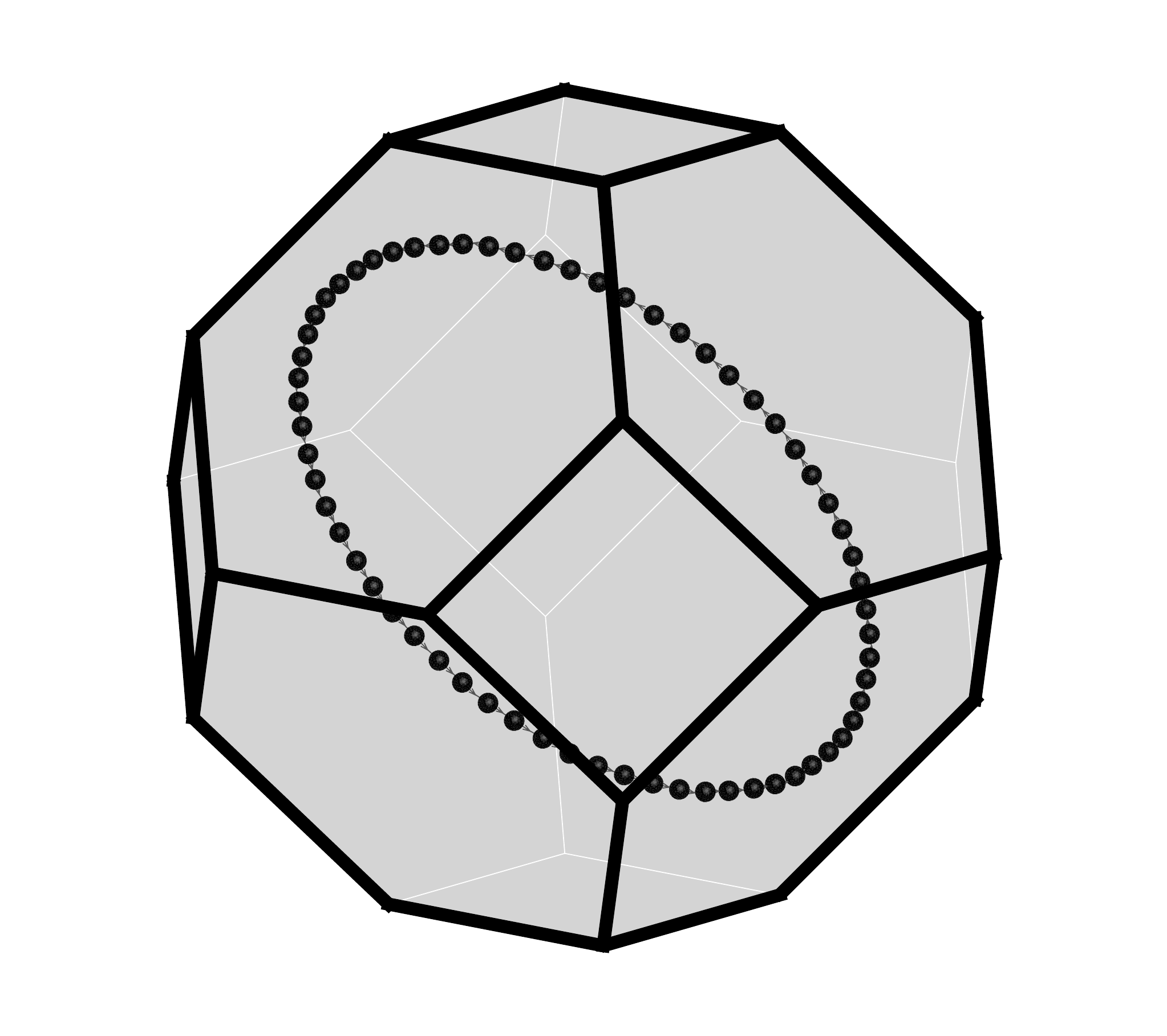}
	\end{subfigure}
	\begin{subfigure}{0.30\textwidth}\caption{}
        \includegraphics[width=0.99\linewidth]{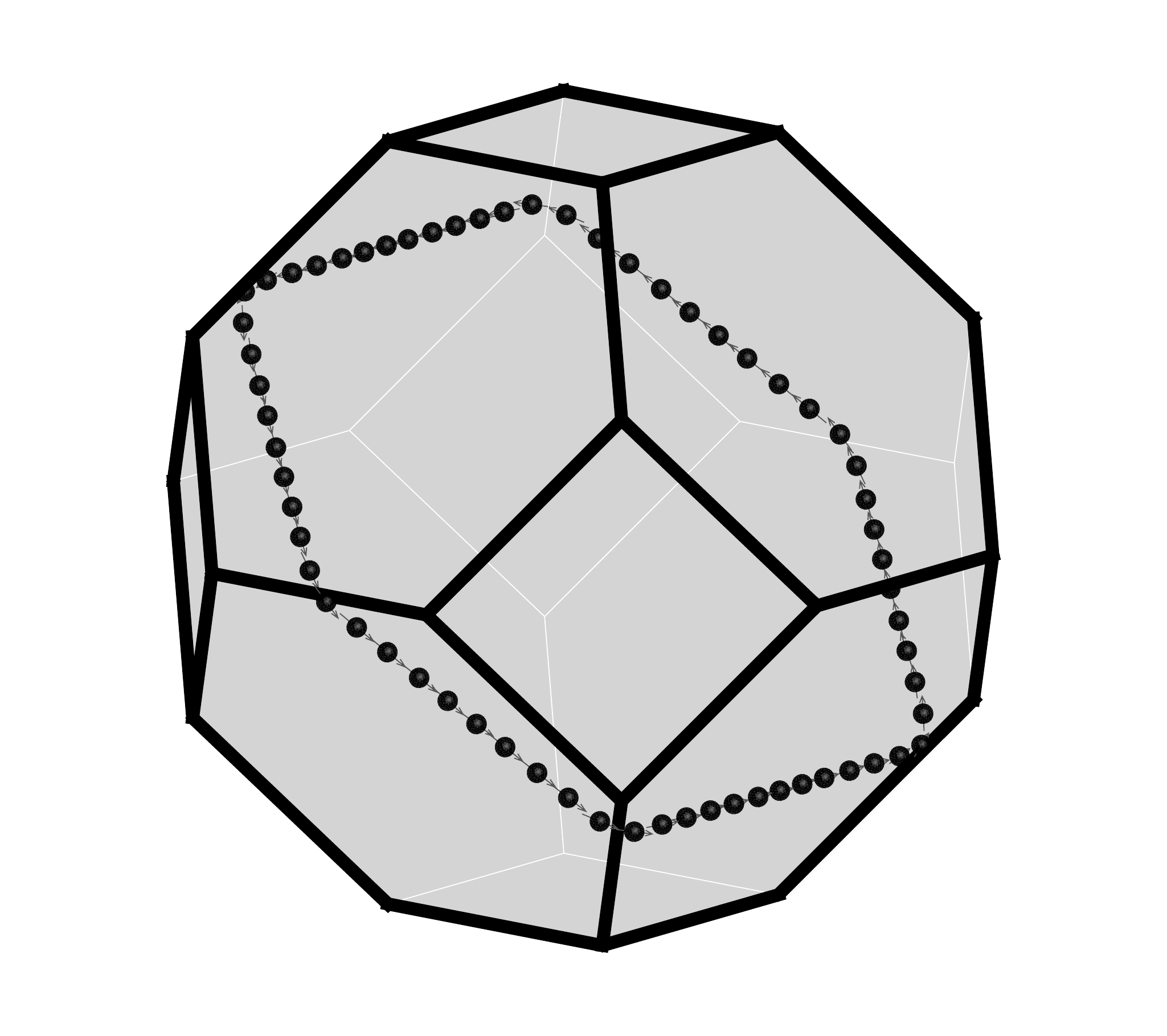}
	\end{subfigure}
\caption{\small BCC grain in elastic matrix. Snapshots of single loop nucleating from a randomly-located source and expanding under the action of an applied uniaxial stress.}
\label{VkCD6A}
\end{center}
\end{figure}

We begin by considering the simplest case of a single loop nucleated at a source on a slip plane of arbitrary locations. A sequence of snapshots of the expanding loop are shown in Fig.~\ref{VkCD6A}. The loop initially expands unimpeded and eventually arrests at the grain boundary. The example serves to illustrate how, despite the line-free character of the calculations, the monopoles nevertheless align themselves in order to attain low-energy configurations. The effectiveness of the geometrical update of the monopole line elements is also evident in the figure. In particular, the monopoles march 'head-to-toe' in order to maintain a closed-loop, hence divergence-free, configuration.

\begin{figure}
\begin{center}
	\begin{subfigure}{0.30\textwidth}\caption{}
        \includegraphics[width=0.99\linewidth]{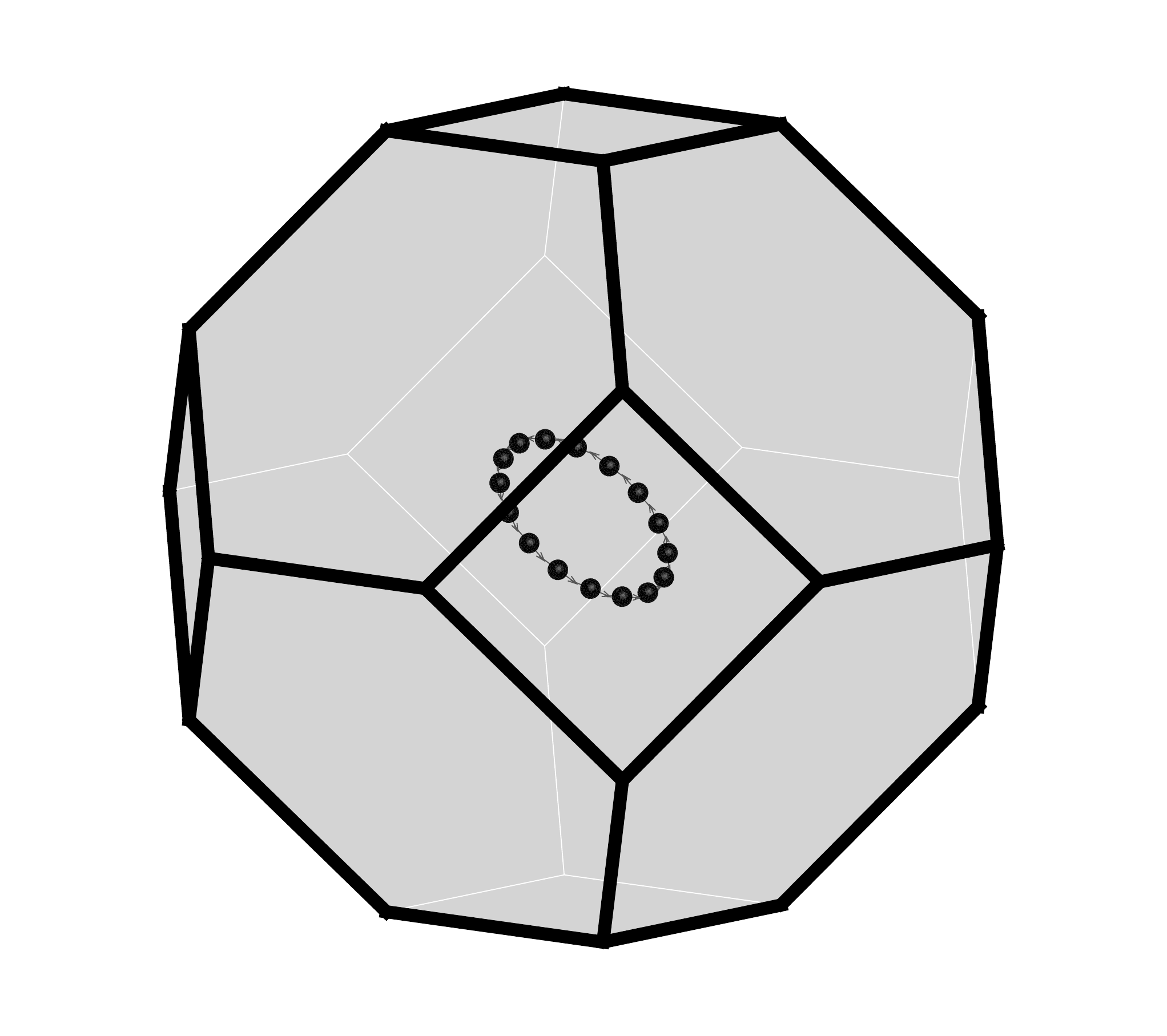}
	\end{subfigure}
	\begin{subfigure}{0.30\textwidth}\caption{}
        \includegraphics[width=0.99\linewidth]{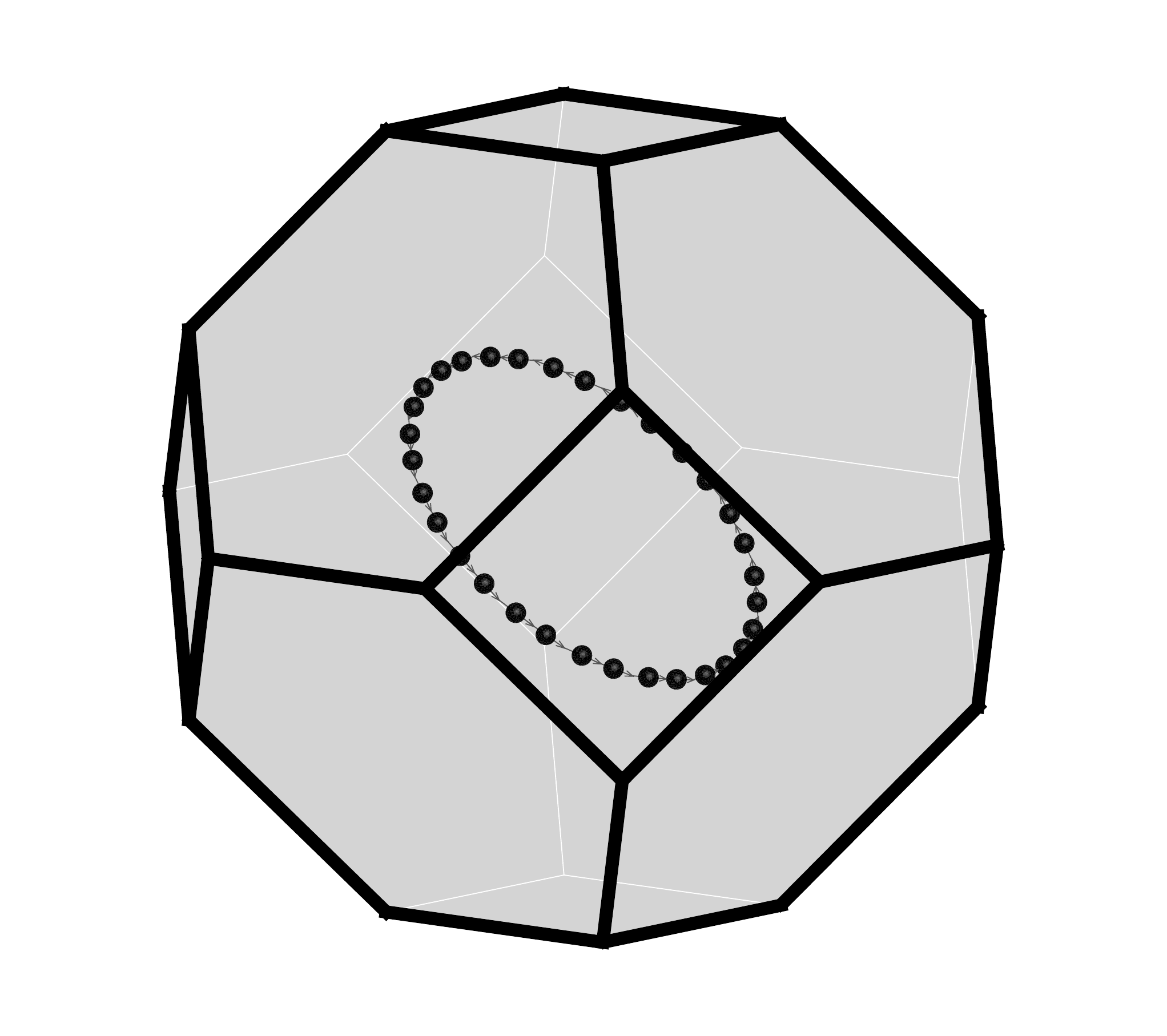}
	\end{subfigure}
	\begin{subfigure}{0.30\textwidth}\caption{}
        \includegraphics[width=0.99\linewidth]{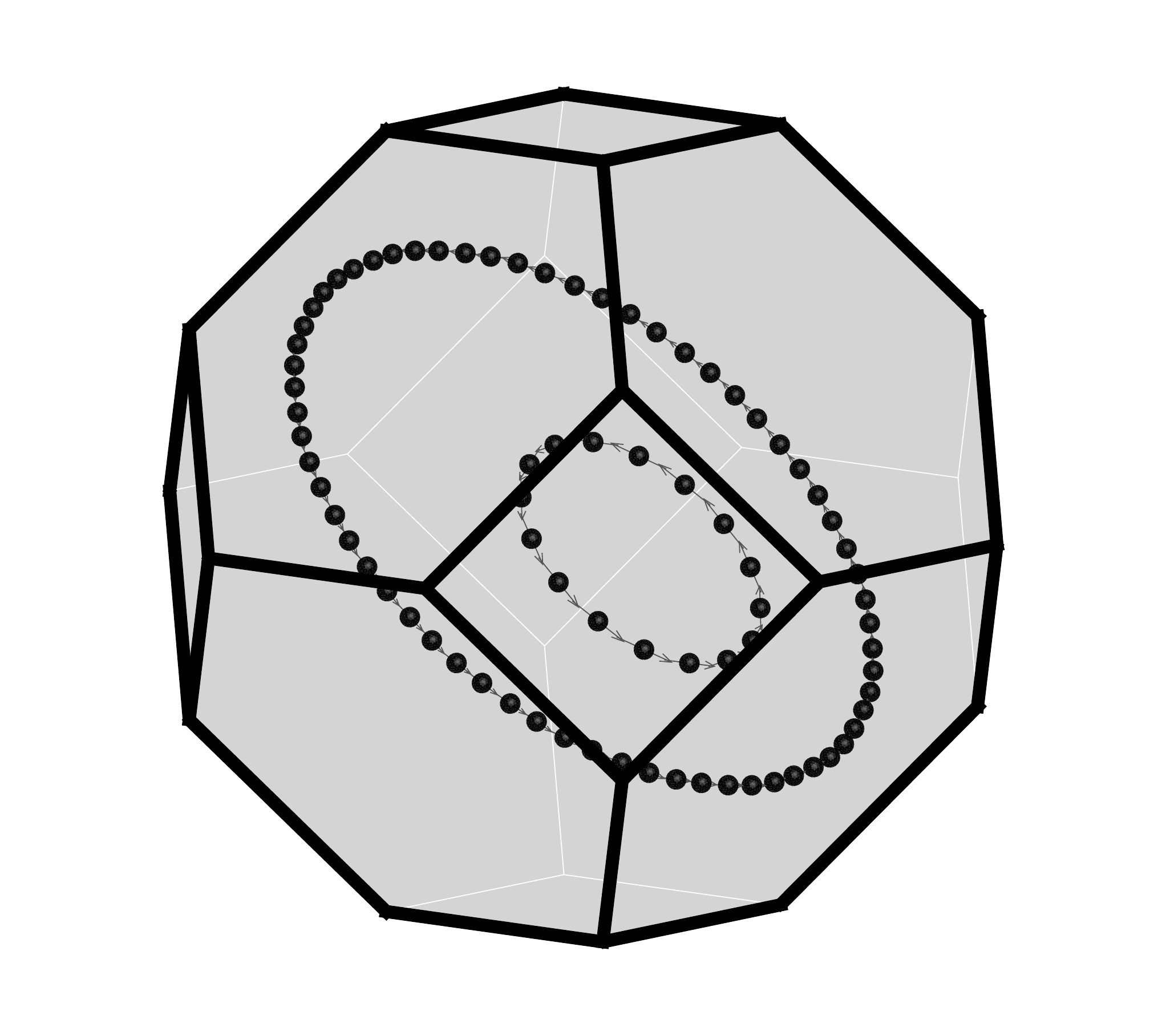}
	\end{subfigure}
	\begin{subfigure}{0.30\textwidth}\caption{}
        \includegraphics[width=0.99\linewidth]{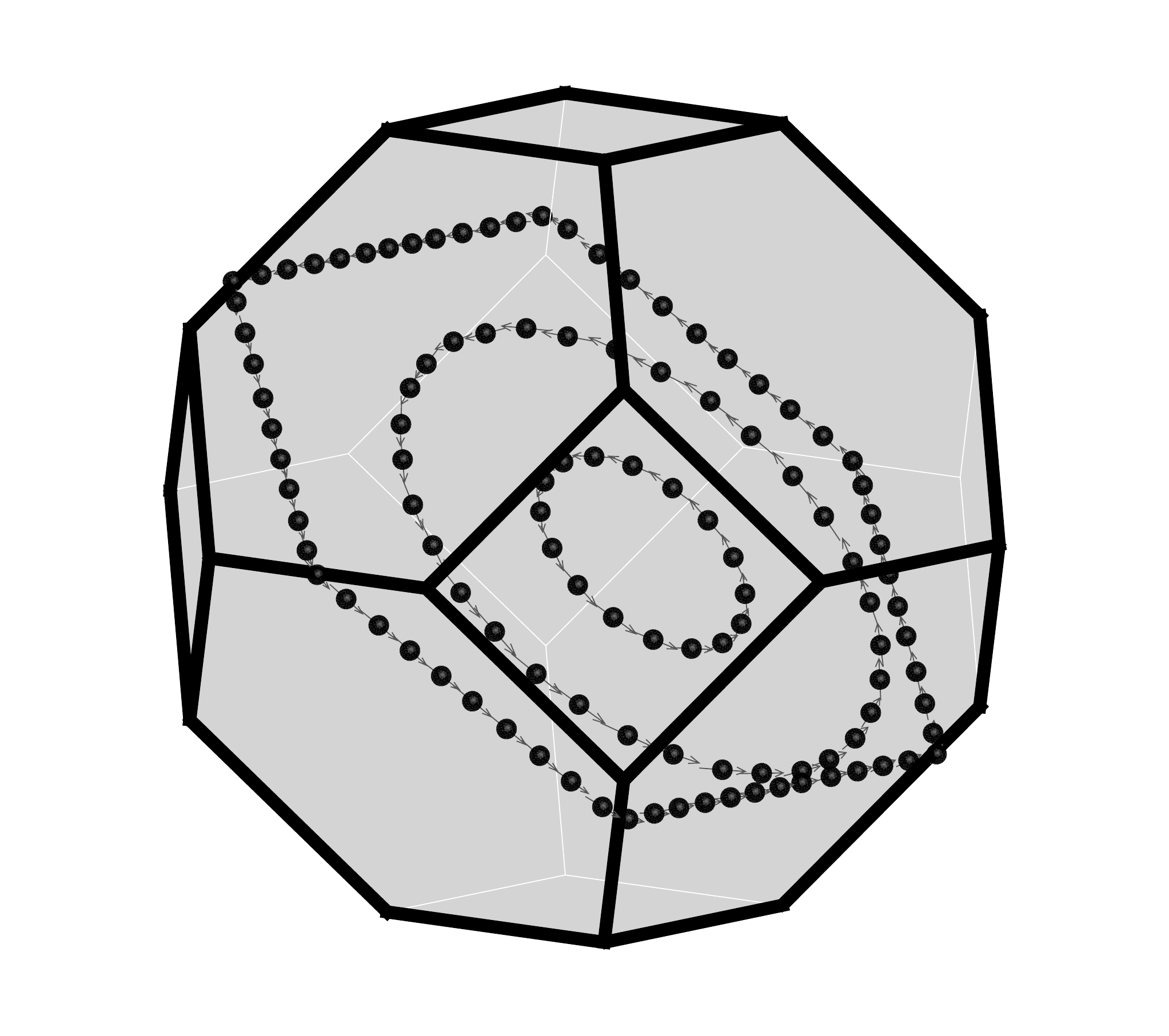}
	\end{subfigure}
	\begin{subfigure}{0.30\textwidth}\caption{}
        \includegraphics[width=0.99\linewidth]{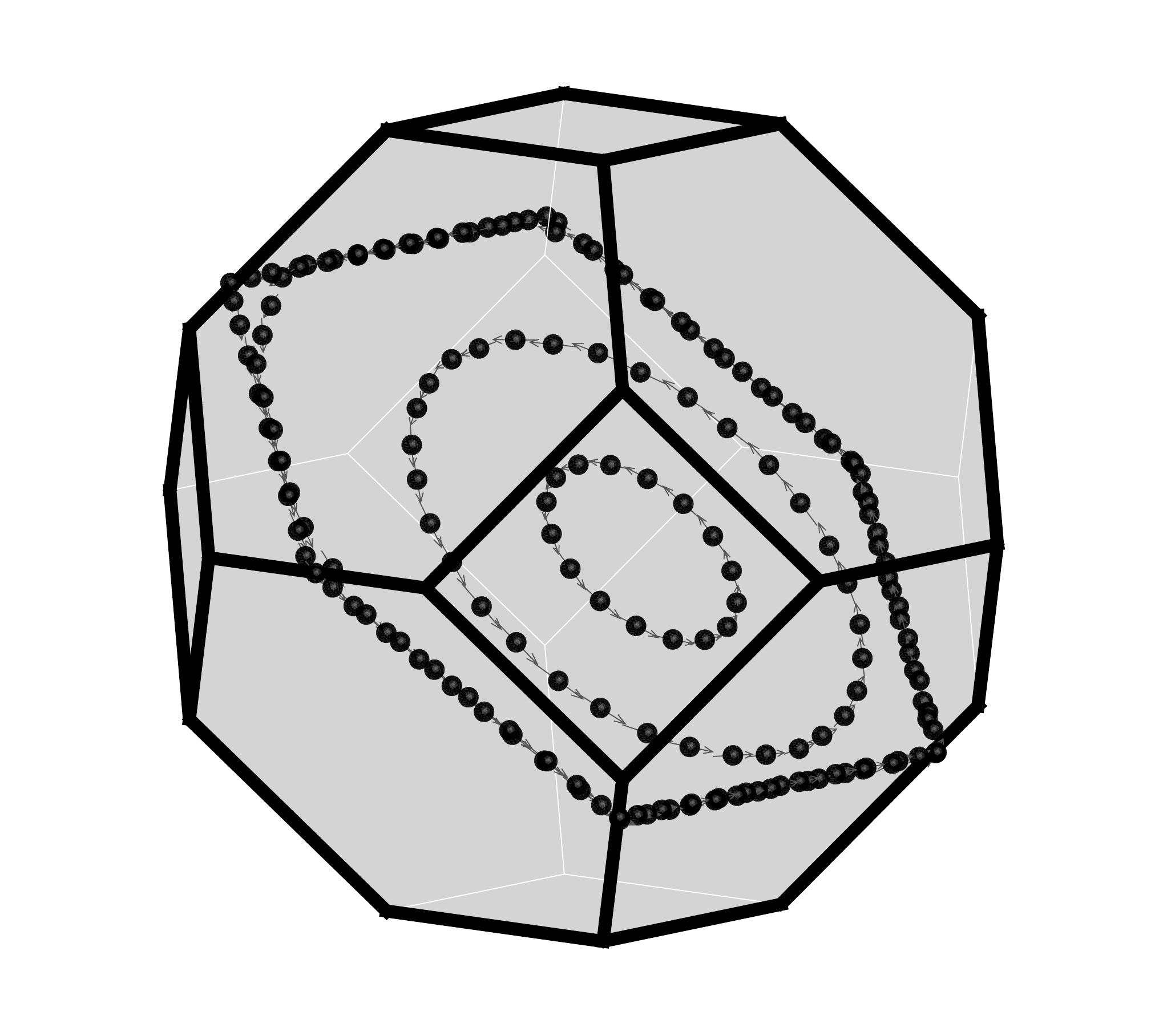}
	\end{subfigure}
	\begin{subfigure}{0.30\textwidth}\caption{}
        \includegraphics[width=0.99\linewidth]{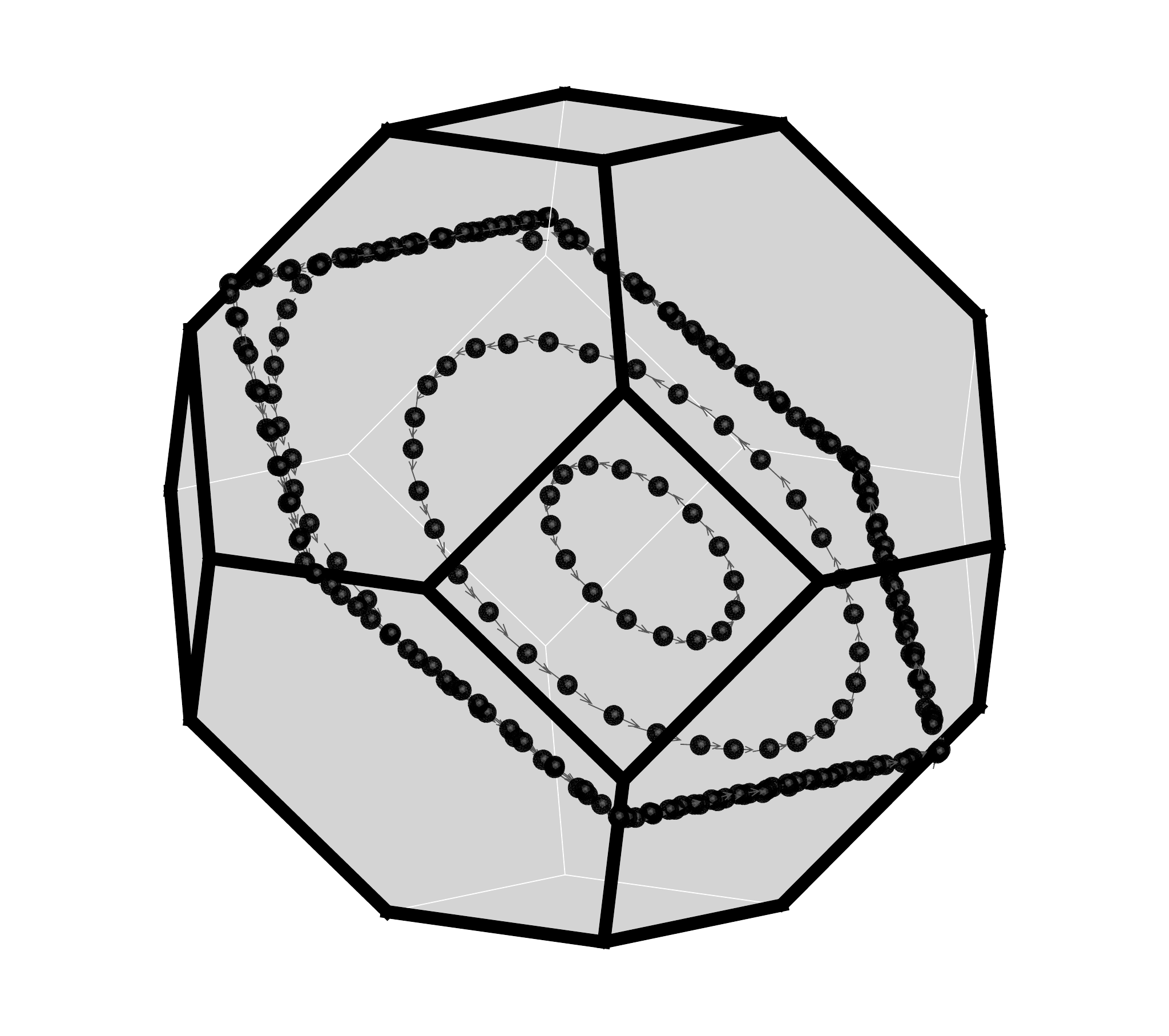}
	\end{subfigure}
\caption{\small BCC grain in elastic matrix. Snapshots of multiple loops nucleating from a common randomly-located source and expanding under the action of an applied uniaxial stress.}
\label{YCT8Ol}
\end{center}
\end{figure}

Fig.~\ref{YCT8Ol} shows a sequence of snapshots corresponding to the case in which the source is allowed to operate repeatedly, as described in Section~\ref{u7HnBd}. As may be seen from the figure, the leading loop is followed at regular intervals by trailing loops. As multiple loops are arrested at the grain boundary, they form a pile up. The example thus demonstrates capability for repeated nucleation, loop-to-loop interaction and dislocation pile-up at grain boundaries, all of which constitute important mechanisms of dislocation multiplication and interaction.

\subsection{Activation of a single slip system}

\begin{figure}
\begin{center}
	\begin{subfigure}{0.30\textwidth}\caption{}
        \includegraphics[width=0.99\linewidth]{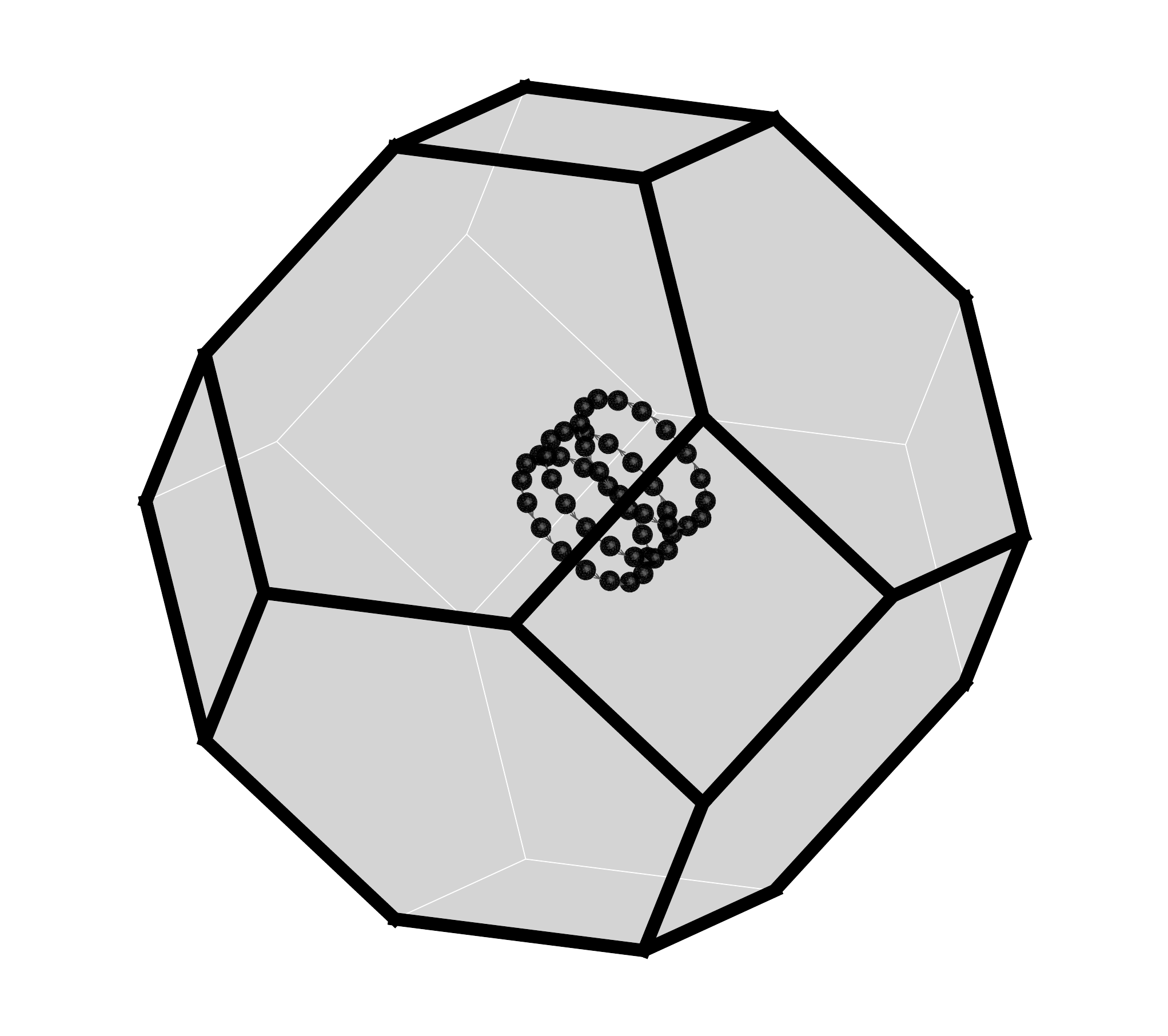}
	\end{subfigure}
	\begin{subfigure}{0.30\textwidth}\caption{}
        \includegraphics[width=0.99\linewidth]{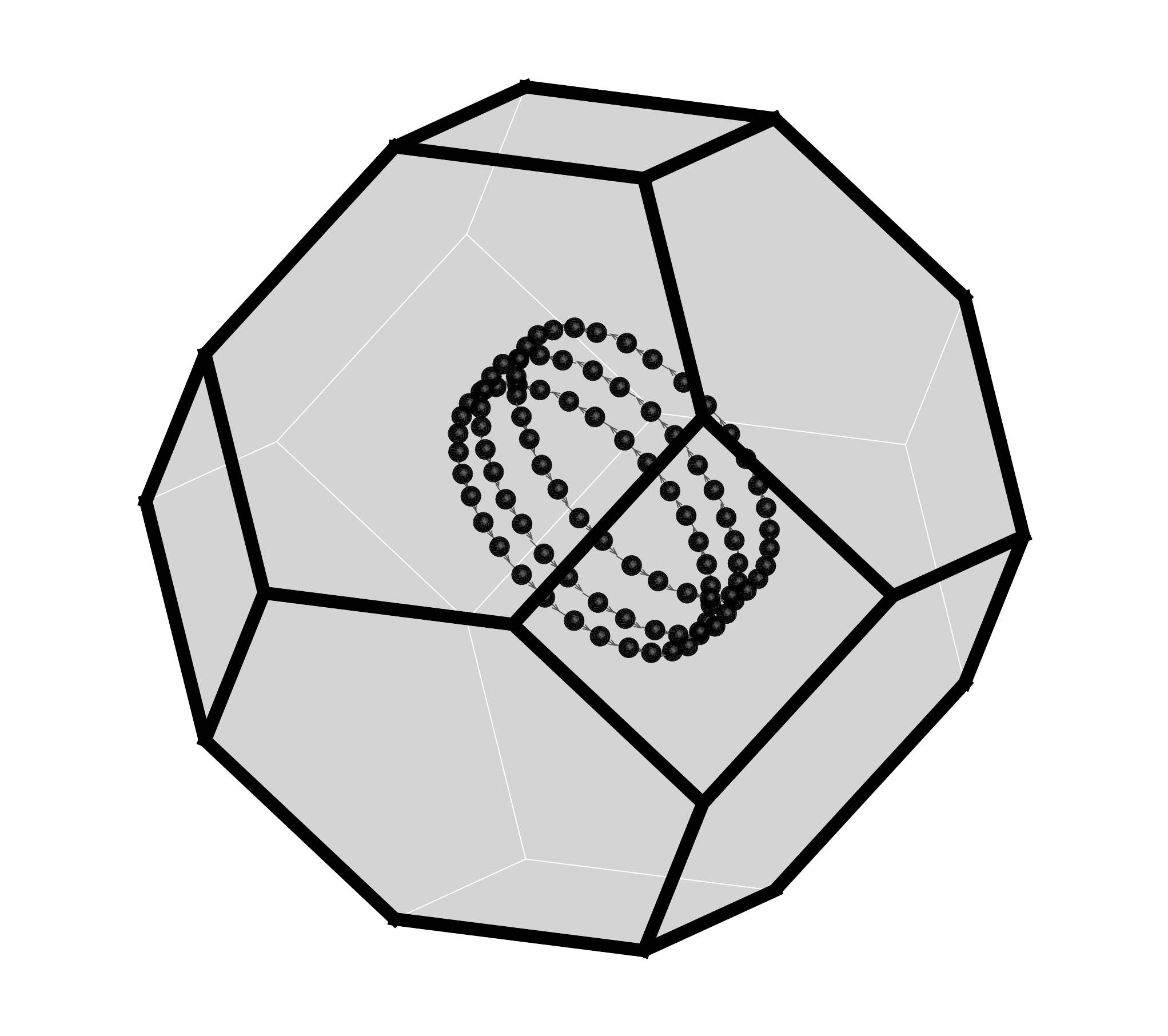}
	\end{subfigure}
	\begin{subfigure}{0.30\textwidth}\caption{}
        \includegraphics[width=0.99\linewidth]{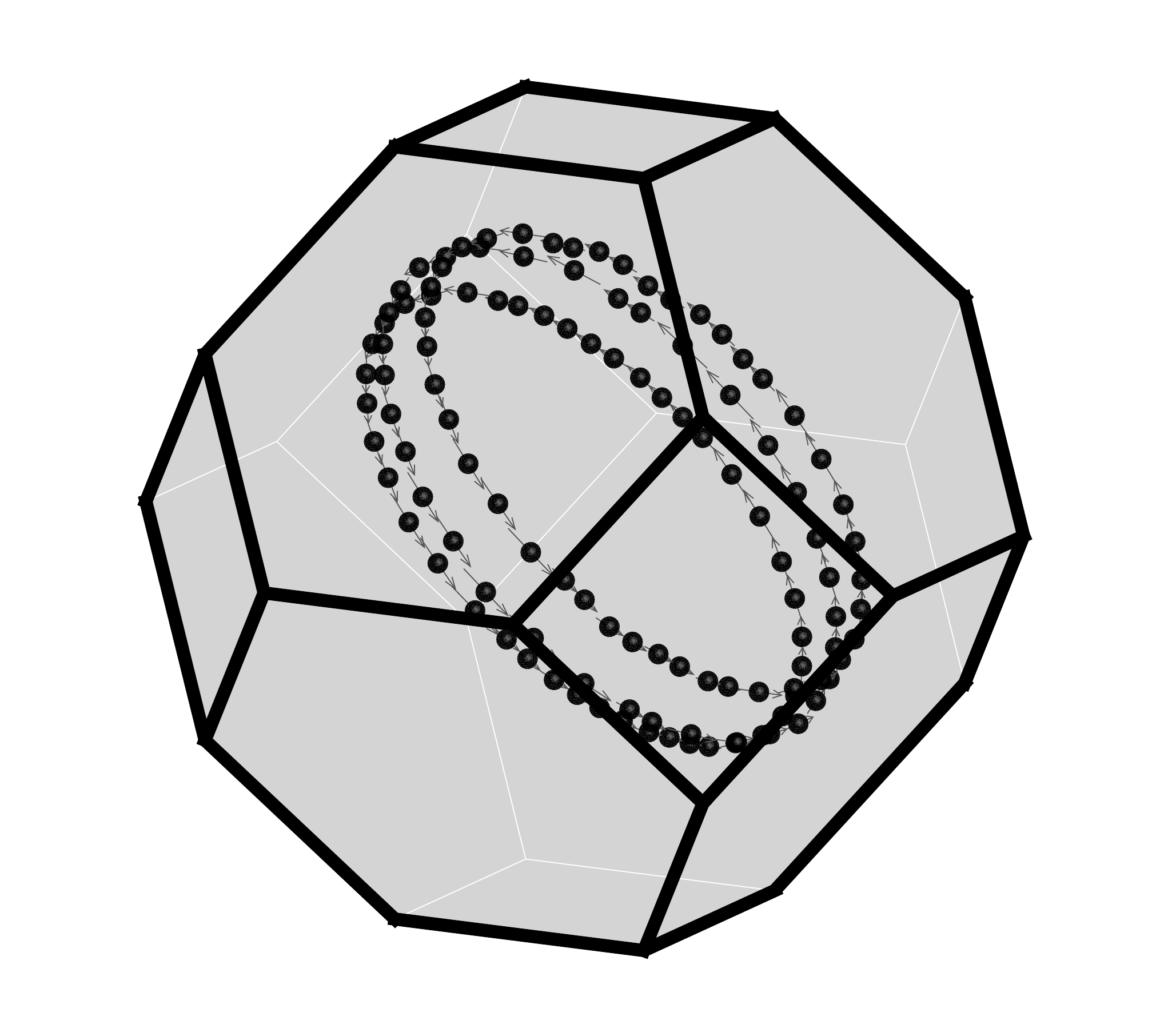}
	\end{subfigure}
	\begin{subfigure}{0.30\textwidth}\caption{}
        \includegraphics[width=0.99\linewidth]{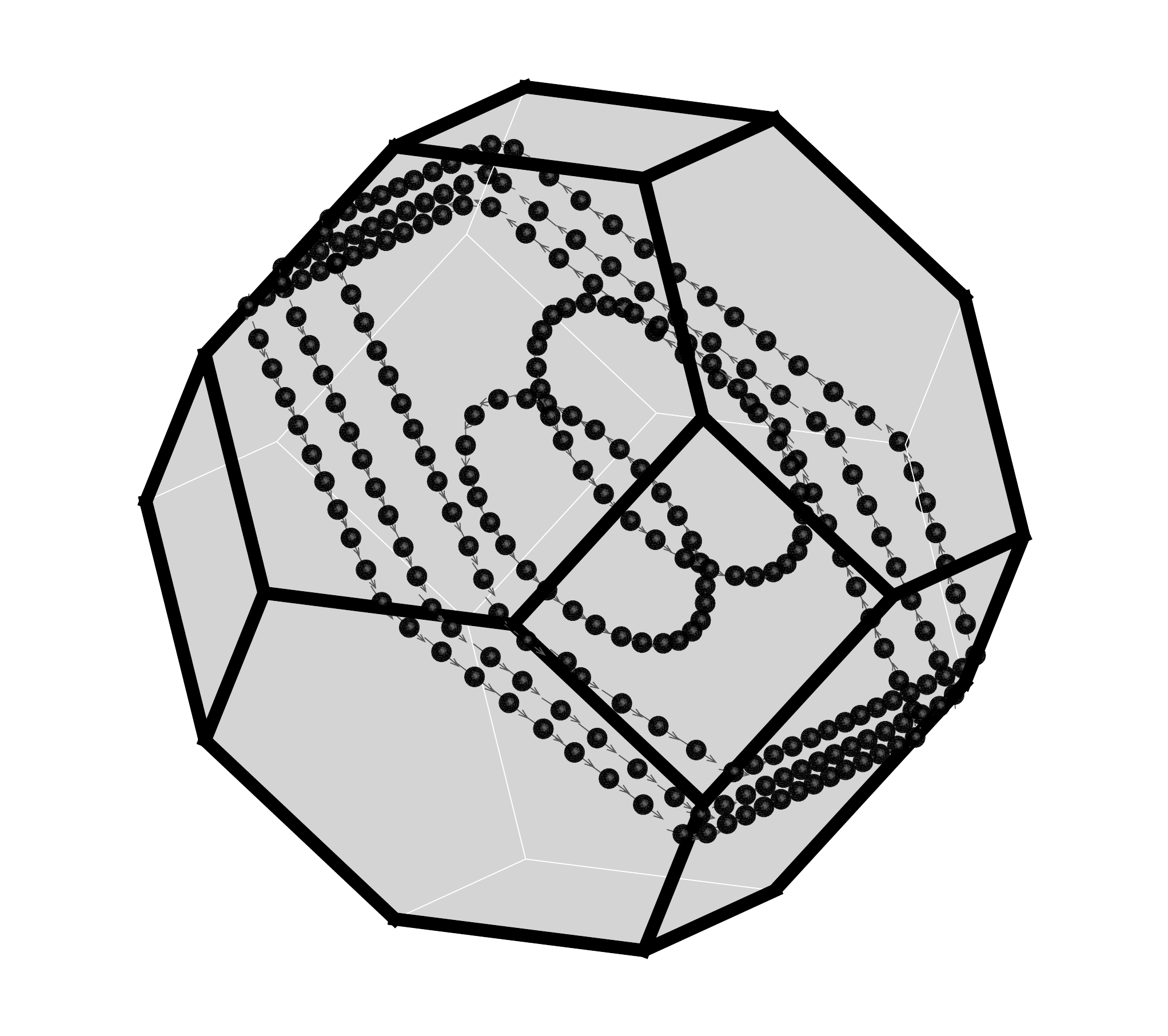}
	\end{subfigure}
	\begin{subfigure}{0.30\textwidth}\caption{}
        \includegraphics[width=0.99\linewidth]{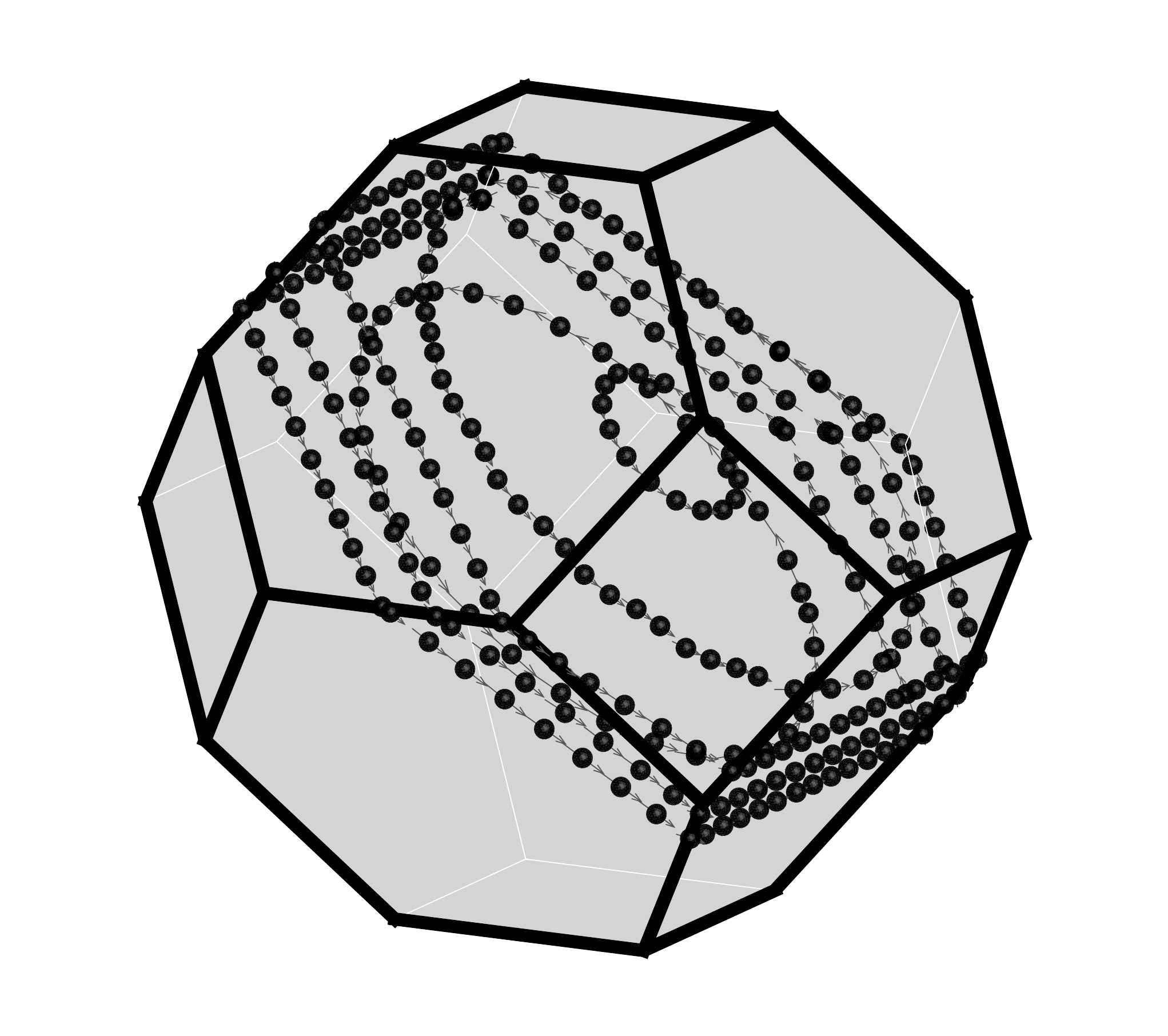}
	\end{subfigure}
	\begin{subfigure}{0.30\textwidth}\caption{}
        \includegraphics[width=0.99\linewidth]{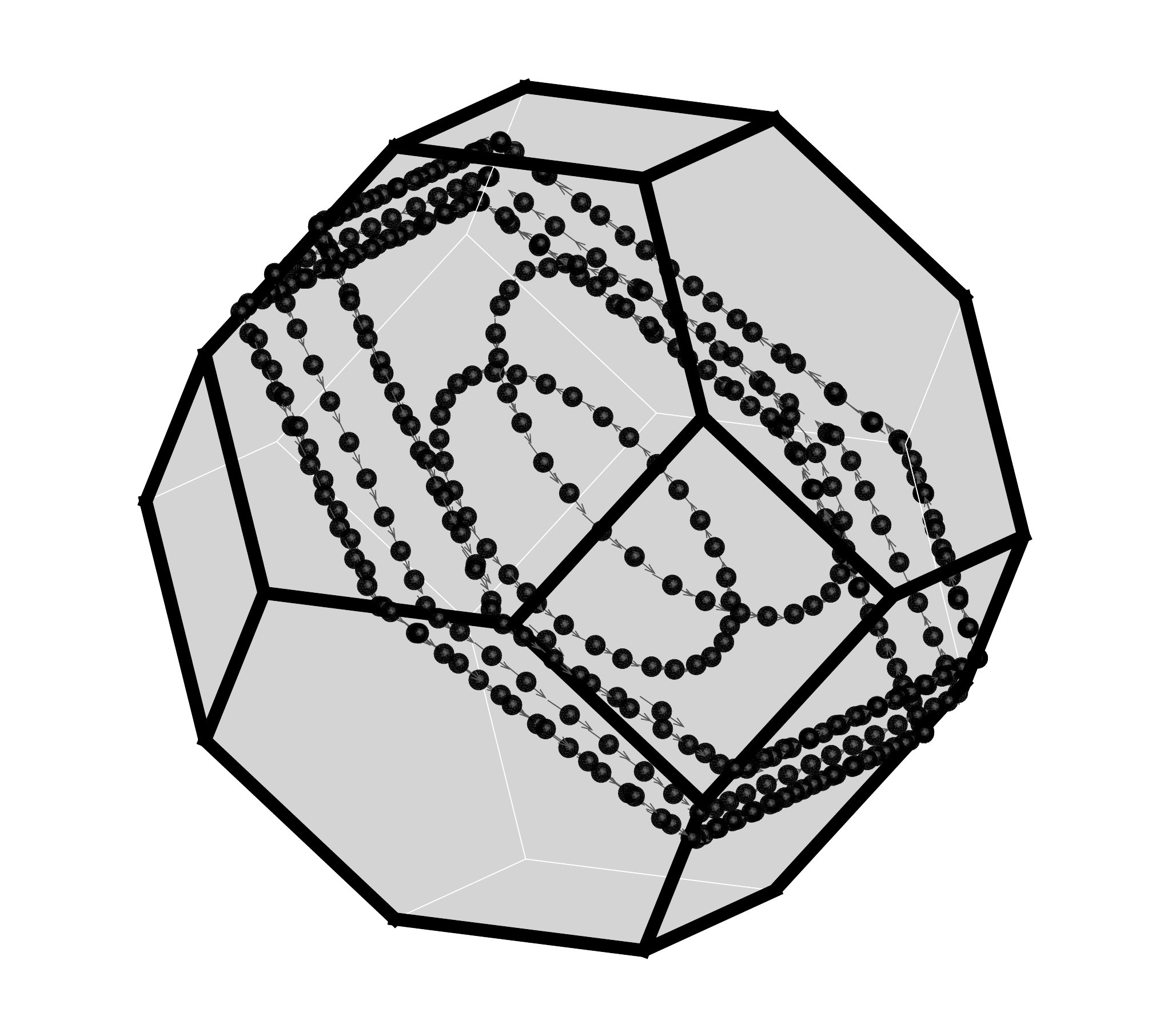}
	\end{subfigure}
\caption{\small BCC grain in elastic matrix. Snapshots of multiple loops nucleating from randomly-located sources on multiple parallel slip planes and expanding under the action of an applied uniaxial stress.}
\label{dPA1ds}
\end{center}
\end{figure}

Fig.~\ref{dPA1ds} shows a further sequence of snapshots corresponding to the case in which several slip planes in a slip system are allowed to operate simultaneously. The location of the sources and slip planes is chosen at random. As in the preceding case, the sources operate repeatedly to nucleate multiple dislocation loops that expand under the action of the applied load and eventually pile up at the boundary. The example serves to illustrate the full three-dimensional character of the formulation, which allows for coplanar dislocations as well as fully-interacting dislocations on multiple planes.

\subsection{Activation of multiple slip systems}

\begin{figure}
\begin{center}
	\begin{subfigure}{0.30\textwidth}\caption{}
        \includegraphics[width=0.99\linewidth]{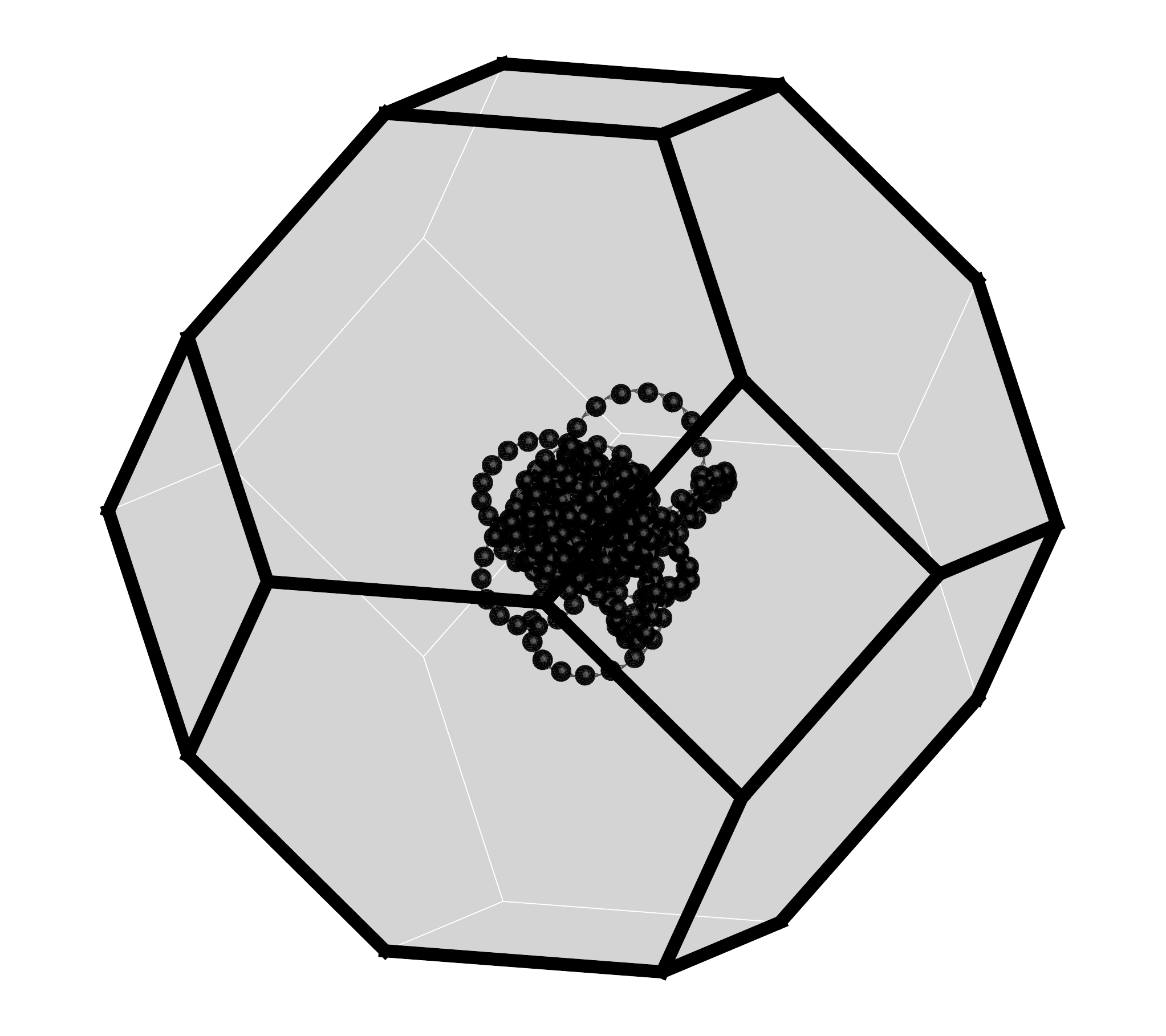}
	\end{subfigure}
	\begin{subfigure}{0.30\textwidth}\caption{}
        \includegraphics[width=0.99\linewidth]{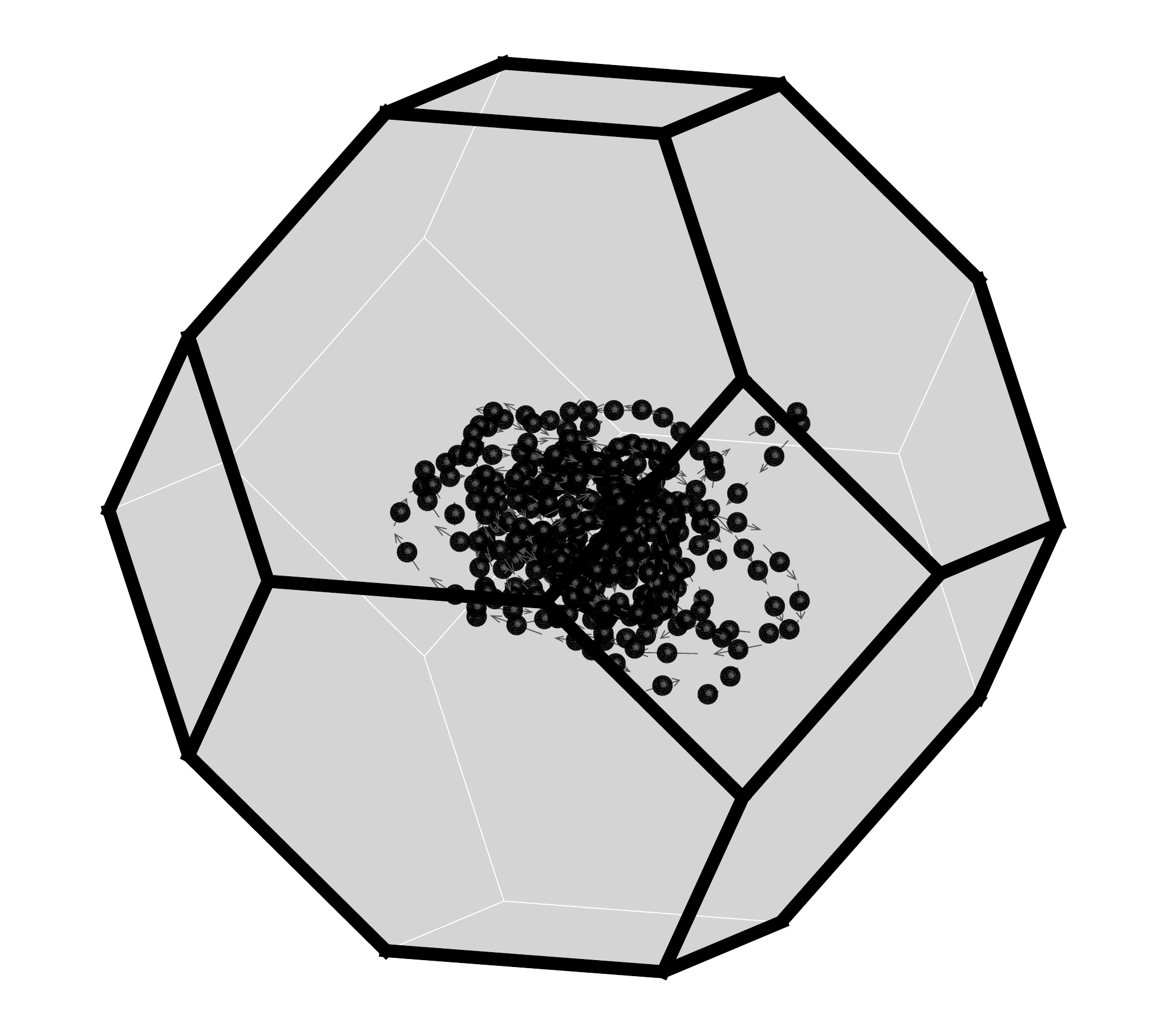}
	\end{subfigure}
	\begin{subfigure}{0.30\textwidth}\caption{}
        \includegraphics[width=0.99\linewidth]{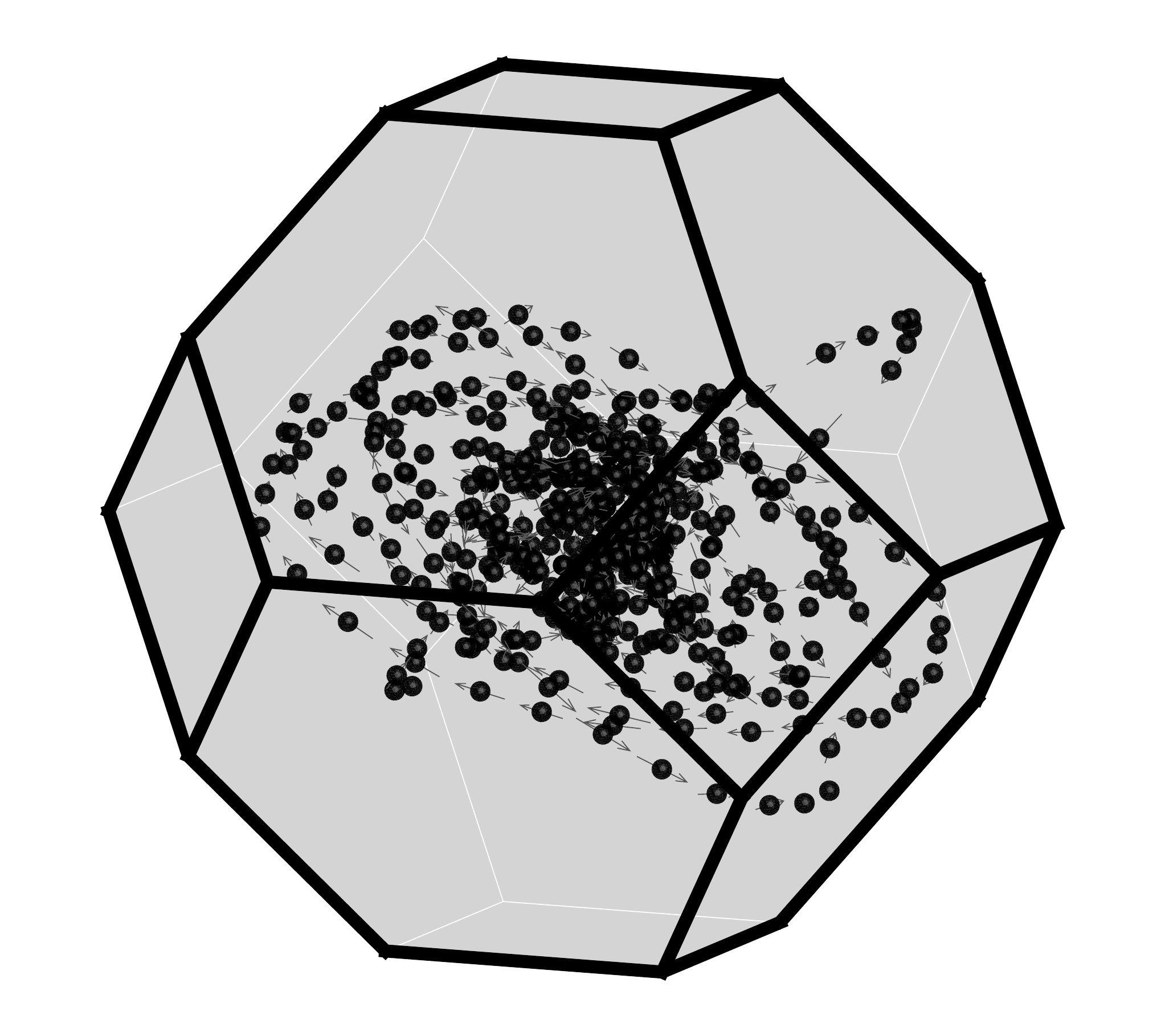}
	\end{subfigure}
	\begin{subfigure}{0.30\textwidth}\caption{}
        \includegraphics[width=0.99\linewidth]{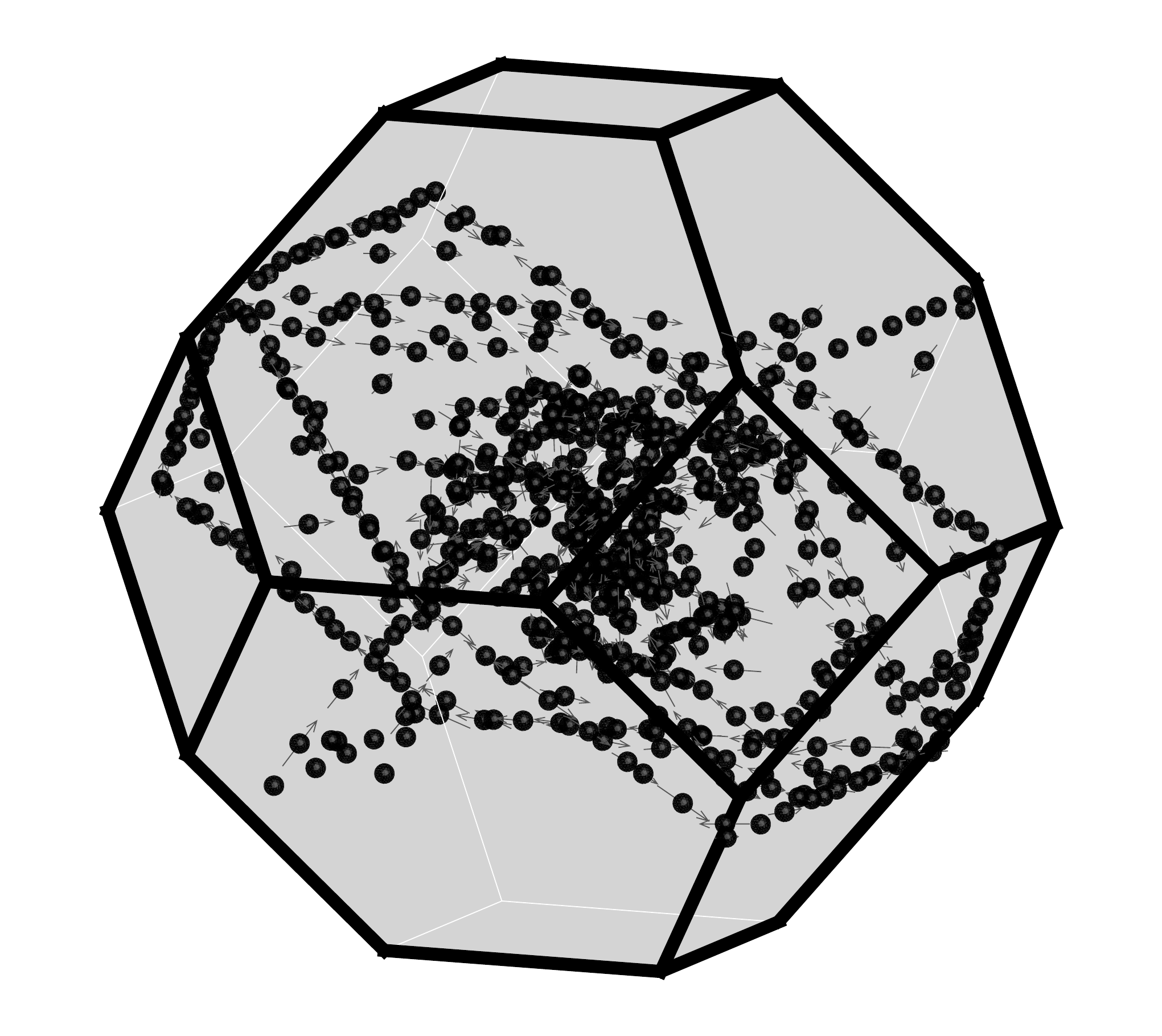}
	\end{subfigure}
	\begin{subfigure}{0.30\textwidth}\caption{}
        \includegraphics[width=0.99\linewidth]{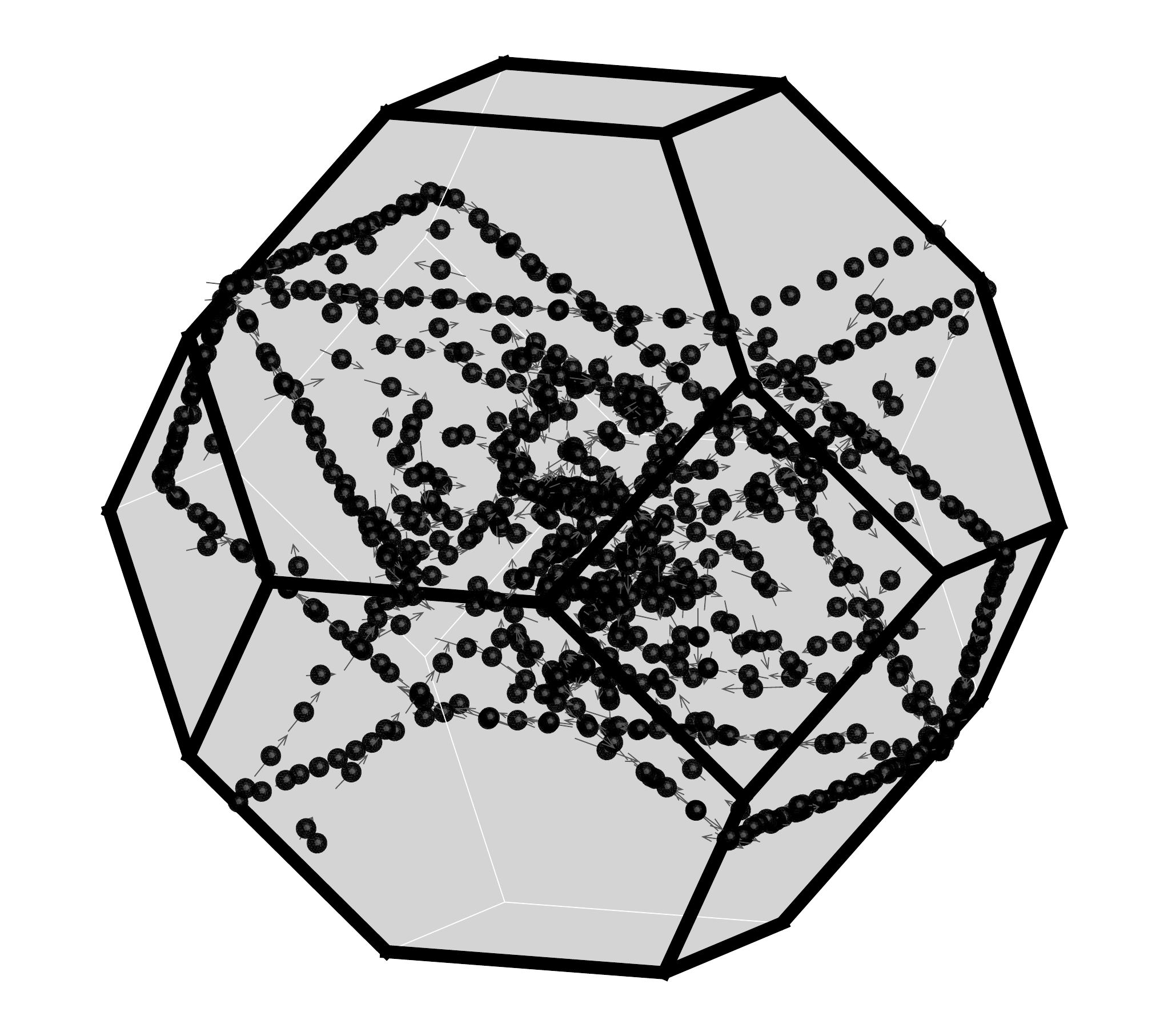}
	\end{subfigure}
	\begin{subfigure}{0.30\textwidth}\caption{}
        \includegraphics[width=0.99\linewidth]{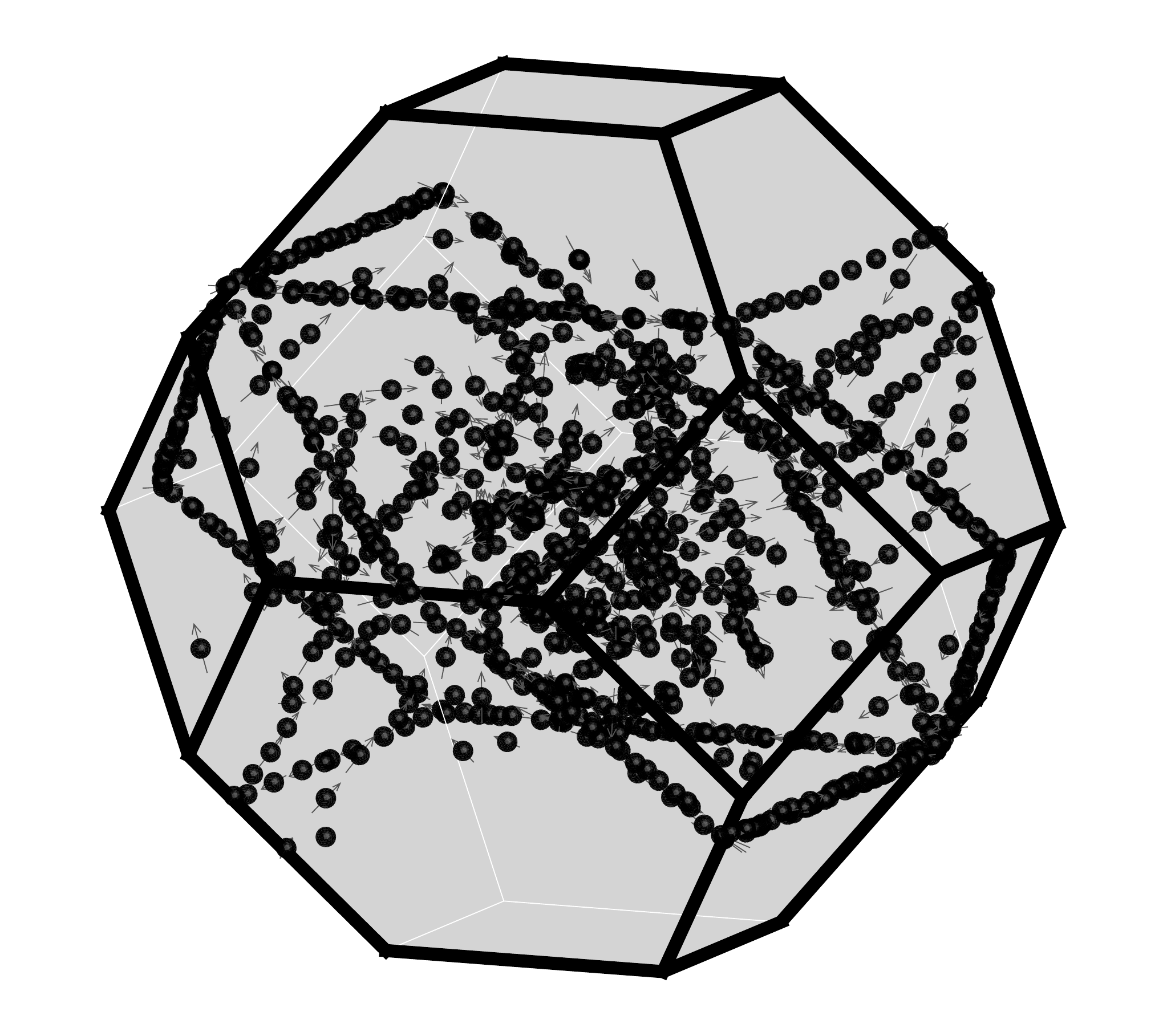}
	\end{subfigure}
\caption{\small BCC grain in elastic matrix. Snapshots of multiple loops nucleating from randomly-located sources on multiple slip planes and expanding under the action of an applied uniaxial stress.}
\label{3fOOTm}
\end{center}
\end{figure}

Fig.~\ref{3fOOTm} finally displays the complex evolution of the dislocation ensemble that ensues when multiple sources, slip planes and slip systems are allowed to operate simultaneously. In particular, the sequence of snapshots shown in the figure illustrate the ability of loops in different slip planes to interact at close range, cross each other or form structures. The robust ability of the method to account for---and negotiate---the complex dislocation interactions and evolutions exemplified by the example is noteworthy and bodes well for the general application of the method to a broad range of applications.

\section{Summary and conclusions}

We have developed an approximation scheme for three-dimensional dislocation dynamics in which the dislocation line density is concentrated at points, or {\sl monopoles}. Every monopole carries a Burgers vector and an element of line. Since monopoles are Dirac masses, the monopole representation requires an extension of the classical dislocation transport problem (cf., e.~g., \cite{mura:1987}), which is restricted to 'continuously distributed dislocations', to general measures. This extension requires: i) expressing the transport equations (\ref{3kgrKx}) in weak form, eq.~(\ref{09ZVMn}); ii) expressing the elastic energy in terms of Airy stress potentials, eq.~(\ref{cHiaj8}); and iii) regularizing the elastic energy, eq.~(\ref{tRi9gL}). By virtue of these manipulations, the dislocation density appears {\sl linearly} in all expressions, which thus make sense for general measures. In addition, the regularization of the elastic energy eliminates the logarithmic divergence of Volterra dislocations and assigns finite energies and Peach-K\"ohler forces to general dislocation measures, including monopoles.

Following concepts from optimal transportation theory (cf., e.~g., \cite{Villani:2003}), we discretize the dislocation transport problem in time by introducing incremental transport maps. These maps push forward the dislocation density from one configuration to the next, eq.~(\ref{3rouPi}). For dislocation densities in the form of monopoles, this push-forward operation takes a particularly simple form, eq.~(\ref{0LetiE}): i) the Burgers vectors of the monopoles remain constant; and ii) the elements of line of the monopoles are updated according to the local gradient of the incremental transport map. It bears emphasis that these operations are geometrically exact to within the interpolation accuracy of the transport map. In particular, the geometrical update preserves the null-divergence constraint and results in line stretching (resp. shortening) for divergent (resp. convergent) geometries, cf.~Fig.~\ref{qSJ46d}. Further adapting concepts from optimal transportation theory and, in particular, from the pioneering work of Jordan, Kinderlehrer and Otto \cite{JKO:1998, OTM:2010, FPO:2017}, we formulate an incremental minimum principle for the transport map, eq.~(\ref{T4iUsw}), that encodes the energetics and mobility kinetics of the system. In particular, the motion of the monopoles is the result of a competition between energy, which drives the monopoles to low-energy configurations, and mobility, which opposes motion. Finally, the requisite interpolation of the transport maps is effected by means of mesh-free max-ent interpolation \cite{ArroyoOrtiz:2006}.

A distinguishing attribute of the proposed method of monopoles relative to traditional approximation schemes based on segments is that an explicit linear connectivity, or 'sequence', between the monopoles need not be defined. In this sense, the method is 'line-free'. The satisfaction of the requisite null-divergence constraint is ensured by the geometric exactness of the incremental updates, eq.~(\ref{0LetiE}). In addition, the monopoles tend to align 'head-to-toe' spontaneously in order to minimize the elastic energy. The result, which is clearly evident in the numerical examples, is that, while not explicitly enforced, the monopole ensemble approximates a collection of lines at all times. The examples also attest to the remarkable robustness of the method and, in particular, to its ability to negotiate complex dislocation dynamics including nucleation, close-range interactions, pileups, intersections and other mechanisms.

We close by remarking that the present work has been primarily concerned with the mathematical framework, implementation and numerical testing of the proposed method of monopoles. In particular, we have not attempted to model specific material systems or make quantitative predictions thereof with any degree of physical fidelity. There is extraordinarily extensive experience in applying dislocation dynamics to the elucidation of a vast array of physical phenomena that we believe can be combined with the proposed method of monopoles to great effect. In particular, as already noted, we believe that the method provides an effective avenue for extending to three dimensions the wealth of point-dislocation methods that have been developed and extensively applied in two dimensions, starting with the seminal paper of Lubarda, Blume and Needleman \cite{lubarda:1993d}. These connections and extensions suggest themselves as worthwhile directions of future research.

\section*{Acknowledgements}

AD and MO are grateful for support from the Stanback STEM program in Aerospace at Caltech. MPA is grateful for support from the
Ministerio de Economía y Competitividad of Spain under grant number DPI2015-66534-R.

\begin{appendix}

\section{Max-ent interpolation}\label{riePr8}

The zeroth-order consistent max-ent shape functions at $\mbs{x}$ are the solutions of the constrained optimization problem \cite{ArroyoOrtiz:2006}
\begin{subequations}\label{Wr5aYo}
\begin{align}
    &
    \text{Minimize:}
    \quad
    \sum_{a=1}^M \beta_a N_a(\mbs{x})  \vert \mbs{x}- \mbs{x}_a \vert^2
    +
    \sum_{a=1}^M N_a(\mbs{x}) \log N_a(\mbs{x}) ,
    \\ &
    \text{subject to:}
    \quad N_a(\mbs{x}) \geq 0, \ a=1,\dots,M,
    \qquad \sum_{a=1}^M N_a(\mbs{x}) = 1 .
\end{align}
\end{subequations}
where $\{\mbs{x}_a\}_{a=1}^M$ are the nodes of the interpolation and $\{\beta_a\}_{a=1}^M$ are adjustable parameters. The shape functions thus defined supply the least biased and most local reconstruction of a function whose values are known on the node set \cite{ArroyoOrtiz:2006}. Problem (\ref{Wr5aYo}) can be solved explicitly, with the result
\begin{equation}\label{b4lEfR}
    N_{a}(\mbs{x})
    =
    \frac{1}{Z}
    \exp\left(-\frac{\beta_a}{2} |\mbs{x}-\mbs{x}_{a}|^2\right) ,
\end{equation}
where
\begin{equation}\label{7touwI}
    Z
    =
    \sum_{a=1}^M
        \exp\left(-\frac{\beta_a}{2} |\mbs{x}-\mbs{x}_{a}|^2\right)
\end{equation}
is the partition function. Suppose that the nodes moves to new positions $\{\mbs{y}_a\}_{a=1}^M$. We then define an interpolated transport map as
\begin{equation}
    \mbs{\varphi}(\mbs{x})
    =
    \mbs{x}
    +
    \sum_{a=1}^M (\mbs{y}_{a}-\mbs{x}_{a}) N_{a}(\mbs{x}) ,
\end{equation}
with gradient
\begin{equation}
    \nabla \mbs{\varphi}(\mbs{x})
    =
    \mbs{I}
    +
    \sum_{a=1}^M (\mbs{y}_{a}-\mbs{x}_{a}) \nabla N_{a}(\mbs{x}) .
\end{equation}
Suppose that $\mbs{y}_{a} = \mbs{x}_{a}+\mbs{u}$, i.~e., the nodal set translates by $\mbs{u}$. From the zeroth-order condition, we find
\begin{equation}
    \mbs{\varphi}(\mbs{x})
    =
    \mbs{x}
    +
    \left(\sum_{a=1}^M N_{a}(\mbs{x})\right) \mbs{u}
    =
    \mbs{x} + \mbs{u} ,
\end{equation}
and
\begin{equation}
    \nabla \mbs{\varphi}(\mbs{x})
    =
    \mbs{I}
    +
    \left(\sum_{a=1}^M \nabla N_{a}(\mbs{x})\right) \mbs{u}
    =
    \mbs{I} ,
\end{equation}
as required.

\end{appendix}

\bibliography{biblio}
\bibliographystyle{unsrt}

\end{document}